\documentclass[aps,rmp,reprint,amsmath,amssymb,groupaddress,longbibliography]{revtex4-1}

\usepackage{graphicx}
\usepackage{dcolumn}
\usepackage{bm}
\usepackage{natbib}
\usepackage[hidelinks]{hyperref}
\usepackage{placeins}  

\usepackage{color}
\definecolor{orange}{rgb}{1,0.5,0}
\definecolor{violett}{rgb}{0.8,0.,0.7}
\newcommand{\REV}[1]{{\color{black} #1}}

\newcommand{\REVV}[1]{{\color{black} #1}}
\usepackage[normalem]{ulem}  

\graphicspath{{../Figs_small/}{./}}

\begin{document}

\title[Physics of Microswimmers]{Physics of Microswimmers --
         Single Particle Motion and Collective Behavior}

\author{Jens Elgeti, Roland G. Winkler, Gerhard Gompper}

 \altaffiliation[]{Theoretical Soft Matter and Biophysics, Institute of Complex
Systems and Institute for Advanced Simulation,
Forschungszentrum J\"{u}lich, D-52425 J\"{u}lich, Germany}

\date{\today}


\begin{abstract}
Locomotion and transport of microorganisms in fluids is an essential aspect of life.
Search for food, orientation toward light, spreading of off-spring, and the
formation of colonies are only possible due to locomotion.  Swimming at the
microscale occurs at low Reynolds numbers, where fluid friction and viscosity
dominates over inertia. Here, evolution achieved propulsion mechanisms, which
overcome and even exploit drag. Prominent propulsion mechanisms are rotating
helical flagella, exploited by many bacteria, and snake-like or whip-like motion
of eukaryotic flagella, utilized by sperm and algae. For artificial microswimmers,
alternative concepts to convert chemical energy or heat into directed motion
can be employed, which are potentially more efficient. The dynamics of
microswimmers comprises many facets, which are all required to achieve locomotion.
In this article, we review the physics of locomotion of biological and synthetic
microswimmers, and the collective behavior of their assemblies.  Starting from
individual microswimmers, we describe the various propulsion mechanism of
biological and synthetic systems and address the hydrodynamic aspects of
swimming. This comprises synchronization and the concerted beating of flagella
and cilia.  In addition, the swimming behavior next to surfaces is examined.
Finally, collective and cooperate phenomena of various types of isotropic and
anisotropic swimmers with and without hydrodynamic interactions are discussed.

\end{abstract}

\maketitle

\tableofcontents


\section{Introduction}
\label{sec:intro}

Cell motility is a major achievement of biological evolution and
is essential for a wide spectrum of cellular activities.
Microorganisms, such as spermatozoa, bacteria, protozoa, and algae,
use flagella---whip-like structures protruding from their
bodies---for their propulsion. Swimming of uni- and multi-cellular
organisms is essential for their search for food (chemotaxis), the
reaction to light (phototaxis), and the orientation in the
gravitation field (gravitaxis). Furthermore, flagellar motion
plays a major role in higher organisms, where they transport fluid in
the respiratory system in form of cilia, are involved in cellular
communications, and even determine the morphological left-right
asymmetry in the embryo.

Unicellular swimmers, e.g., bacteria like {\em Escherichia coli},
spermatozoa, and {\em Paramecia} are typically
of a few to several ten micrometers in size. The physics ruling the
swimming on this micrometer scale is very different from that
applying to swimming in the macro-world. Swimming at the
micrometer scale is swimming at low Reynolds numbers
\cite{purc77}, where viscous damping by far dominates over
inertia. Hence, swimming concepts of the high Reynolds-number
macro-world are ineffective on small scales. In the evolutionary
process, microorganisms acquired propulsion strategies, which
successfully overcome and even exploit viscous drag.

The design of artificial nano- and microswimmers is highly
desirable to perform a multitude of tasks in technical and medical
applications. Two general design strategies are currently
followed, each posing particular challenges. First, successful
concepts realized in nature can be adopted or underlying
principles and mechanisms can be exploited. Second, novel
construction principles can be invented, which are simpler but
potentially more efficient, or plainly more practical from an
engineering perspective. Major obstacles in such an endeavor
are the availability of sustainable energy sources for artificial
microswimmers, and physical concepts for efficient energy
conversion into a propulsive force. Another issue is the active
control of artificial microswimmers, such that they perform tasks
or respond to external stimuli. The design and fabrication of a
synthetic swimmer with such features would be extremely valuable
in a diversity of fields like medicine, biology, material science,
and environmental science. Such machines might transport cargo,
e.g., in medicine or microfluidic chips, conduct operations in
cells, remove toxic materials from human bodies or toxic water
streams, or actively control material behavior as, e.g.,
viscoelastic properties.

Microswimmers hardly ever swim alone. Sperm cells are released by
the millions to compete in the run for the egg. Bacteria grow by
dividing and invading their surroundings together. Artificial
microswimmers will only be able to deliver useful quantities of
pharmaceuticals  or modify material properties when present in
large numbers. Indeed, in assemblies of motile microorganisms,
cooperativity reaches a new level of complexity as they exhibit
highly organized movements with remarkable large-scale patterns
such as networks, complex vortices, or swarms.

In this article, we review the physics of locomotion of biological and
synthetic microswimmers, and the emergent collective behavior of their
assemblies. Several previous review articles concerning microswimmers
have addressed different aspects of their motility and collective
behavior.
Generic aspects of the emergent large-scale behavior
of self-propelled particles and active soft matter have been reviewed
by \citet{tone05b}, \citet{Ramaswamy2010}, \citet{vics12}, \citet{marc:13},
and \citet{sain13}.
The hydrodynamics of swimming has been reviewed by \citet{laug09},
\citet{ishi09}, \citet{koch:11}, and \citet{gole:11}.
Aspects of bacterial motility have been discussed by
\citet{hars:03} and \citet{cate12}. Sperm motility and chemotaxis
has been reviewed by \citet{alva14}.
The dynamical properties of active
Brownian particles have been discussed by \citet{roma:12}, with emphasize
on the stochastic dynamics in the framework of statistical
physics.
The propulsion of synthetic swimmer on the nanoscale and the
development of nanomachines have been addressed by \citet{ozin05},
\citet{seng12}, and \citet{ebbens2010}.

\subsection{Biological Microswimmers} \label{sec:bio-swimmers}

\begin{figure}[h]
\centering
\includegraphics[width=0.48\textwidth]{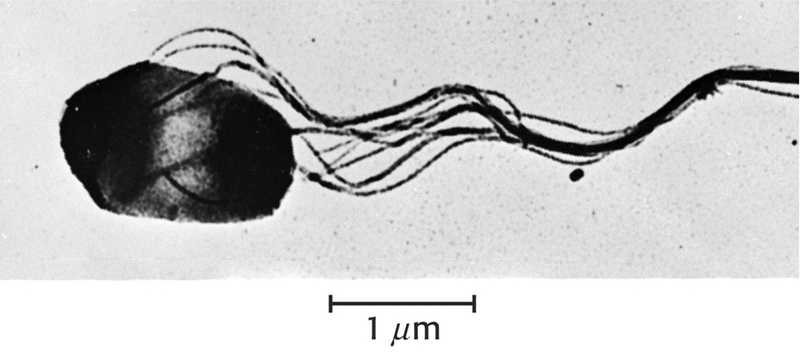}
\caption{Salmonella swim by a bundle of rotating helical flagella.
From \citet{bren77}.
}
\label{fig:salmonella}
\end{figure}

\begin{figure}
\centering
\includegraphics[width=0.48\textwidth]{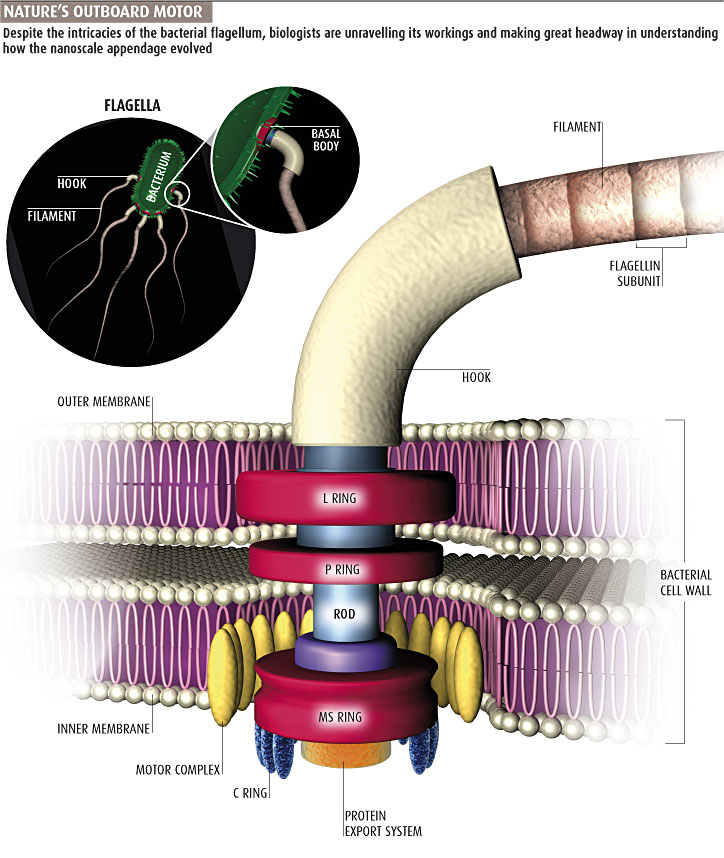}
\caption{Each bacterial flagellum is driven by
a rotary motor embedded in the bacterial cell wall. The motor has a
series of rings, each about 20 nanometers in diameter, with a rod
inside. Attached to the rod is a curved ``hook" protein linked to the
flagellum. The flagellum is 5 to 15 micrometers long and made of
thousands of repeating units of the protein flagellin. The motor is
powered by the flow of sodium or hydrogen ions across the cell wall.
\REV{From \citet{jone08}}.
}
\label{fig:flagellum_motor}
\end{figure}

{\bf Flagellated Bacteria.}  A wide variety of bacteria
exploit helical filaments, called flagella, for their propulsion.
Different species possess various numbers and different
arrangements of flagella. According to the arrangement,
flagellated bacteria are classified as {\em monotrichous} bacteria
which possess a single flagellum only, {\em lophotrichous},
bacteria with multiple flagella located at a particular spot on
their surface, {\em  amphitrichous} bacteria grow a single
flagellum on each of the two opposite ends, and {\em peritrichous}
bacteria are covered by multiple  flagella pointing in all
directions \cite{jans:11}. Prominent examples of {\em
peritrichous} bacteria are {\em Escherichia coli} \cite{berg04},
{\em Salmonella typhimurium}, see Fig.~\ref{fig:salmonella}, {\em
Rhizobium lupini}, or {\em Proteus mirabilis} bacteria  to name
just a few. A flagellum is rotated by a motor complex, which
consists of several proteins, and  is anchored in the bacterial
cell wall \cite{bren77,berg03,berg04}, see Fig.~\ref{fig:flagellum_motor}.
The motor itself is connected to the flagellum by a flexible hook.

\begin{figure}
\centering
\includegraphics[width=0.48\textwidth]{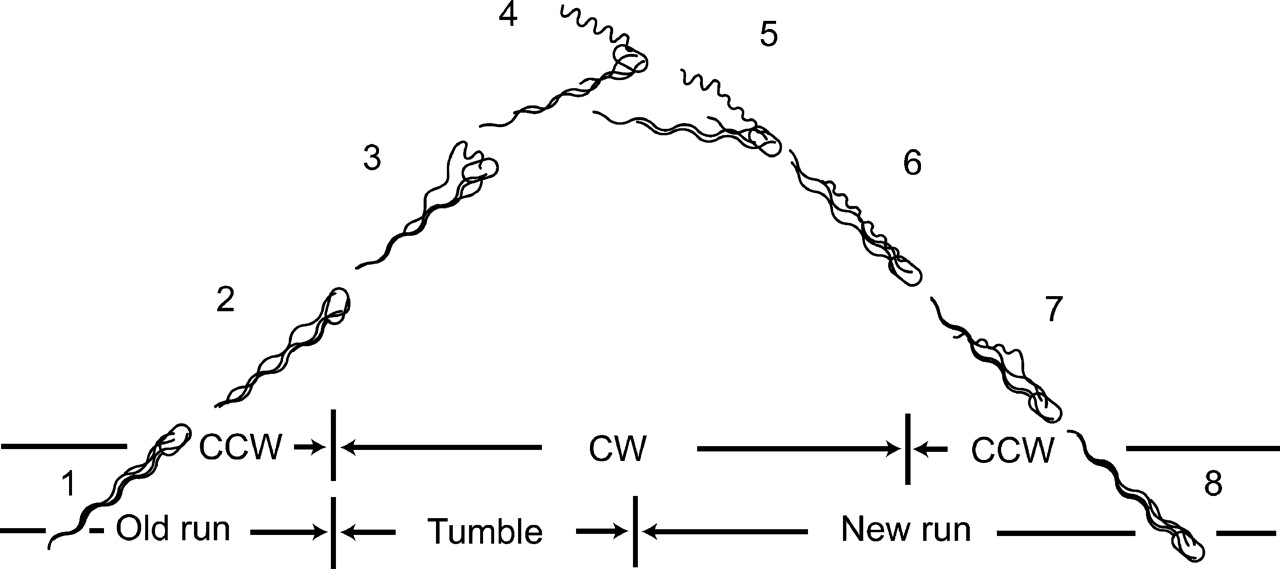}
\caption{Bacteria like {\em E.~coli} and {\em salmonella} move by a
  ``run-and-tumble"
motion. During the ``run''  phase, the flagella form a bundle, and the
bacterium moves forward in one direction. In the ``tumble" phase, one
or a few flagella reverse their rotational direction and leave the
bundle. This induces a tumbling motion which changes the orientation
of the bacterium randomly. CW denotes clockwise, CCW counterclockwise
rotation. From~\citet{darn07}.
}
\label{fig:run_and_tumble}
\end{figure}

Bacteria like {\em E.~coli} and {\em Salmonella} swim in a
``run-and-tumble" motion illustrated in
Fig.~\ref{fig:run_and_tumble} \REV{\cite{macn:77,berg04,turn:00,hyon12}}. 
In the ``run" phase (stage 1 in Fig.~\ref{fig:run_and_tumble}), the
helical winding of all flagella is left-handed, and they rotate
counterclockwise. The flagella form a bundle (see also
Fig.~\ref{fig:salmonella}), and the bacterium moves forward in a
direction determined by its long axis. At the beginning of the
``tumble" phase, one flagellum reverses its rotational direction
to clockwise (stage 2 in Fig.~\ref{fig:run_and_tumble}). The
flagellum leaves the bundle which implies a random reorientation
of the bacterium (stage 3 and 4). The reversal of the rotational
direction is accompanied by a change of the helical handedness
from left-handed to right-handed and the flagellum undergoes a
polymorphic transition
\cite{call:75,macn:77,armi:87,shah:00,darn07,darn:07,call:13,voge:10,voge:13}.
At the end of the ``tumbling" phase, all flagella start to rotate
again in the same counterclockwise direction (stage 5), the bundle
reforms (stage 6), and the bacterium returns to a directional
motion (stage 7 and 8).

The flagella of bacteria like {\em Rhizobium meliloti} or {\em
Rhizobium lupini} are only capable of limited polymorphic
transitions and their motors are unidirectional
\cite{plat:97,scha:02,schm:02}. These bacteria modulate the
rotation speed of individual motors to induce tumbling
\cite{plat:97,scha:02}.

\REV{Uni-flagellated bacteria, e.g., {\em Vibrio alginolyticus},  tumble 
by motion reversal and by exploition of a buckling instability of 
their hook \cite{son:13}. This mechanism represents an example of
one of the smallest engines in nature where a biological function 
arises from controlled mechanical failure, and reveals a new role 
of flexibility in biological materials.}

The purpose of the ``run-and-tumble'' motion is to detect gradients
in the concentration of chemicals (e.g. food) or temperature (to
avoid regions of too high or too low temperature).
This is achieved by extending the ``run'' phase in
case of improving environmental conditions, and by shortening
it in case of worsening conditions.

Bacteria not only swim as individuals, but exhibit collective
behavior at a moist surface or in a thin liquid film in form of
swarming \cite{cope:09,darn:10,kear:10,part:13}. Swarming was
distinguished from other forms of surface translation
\cite{hein:72,kear:10,darn:10}, e.g., swimming, due to particular
shape changes which swarming bacteria undergo. During the
transition from swimming to swarming cells, the number of flagella
increase and the cells become often more elongated by suppression
of cell division \cite{stah:83,jone:04,kear:10,darn:10}. This
points toward the significance of flagella and flagella
interactions between adjacent cells \cite{stah:83,jone:04} for
swarming,  aside from possible shape induced physical
interactions.
\\

{\bf Listeria.}
Listeria is a facultative anaerobic bacterium, capable of
surviving in the presence or absence of oxygen. It can grow and
reproduce inside the host cell.
Outside the body, Listeria use flagella for swimming, but at body
temperature flagellin production is turned off, and Listeria hijack
the actin cytoskeleton of the cell for its mobility. By
expressing actin promotors on
their surface, a thick actin gel develops on the surface. The gel
formes a comet and pushes the bacterium through the cell.

In vitro, this mechanism can be mimicked by micrometer-size
colloid beads
coated with nucleation promotion factors. The gel initially formes
isotropically, but due to tension in the growing gel layer, the symmetry
is broken spontaneously, and an actin comet formes \cite{Kawska2012}.
The physics of Listeria motility has been reviewed recently by
\citet{Prost2008}.
\\

\begin{figure}
\centering
\includegraphics[width=0.48\textwidth]{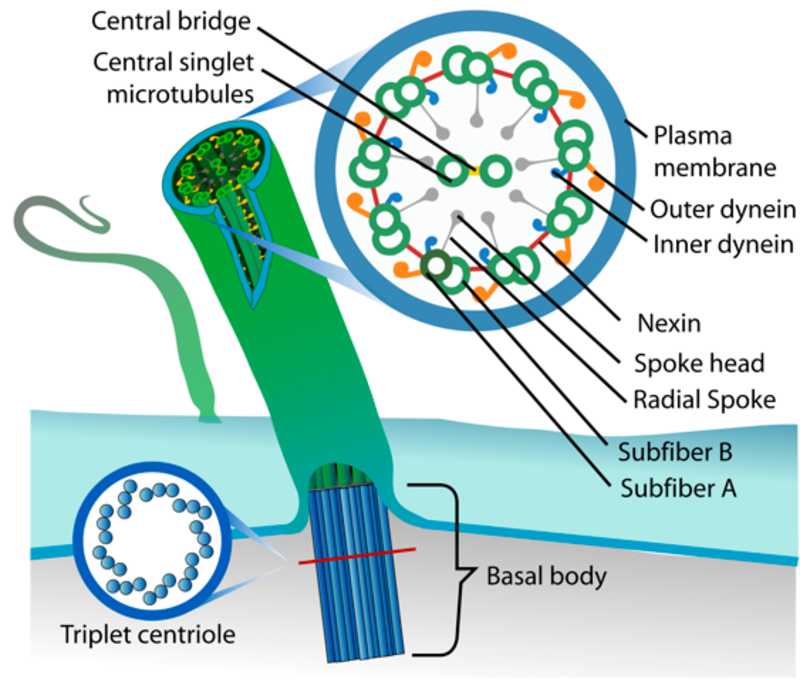}
\caption{The axoneme consists of nine double microtubules, arranged
around the perimeter, and two central microtubules. The microtubules
are connected by many linker proteins. Motor proteins connecting
neighboring double microtubules lead to active bending.
\REV{From \url{http://en.wikipedia.org/wiki/Cilium}}.
}
\label{fig:axoneme}
\end{figure}

{\bf Eukaryotic Cells.}
Eucaryotic swimmers are usually propelled by cilia or (eucaryotic-)flagella.
These motile hairlike extensions of the cell  consist  of
a bundle of microtubules, which are connected by motor proteins
and other connecting proteins,
see Fig.~\ref{fig:axoneme}. The underlying structure is called the
axoneme. It consists of two central microtubules,
which are surrounded by nine double microtubules (rather stiff filaments
with a persistence length of about 1 mm).
The microtubules are connected by many proteins (nexin links,
central spokes, ...), which stabilize the structure.
Motor proteins (dynein) connecting neighboring double microtubules
cause a local active bending force by sliding the microtubules
relative to each other. Thus, motor activity is spread out over the
whole length of the eukaryotic flagellum. The axoneme structure
is well conserved across all eukaryotes. Hence, it seems very likely
that it has evolutionary already been present in the very first eukaryotic
cells.

\begin{figure}
\centering
\includegraphics[width=0.35\textwidth]{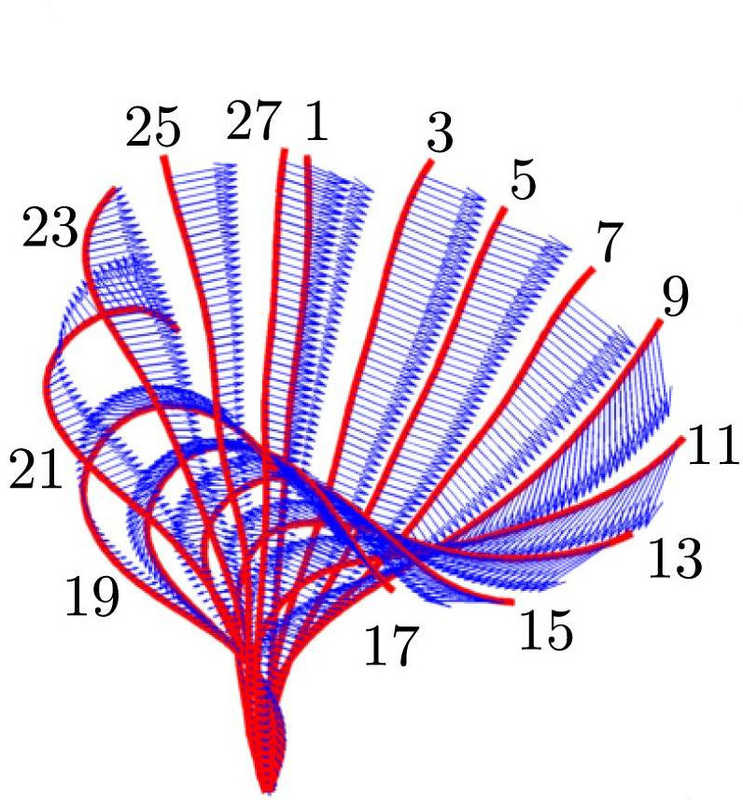}
\caption{
\REVV{
The beat of cilia has two distinct phases, the power stroke and the recovery stroke. 
Snapshots are shown from Volvox somatic cells, imaged at 1000 frames per second. 
During the power stroke, the cilium is stretched out straight and moves rather fast 
in one direction (frames 1 - 11), while during the recovery stroke, it bends and 
slowly retracts (frames 13-27). Adapted from \citet{brum:14}.}
\label{fig:ciliumbeat}
}
\end{figure}

\begin{figure}
\centering
\includegraphics[width=0.48\textwidth]{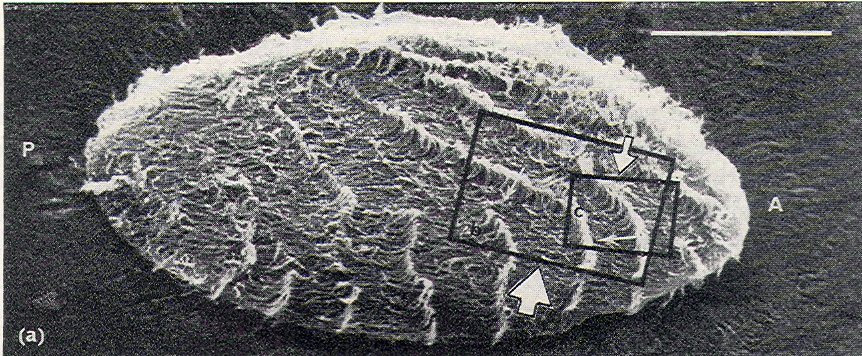}
\caption{The surface of {\em Opalina} is covered by many hairlike
filaments called cilia. {\em Opalina} swims forward by a stroke-like
wiggling of the cilia. The formation of metachronal waves is
clearly visible. The scale bar is $100 \mu m$. From \citet{tamm70}.}
\label{fig:opalina}
\end{figure}

The main structural difference between {\em cilia} and {\em flagella}
is their length.
A typical cilium is $10 \mu m$ long, while sperm flagella are about
$50 \mu m$ long. The second large difference concerns their beat
patterns.
While flagella beat with an almost perfect propagating sinusoidal
bending wave,
the ciliar beat has two distinct phases.
During the {\em power stroke}, the cilium is stretched out straight and moves
rather fast in one direction, while it bends, twists a little sideways
and slowly retracts in the {\em recovery stroke}, as shown in
Fig.~\ref{fig:ciliumbeat}. A particularly interesting feature of the beat
pattern of cilia arrays is the formation of ``metachronal waves", as shown in 
Fig.~\ref{fig:opalina}; the cilia do not beat synchronously,
but in a pattern resembling a wheat field in the wind.

Different physical mechanisms for the beat of eukaryotic flagella and cilia
have been suggested. Either the beat arises from a coupling
of the activity of motor proteins with the curvature of the flagellum,
where large curvature implies detachment of motors from the neighboring
filament \REV{\cite{lind94,lind10b,lind14,bayl14}}, or it arises from 
the cooperative behavior of
several motors pulling in opposite directions, with the ``winners"
pulling along the ``losers" for a while.
This tug-of-war of the molecular motors results in a negative
force-velocity relation at zero velocity. The system is thus intrinsically
unstable, and starts moving in one direction. As the filaments move
relative to each other and deform,
elastic forces build up, eventually causing stalling and motion reversal.
The system thus starts to oscillate in time. This
oscillation explains how the axoneme's internal machinery self-organizes
to generate the flagella beat \cite{juli97,cama99,Kruse07,Hilfinger2009}.

The domain of eukaryotes is home to an extraordinary number of
different microswimmers. While all employ axonemes as motors, they
use them in all kinds of ways and arrangements to propel themselves.
In the following, we present a few examples of eukaryotic microswimmers,
which have
received particular attention in the biophysical community.
\\

\begin{figure}
\centering
\includegraphics[width=0.48\textwidth]{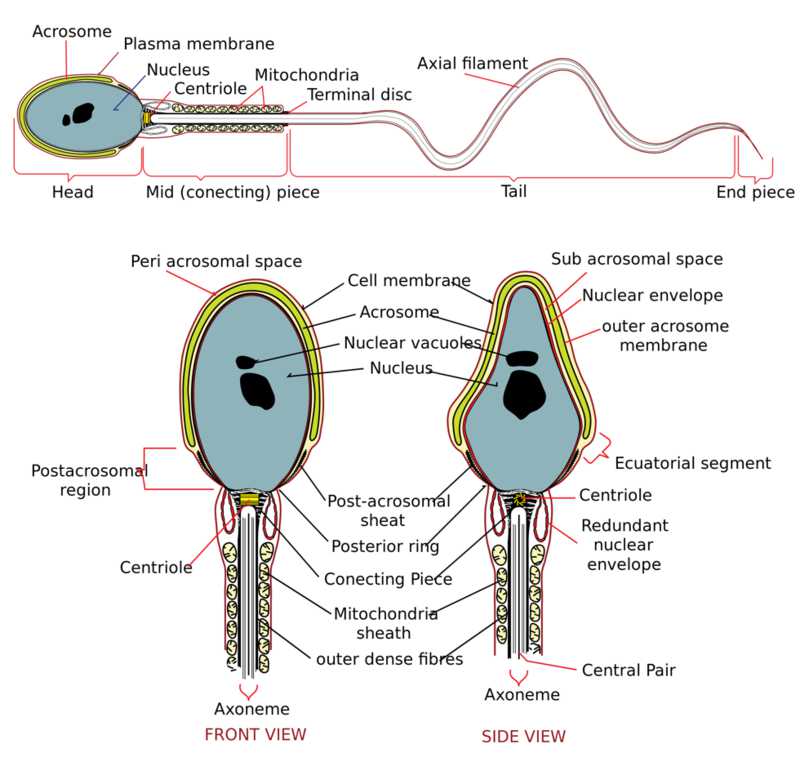}
\caption{Schematic presentation of the structure of human sperm.
Sperm cells consist of a head 5 $\mu$m by 3 $\mu$m and a tail 41 $\mu$m
long.
Spermatozoa beat at $10-60$ Hz, which generates swimming velocities
  on the order of $100 \ \mu$m/s.
From \url{http://en.wikipedia.org/wiki/Spermatozoon}.
\label{fig:human_sperm}
}
\end{figure}

\begin{figure}
\centering
\includegraphics[width=0.40\textwidth]{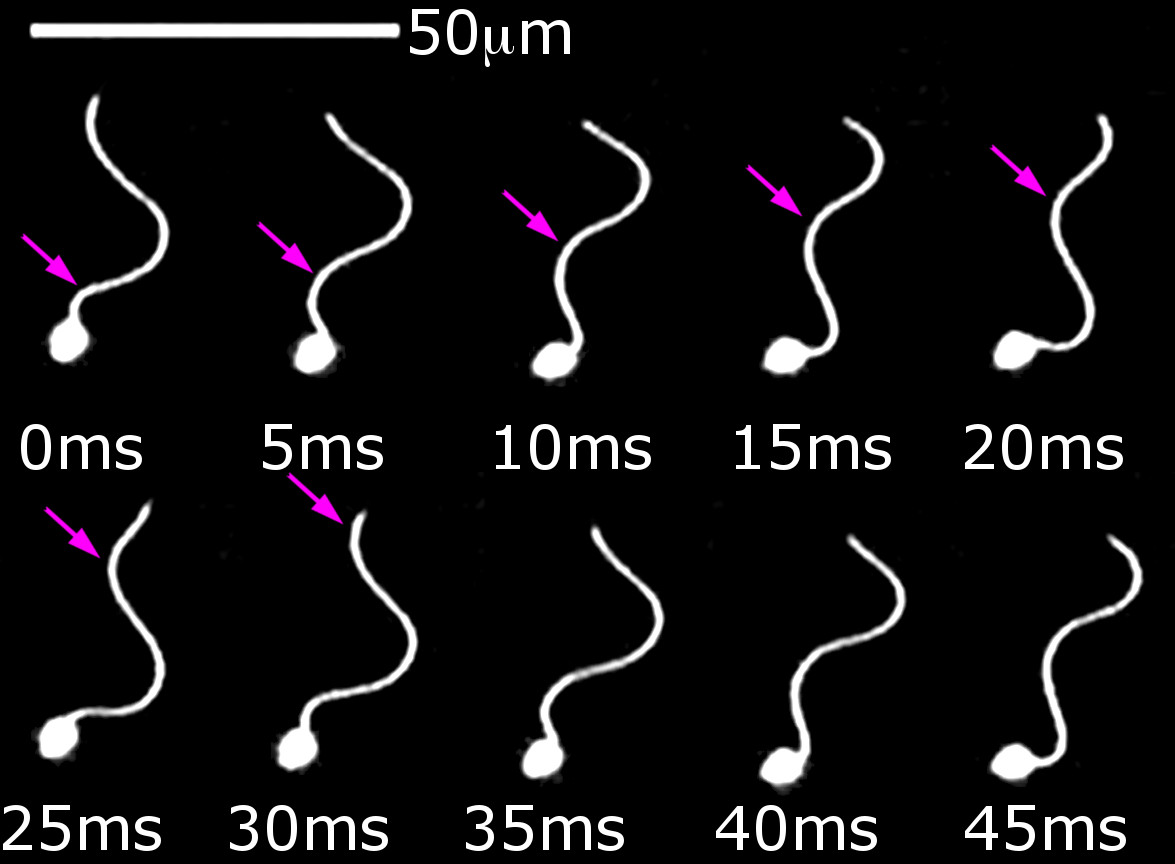}
\caption{Sperm swim by a snake-like motion of their flagellum.
The time sequence (from left to right) of snapshots of the beat of
sea-urchin sperm shows a sinusoidal traveling wave on the flagellum.
Arrow indicate the wave propagation from the head to the tip.
Images courtesy of J. Jikeli, L. Alvarez and U.B. Kaupp
(Forschungszentrum caesar, Bonn).}
\label{fig:sperm_frames}
\end{figure}

{\bf Sperm.}
Many sperm cells consist of a (roughly spherical) head, which
comprises the genetic material, a midpiece, which contains many
mitochondria for energy production, and the eukaryotic flagellum, see
Fig.~\ref{fig:human_sperm}.
A sperm cell is propelled though a fluid by a snake-like wiggling
of its flagellum, as shown in Fig.~\ref{fig:sperm_frames}.
The flagellar beat is a propagating bending wave, with wave lengths
smaller than the flagellar length.
The beat can either be planar, as in Fig.~\ref{fig:sperm_frames},
in particular for sperm swimming near surfaces \cite{frie10}, or
be three-dimensional with a conical envelope \cite{coss03,wool03}.
The purpose of sperm motility is of course the transport of the
genetic material to the egg. Here, chemotaxis is important to guide
the spermotozoa to find their target
\cite{kaup03,kaup08,frie07,eise06b,alva14}.
\\

\begin{figure}
\centering
\includegraphics[width=0.45\textwidth]{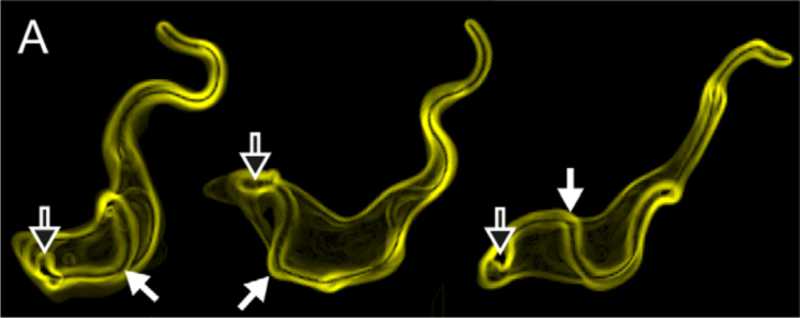}
\caption{Swimming motion of trypanosomes. The contour of the flagellum
attached to the cell body (closed arrows) is clearly visible from the
flagellar pocket (open arrows) to the anterior free end.
From \citet{Heddergott2012}.
}
\label{fig:trypanosome}
\end{figure}

{\bf Trypanosomes.} Trypanosomes ({\em T. brucei}) are motile
parasites responsible for sleeping sickness. Their motility is
mediated by a single flagellum. However, unlike in sperm the flagellum
of a trypanosome emerges from the
flagellar pocket near the base of the cell and runs along the
length of the entire body, as illustrated in
Fig.~\ref{fig:trypanosome}. The cell surface hydrodynamic drag is
used by trypanosomes to sweep antibodies to the flagellar pocket,
the ``cell mouth'', for host immune evasion. Trypanosomes are
pulled forward by the planar beat of the flagellum, while the
asymmetrically shaped body induces also a rotational motion
\cite{Uppaluri2011,Babu2012,Heddergott2012}.
\\

{\bf Ciliates.} Ciliates are a phylum of microorganisms characterized by motile
cilia on their surface. Among them, {\em Paramecia} and {\em Opalina}
have received significant attention in the biophysics community as
model systems.
{\em Paramecia} is a group of unicellular ciliate protozoa, which range
from about 50 $\mu$m to 350 $\mu$m in length and move with a speed of
approximately $10^3 \mu$m/s (10 body lengths per second).
They generally feed on bacteria and other small cells.
{\em Opalina} is a genus of protozoa found in the intestines
of frogs and toads. It is without a mouth or contractile vacuole. The
surface of Opalina is covered uniformly with cilia.

Both in {\em Paramecia} and {\em Opalina}, cilia beat neither independently nor
completely synchronously, but instead (as explained above) in the form
of metachronal waves \cite{tamm70,mach72,okam94}, see Figs.~\ref{fig:ciliumbeat}
and \ref{fig:opalina}.
\\

\begin{figure}
\centering
\includegraphics[width=0.35\textwidth]{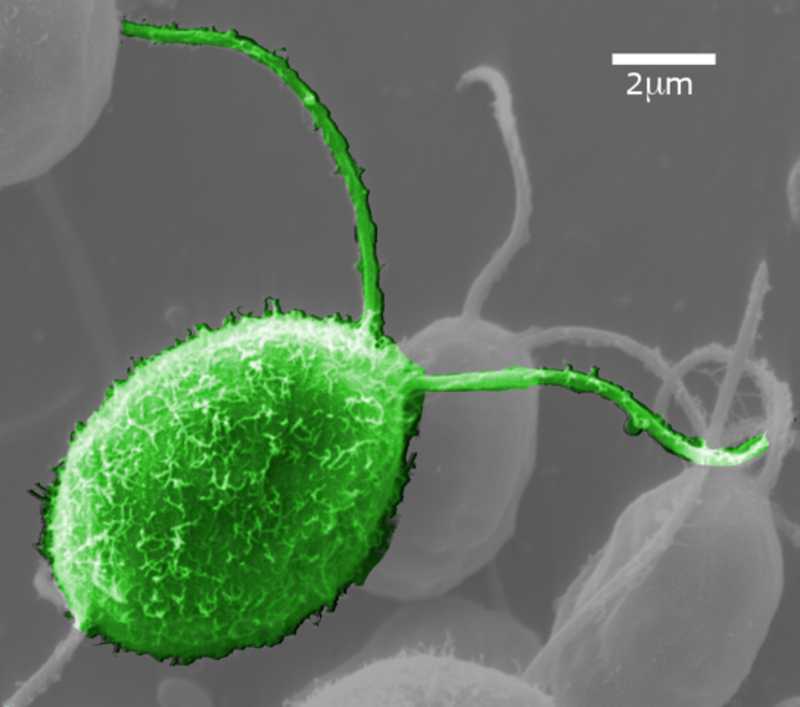}
\caption{
{\em Chlamydomonas} swims by a breast-stroke-like motion of its two
flagella. Steering in reaction to a light stimulus is facilitated by
additional beats of one flagellum.
\REV{Adapted from \url{http://remf.dartmouth.edu/images/algaeSEM}}.
\label{fig:clamydomonas}
}
\end{figure}

{\bf {\em Chlamydomonas reinhardtii.}}
{\em Chlamydomonas reinhardtii} is a single-celled green algae of
about 10 $\mu$m in diameter, which swims with two flagella, see
Fig.~\ref{fig:clamydomonas}. Chlamydomonae are equipped with a
light-sensing  ``eye-spot".
Widely distributed worldwide in soil and fresh water,
{\em C. reinhardtii} is used as a model organism in biology in a wide range
of subfields. The swimming motion of
{\em Chlamydomonas} resembles the human breast-stroke: the flagella are
pulled back in a nearly straight shape, and are then bend over and pushed
forward again.
The oscillatory velocity field induced by swimming {\em C. reinhardtii}
has been observed in
time-resolved measurements recently \cite{dres10a,guas10}.
\\

{\bf Volvox.}
Volvox is another green algae. It forms spherical colonies of up
to 50,000 cells.
Each mature Volvox colony is composed of numerous flagellate cells
similar to {\em Chlamydomonas}, embedded in the surface of a hollow sphere.
The cells move their flagella in a coordinated fashion, with distinct
anterior and posterior poles of the colony. Each individual cell has an
eye-spot, more developed near the
anterior, which enable the colony to swim towards light \cite{Solari2011}.
Recently, the flow field around freely swimming Volvox has been
measured directly \cite{dres09,dres10a}.
The order of Volvocales contains many more microswimmers, in
particular spherical swimmers of many different sizes.
\\

\begin{figure}
\centering
\includegraphics[width=0.40\textwidth]{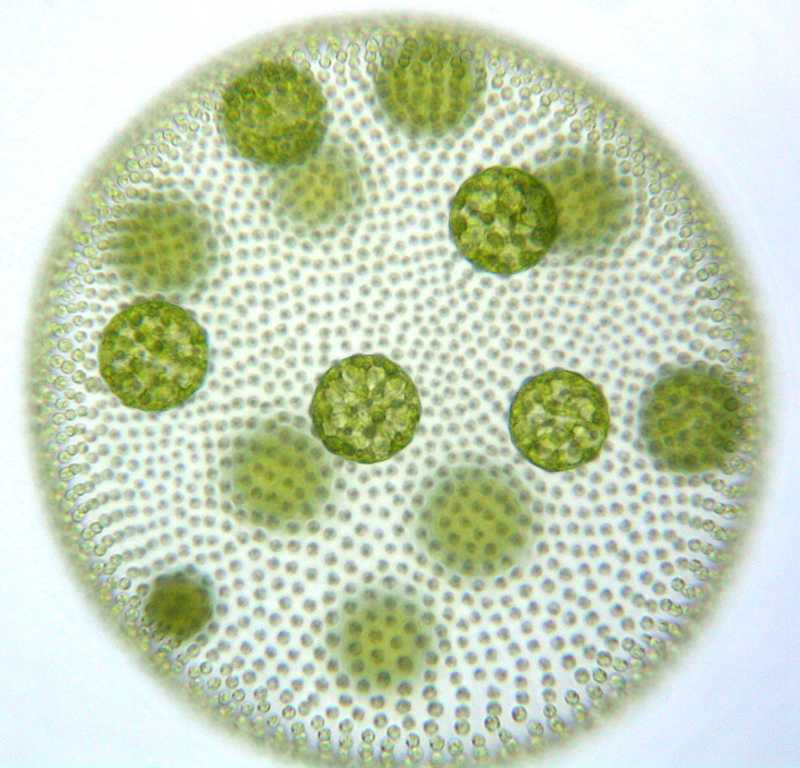}
\caption{
\REV{Volvox carteri is a colony of several thousand cells organized around a 
spheroidal structure made of a gel-like extracellular matrix. The small 
cells are somatic cells; the larger cells visible inside the colony are 
the reproductive cells that will form new colonies. Each somatic cell 
has two flagella, which beat in a coordinated fashion to rotate and 
move the colony.}
From \url{http://www2.unb.ca/vip/photos.htm} (A. Nedelcu, Biology Dept.,
University of New Brunswick)
\label{fig:volvox}
}
\end{figure}

{\bf Gliding Motility of Cells.}
Another form of cellular motility is gliding or crawling. In this case,
cells do not swim in a fluid, but move on a substrate or through
a gel or porous material. Because this form of locomotion on a
substrate is not swimming, it goes somewhat beyond the scope of this
review. However, the investigation of crawling eukaryotic
cells and gliding bacteria can be a way to separate hydrodynamic
contributions from generic self-propulsion effects, and provides new
insights into the collective behavior of many self-propelled
particles.
\\
Crawling eukaryotic cells propel themselves with the help of their
actin cytoskeleton. Velocities are of the order of a few tens of
microns per hour, two to four orders of magnitude slower than swimming
cells.
Model organisms studied experimentally include Madin-Darby Canine Kidney
(MDCK) cells, which display intriguing collective behavior
\cite{Poujade2007,Trepat2009,Basan2013},
and lymphocytes, keratocytes\cite{Kim2012} and keratinocytes for single-cell motility.
\\

Also many bacteria can move actively by gliding on a
substrate.  A particularly well studied example is {\em myxobacteria},
which typically travel in swarms.
{\em Myxococcus xanthus} has a signaling system that requires cell-to-cell
contact, coordinates cell movements and gives rise to rippling patterns
\cite{Kaiser03}.  It has an elongated, spherocylinder-like body with
an aspect ratio of about 10. After collisions, the bacteria acquire
nearly identical nematic orientation.
This makes myxobacteria a very interesting model system to study the
collective behavior of self-propelled rod-like particles
\cite{Peruani2012,Harvey13,harv11,gejj12}.

\subsection{Synthetic Microswimmers}

Locomotion on the nanoscale through a fluid environment is one of the
grand challenges confronting nanoscience today \cite{ozin05}.
The vision is to synthesize, probe, understand, and utilize a new
class of motors made from nanoscale building blocks that derive
on-board or off-board power from in-situ chemical reactions.
The generated mechanical work allows these motors to move through
a fluid phase while simultaneously or sequentially performing
a series of tasks. A large variety of such swimmers have been constructed
recently, from bimetallic nanorods \cite{paxt04,four05,ozin05}, to 
\REV{rotating filaments \cite{mang:06},} and 
artificial sperm \cite{drey05,will14}.  Some examples are given below.
\\

{\bf Bimetallic Nanorods and Microspheres.}
A simple class of synthetic nanoscale motors are made from
bimetallic Pt-Au nanorods immersed in a $\mathrm{H}_2\mathrm{O}_2$ solution
\cite{paxt04,Paxton2005,four05}. The catalytic reaction
$2 \, \mathrm{H}_2\mathrm{O}_2 \to 2 \, \mathrm{H}_2\mathrm{O} + \mathrm{O}_2$ occurs
at the Pt end of the rod and is the power source for the
motion. One plausible mechanism for motion involves the
surface tension gradient due to $\mathrm{O}_2$ adsorption on the nonreactive
Au end.  The molecular-level details
of how $\mathrm{O}_2$ generated at the Pt end of the nanorod leads to
the propulsive force remain to be elucidated.
\\

{\bf Catalytic Janus Colloids.}
Similarly, spherical particles (like polystyrene or silica beads
with metallic caps), which catalyze a chemical reaction inside
the fluid, display self-propulsion \cite{hows07,erbe08}.
The catalytic reaction implies an asymmetric, non-equilibrium
distribution of reaction products around the colloid, which generates
osmotic or other phoretic forces
\cite{gole05,ruec07,gole09,pope09,pope10,thak11b,saba12}.
These objects are denoted ``diffusio-phoretic swimmers".

The concept of diffusio-phoretic swimmers can be taken one step
further by constructing  self-assembled, photoactivated colloidal
microswimmers. An example is a polymer sphere, which includes a smaller
hematite cube in dilute $\mathrm{H}_2\mathrm{O}_2$ solution. The
decomposition reaction
is catalyzed by the hematite cubes, but only under illumination.
Pairs of spheres and cubes then self-assemble into self-propelled
microswimmers at a surface \cite{palacci2013}.
The dynamic assembly results from a competition
between self-propulsion of particles and an attractive interaction
induced respectively by osmotic and phoretic effects activated by light.
\\

{\bf Thermophoretic Janus Colloids.}
Janus colloids with a metallic cap can also display self-propulsion
due to self-thermophoresis. In this case, the cap is heated by a laser
beam, which generates a temperature difference between the two sides
of the Janus particle; the colloid then diffuses in this temperature
gradient \cite{Jiang2010}. This approach has been extended to thermophoretic
Janus colloids in binary fluid mixtures (with an upper miscibility gap)
near the demixing critical point \cite{volpe2011,buttinoni2012}. In this case,
heating by the laser beam leads to the formation of a droplet of
one phase which adheres to the cap. This approach has the advantage that
a small laser power suffices to induce self-propulsion.
\\

{\bf Bubble Jets.}
Another type of catalytic microswimmer consists of a hollow micro-tube
with functionalized surface. Such a microswimmer has been developed and
fabricated by templated electrodeposition on a pre-stressed thin polymer
film, which generates strain and leads to detachment from the substrate
and roll-up into a tube. The catalytic reaction decomposes $\mathrm{H}_2\mathrm{O}_2$
into $\mathrm{O}_2$ and water inside the tube, which leads to the formation of
nano- and microbubbles. By spontaneous symmetry breaking, the bubbles leave
the tube on one side; this implies a jet-like propulsion. Such tubular
micro-engines offer several advantages, like easy motion control,
integration of various functions, scalable size in diameter and length,
and straight trajectories \cite{Sanchez2009,Sanchez2011,Mei2011}.
\\

{\bf Rotators.}
Rotating discs \cite{grzy00,Llopis2008}, rotating spheres or
dumbbells \cite{tier08} are another important class of swimmers.
These rotators can be autonomous swimmers, like the Volvox algae
mentioned above, or actuated synthetic particles, like
super-paramagnetic particles rotated by an external magnetic field
\cite{blei06}.
Rotators show an interesting collective behavior, like a circling
motion of Volvox around each other \cite{dres09}, or a lane formation
of rotating magnetic discs in microfluidic channels \cite{goet:10.1,goet:11}.
\\

{\bf Self-propelled Droplets.}
Aqueous droplets suspended in an oil phase containing surfactants
can also be made to swim. Propulsion arises for example due to the spontaneous
bromination of mono-olein as the surfactant. The droplet surface
is covered by a dense surfactant monolayer, which reacts with
the bromine fuel that is supplied from inside the droplet, such
that bromination proceeds mainly at the droplet
surface, and reduces the amphiphilicity of the surfactant.
This results in a self-sustained bromination gradient
along the drop surface, which propels the droplet due to Marangoni
stresses \cite{Thutupalli2011}.
\\

{\bf Biomimetic Microswimmers.}

Artifical microswimmers can be constructed by using similar design
principles as those found in biological systems. A by now classical
example is a swimmer which mimics the propulsion mechanism of a
sperm cell \cite{drey05}. The flagellum is constructed from a chain
of magnetic colloid particles and is attached to a red
blood cell, which mimics the sperm head. This artificial
swimmer is set into motion by an alterating magnetic field, which
generates a sidewise oscillatory deflection of the flagellum.
However, the swimming motion is not the same as for a real sperm cell;
the wiggling motion is more a wagging than a travelling sine wave,
and generates a swimming motion toward the tail end, opposite
to the swimming direction of sperm.

A more recent example of a biohybrid swimmer which mimics the
motion of sperm has been presented by \citet{will14}.
In this case, the microswimmer consists of a polydimethylsiloxane
filament with a short, rigid head and a long, slender tail on
which cardiomyocytes (heart-muscle cells) are selectively cultured.
The cardiomyocytes contract periodically and deform the filament
to propel the swimmer. This is a true microswimmer because it requires
no external force fields. The swimmer is about 50 $\mu m$ long and
reaches a swimming velocity up to 10 $\mu m/s$.

\begin{figure}
\centering
\includegraphics[width=0.35\textwidth]{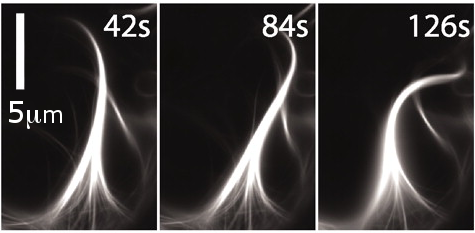}
\caption{A minimal system of microtubules, molecular motors, and
depleting polymer assembles into actively beating bundles.
The sequence of images illustrates the beating pattern.
From \citet{sanc11a}.}
\label{fig:sanc11}
\end{figure}

Another interesting example is artificial cilia \cite{sanc11a}.
The assembly of oscillatory active filament bundles requires three main
components. The first two, kinesin motor clusters and microtubules,
spontaneously organize into aster-like structures \cite{nede97}.
Addition of a third component, a non-adsorbing polymer,
induces attractive interactions between microtubules through the
depletion mechanism, leading to their bundling.
This greatly increases the probability of kinesin clusters to
simultaneously bind and walk along neighboring microtubules, where
the relative motion of microtubules in these bundles
depends on their polarity. This generates a cilia-like beat pattern,
which is, however, symmetric in its forward and backward motions
\cite{sanc11a}, see Fig.~\ref{fig:sanc11}.
\\

\begin{figure}
\centering
\includegraphics[width=0.45\textwidth]{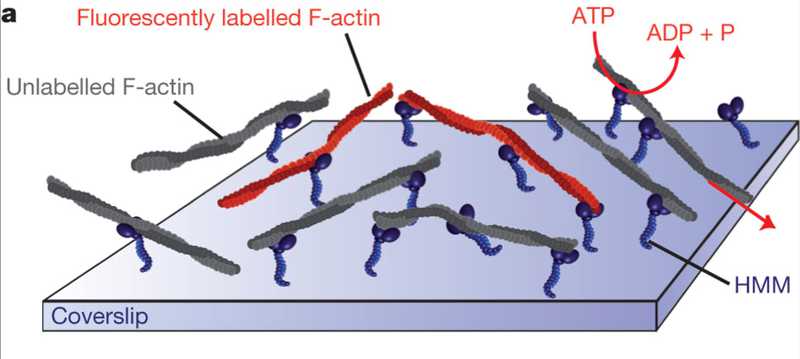}
\caption{Motility assays consist of carpets of motor proteins, on which
polar filaments like actin are propelled along their contour.
From \citet{Schaller10}.}
\label{fig:motility_assay}
\end{figure}

{\bf Motility Assays.}
An in-vitro system to study the interaction of motor proteins with
biological polar filaments like actin and microtubules are motility
assays. In these systems, motor proteins are grafted onto a planar
substrate with their active heads pointing upwards. Polar
filaments, which are lying on such a motor-protein carpet, attach
to the motors and are pushed forward, see Fig.~\ref{fig:motility_assay}.
The filaments in such a system are neither swimmers nor self-propelled
particles, but it is nevertheless a interesting model systems to study the
collective behavior of many moving particles with self-organized
directions of motion, which shows phenomena like density waves and
vortices \cite{Schaller10,Sumino12}.

\subsection{Theoretical Model Microswimmers} \label{sec:theoretical_models}

Several detailed theoretical models have been designed during the
last decades to describe and understand the behavior of flagellated
microswimmers, but also of diffusio-phoretic spheres and rods, etc.
These models and their properties will be reviewed in detail in
the remainder of this article. However, there are also some model
swimmers, which serve as generic models to elucidate the physical
principles of microswimmers. The latter models are briefly introduced
here.
\\

\begin{figure}
\centering
\includegraphics[width=0.40\textwidth]{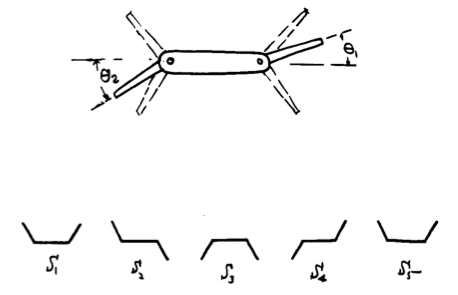}
\caption{The Purcell swimmer consists of three rod-like connected
segments, which can be tilted with respect to each other. When the
two angles $\theta_1$ and $\theta_2$ are varied in the non-reciprocal
temporal sequence shown at the bottom, this micro-machine moves
forward. From \citet{purc77}.}
\label{fig:Purcell_swimmer}
\end{figure}

{\bf Purcell Swimmer.}
It was recognized already more than 30 years ago by \citet{purc77}
that directed forward swimming of micromachines in viscous
fluids---at low Reynolds numbers, see Sec.~\ref{sec:low_Re} below---is
not possible when the angle between {\em two} rigid segments is varied
periodically in time. Therefore, Purcell suggested a swimmer consisting
of {\em three} rigid segments as shown in Fig.~\ref{fig:Purcell_swimmer}.
When the two angles are varied periodically
in time, but in a way which breaks time-reversal symmetry, as shown
at the bottom of Fig.~\ref{fig:Purcell_swimmer}, this machine performs
a directed motion \cite{purc77}.
\\

\begin{figure}
\centering
\includegraphics[width=0.40\textwidth]{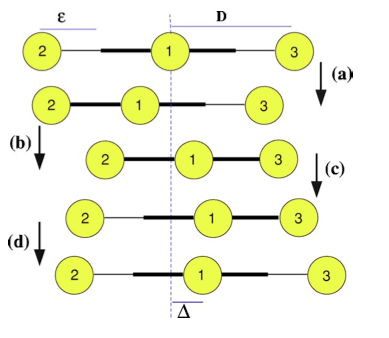}
\caption{Three connected beads swim forward when their distances are
varied cyclically as a function of time. Here, it is important that the
cycle breaks time-reversal symmetry, as shown. In one cycle, the
swimmer moves by a distance $\Delta$. From \citet{naja04}.}
\label{fig:three-bead-swimmer}
\end{figure}

{\bf Three-Bead Swimmer.} \label{swimmer, three-bead}
A very simple one-dimensional swimmer can be constructed from three
spheres that are linked by rigid rods, whose lengths can change
between two values \cite{naja04}. With a periodic motion that
breaks the time-reversal symmetry as
well as the translational symmetry, the model device
swims at low Reynolds number as shown in Fig.~\ref{fig:three-bead-swimmer}.
This swimmer is similar in spirit to the Purcell swimmer, but has
the advantage that it can be treated theoretically much more easily,
so that many of its properties can be analyzed in detail or even
calculated analytically \cite{naja04,gole08,alex08}.
\\

{\bf Squirmers.}
A particular class of microswimmers are almost spherical organisms that are
propelled by active hair-like organelles (cilia) covering the body.
On a mesoscopic length scale, the synchronized beating
of the cilia can be mapped onto a spherical
envelope~\cite{ligh52,blak71b}, and its time average corresponds
to a steady tangential surface velocity.
These objects---called "squirmers"---may also serve
as a simple generic model for other types of microswimmers,
for example the self-propelled droplets mentioned above
\cite{Thutupalli2011}.

The squirmer is modeled as a hard sphere of radius $R$ with a
prescribed tangential surface velocity ${\bm v}_{\rm sq}$,
causing a propulsion
in the direction of the squirmer's instantaneous orientation
${\hat{\bm e}}$.
The relative velocity at a surface point ${\bm r}_{\rm{s}}$
(with respect to the squirmer's center) is given by~\cite{blak71b}
\begin{equation}\label{eq:squirmvel}
{\bm v}_{\rm{sq}}({\bm r}_{\rm{s}},\hat{\bm e})
 =\sum_{n=1}^{\infty} B_n \frac{2}{n(n+1)}
  \left( \frac{\hat{\bm e} \cdot
  {\bm r}_{\rm{s}}}{R} \frac{{\bm r}_{\rm{s}}}{R}
  - \hat{\bm e} \right)
  P'_n\left(\frac{\hat{\bm e} \cdot {\bm r}_{\rm{s}}}{R}\right),
\end{equation}
where $P'_n(\cos\theta)$ is the derivative of the {\textit n}th
Legendre polynomial with respect to the argument and $B_n$ is the amplitude of the
{\textit n}th mode of the surface velocity.
The absolute local surface velocity of the squirmer is given by
\begin{equation}\label{eq:surfvel}
{\bm v}_{\rm s}({\bm r}-{\bm r}_{\rm c},\hat{\bm e})={\bm v}_{\rm c}
+{\bm v}_{\rm sq}({\bm r}-{\bm r}_{\rm c},\hat{\bm e})
+{\bf \Omega} \times ({\bm r}-{\bm r}_{\rm c}),
\end{equation}
where ${\bm r}_{\rm c}$, ${\bm v}_{\rm c}$ and ${\bf \Omega}$ are the
sphere position, velocity and angular velocity, respectively.
The constant $B_1$ sets the average velocity of the squirmer,
$v=\langle \hat{\bm e} \cdot {\bm v}_{\rm s} \rangle=2B_1 /3$
\cite{ishi06}.
To cope with the biodiversity of real microswimmers, the
characteristic surface velocity of the model can be varied by adjusting
the coefficients $B_n$ with $n\ge 2$. In most cases studied so far,
only $B_1$ and $B_2$ are taken to be non-zero, so that $B_n=0$ for $n\ge 3$.

This model was employed, for example, to study
the hydrodynamic interaction of two squirmers without \cite{ishi07}
and with \cite{gg:gomp10l} thermal fluctuations,
monolayers of squirmers without thermal fluctuations \cite{ishi08}, 
or their behavior in external flows 
\REV{\cite{zoet:12,zoet:13,pago:13,gach:13}}.


\section{Life at Low Reynolds Numbers}
\label{sec:low_Re}

\subsection{Hydrodynamics}

The dynamics of the incompressible fluid surrounding a
microswimmers is described by the Navier-Stokes equations
\begin{align} \label{eq:n-s-comp}
\nabla \cdot {\bm v} & = 0 , \\ \label{eq:n-s}
\rho \left(\frac{\partial {\bm v}}{\partial t}
                      + ({\bm v} \cdot \nabla) {\bm v} \right)
& = \eta \nabla^2{\bm v} -\nabla p + {\bm f} ,
\end{align}
where $\rho$ is the fluid density, $\eta$ the fluid viscosity,
${\bm v}({\bm r},t)$ the position- and time-dependent fluid
velocity field, $p({\bm r},t)$ the pressure field, and ${\bm
f}({\bm r},t)$ an applied body force. Scaling length,
velocity, and time by characteristic values $L$, $v_0$, and $T_0$,
respectively, yields
\begin{align}
\label{eq:n-sdimless} Re_{T} \frac{\partial {\bm v}'}{\partial t'}
               + Re({\bm v}' \cdot \nabla) {\bm v'}
= \nabla^2{\bm v}' -\nabla p' + {\bm f}' ,
\end{align}
where the prime denotes dimensionless quantities. Here,
the dimensionless number $Re=\rho v_0 L/\eta$ is the classical Reynolds
number, which is a measure of the importance
of the nonlinear advection term compared to the viscous forces.
A  second dimensionless number is the oscillatory Reynolds number
$Re_{T}=\rho L^2/(\eta T_0)$. It  indicates the importance of the linear
unsteady term, which scales as $\rho v_0/T_0$, compared to the viscous
term \cite{dhon96,laug09,laug:11}.
$Re_T = \tau_{\eta}/T_0$ is the ratio of the viscous time scale
$\tau_{\eta} = \rho L^2/\eta$ for shear wave propagation over the
distance $L$ and the characteristic time $T_0$.
Examples for $T_0$ could be
the self advection time $L/v_0$ (for which $Re_T=Re$), the
rotational period of a bacterial flagellum, or the period of a
beating cycle of a cilia. For a swimmer of length
$L=10\mu\mathrm{m}$ and a velocity of $v_0 = 50 \mu\mathrm{m/s}$,
the Reynolds number in water (with the kinematic viscosity $ \nu =
\eta/\rho=10^{-6} {\mathrm{m}}^2/\mathrm{s}$) is $Re \approx
10^{-3}$. In this case, the nonlinear contributions on the
left-hand-side of Eq.~(\ref{eq:n-sdimless}) can be neglected,
leading to the linearized Navier-Stokes equation
\begin{align} \label{eq:linear_n-s}
\rho \frac{\partial {\bm
v}}{\partial t} = \eta \nabla^2{\bm v} -\nabla p + {\bm f} .
\end{align}
For $Re_T \ll 1$, i.e., $\tau_{\eta} \ll T_0$, equation (\ref{eq:linear_n-s}) turns into the
Stokes equation (creeping flow),
\begin{align} \label{eq:stokes}
\nabla p-\eta\nabla^2 {\bm v} = {\bm f}
\end{align}
and all inertia terms are absent.

The Stokes equation (\ref{eq:stokes}) is linear and time-independent,
and is thus symmetric under time reversal, an aspects
with far reaching consequences for microswimmers undergoing periodic
shape changes, as first realized by \citet{purc77}.
This is expressed by the famous ''scallop theorem'', which can be
stated as: if the shape changes
displayed by a swimmer are identical when viewed in reverse order,
it will generate an oscillatory, but no directed motion
\cite{purc77,laug09,laug:11}. It means that just by opening and
closing its two shells, a mussel (scallop) cannot move forward.
Additional degrees of freedom are required to generate a sequence
of moves, which are not time reversibel. These aspect is discussed
in more detail in  Sec.~\ref{sec:synchronization} below.

\subsection{Solution of Stokes Equation}

The linearized Navier-Stokes equations (\ref{eq:n-s-comp}),
(\ref{eq:linear_n-s}), and (\ref{eq:stokes}), can be solved
analytically for an unbounded system with a external force
field ${\bm f}({\bm r},t)$.  In this case, the velocity field is
given by
\begin{align} \label{eq:linear_n-s_solution}
{\bm v}({\bm r},t) = \int_0^t \int {\mathbf{H}}({\bm r}-{\bm r}',t-t')
\cdot {\bm f}({\bm r}',t') \, d^3r' dt' ,
\end{align}
with the time- and position-dependent hydrodynamic tensor
${\mathbf{H}}({\bm r},t)$
\cite{bede:74,espa:94}. An explicit expression for the tensor
${\mathbf{H}}({\bm r},t)$ has been provided by
\citet{espa:94}, \citet{huan:12,huan:13}, and \citet{thee:13}.
The creeping flow limit, i.e., the solution of Eqs.~(\ref{eq:n-s-comp}) and
(\ref{eq:stokes}), follows when ${\bm f}({\bm r},t)$ change
significantly slower with time than ${\mathbf{H}}({\bm
r},t)$. With time-independent ${\bm f}({\bm r})$,
integration over $t'$ in Eq.~(\ref{eq:linear_n-s_solution}) yields
\begin{align} \label{eq:stokes_solution}
{\bm v}({\bm r}) = \int {\mathbf{H}}({\bm r}-{\bm r}')
\cdot {\bm f}({\bm r}') \ d^3r' .
\end{align}
${\mathbf{H}}({\bm r})$ is the well-know Oseen tensor, with the
Cartesian components \cite{espa:94,dhon96}
\begin{align} \label{eq:oseen}
H_{\alpha \beta} ({\bm r})= \frac{1}{8 \pi \eta r}
\left[ \delta_{\alpha \beta} + \frac{r_{\alpha} r_{\beta}}{r^2}\right] ;
\end{align}
$\alpha, \beta \in \{x,y,z\}$ and $r = |{\bm r}|$. The Oseen
tensor, also denoted as Stokeslet, shows that hydrodynamic
interactions are long range, with a $1/r$ decay like the
Coulomb potential. The Oseen tensor is the Green's function of the
Stokes equation (\ref{eq:stokes}), which is easily seen, when the
point force ${\bm f}({\bm r}) = \delta({\bm r}) \hat {\bm e}$
(where $\hat{\bm e}$ is a unit vector) in the direction $\hat{\bm e}$
is employed.
Then, Eq.~(\ref{eq:stokes_solution}) yields
\begin{align} \label{eq:oseen_green}
{\bm v} ({\bm r})= \frac{1}{8 \pi \eta r} \left[
\hat {\bm e} + \frac{({\bm r} \cdot \hat {\bm e}) {\bm r}}{r^2}\right] .
\end{align}

The presence of confining surfaces modifies the fluid flow field.
At surfaces, the fluid velocity is typically very small, because
the collisions of fluid particles with the surface imply that the
molecules are scattered backwards and thereby transfer momentum
parallel to the wall. Thus, no-slip boundary conditions, with
${\bm v}({\bm r})=0$ at the surface, are usually employed. The
solution of Stokes equations is still given by
Eq.~(\ref{eq:stokes_solution}), however the Ossen tensor has to be
replaced by the Blake tensor \cite{blak71a}, which
satisfies the no-slip boundary condition.

\subsection{Dipole Swimmers}
\label{sec:dipole}

\begin{figure}
\centering
\includegraphics*[width=0.48\textwidth]{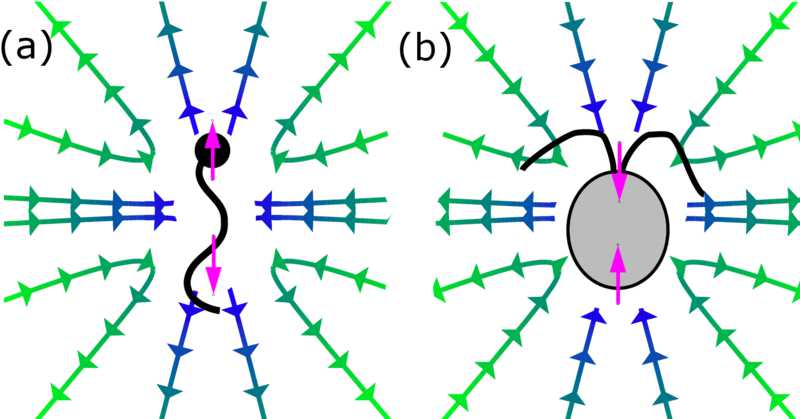}
\caption{Schematics of the flow field dipole swimmers,
(a) pusher and (b) puller.}
\label{fig:dipole_schematic}
\end{figure}

Most swimmers move autonomously, with no external force applied, and
hence the total interaction force of the swimmer on the fluid, and
{\em vice versa}, vanishes. In the simplest case, which actually
applies to many microswimmers like bacteria, spermatozoa, or
algae, the far-field hydrodynamics (at distances from the swimmer
much larger than its size) can well be described by a force
dipole \cite{laug09,ishi09}. This has been confirmed experimentally
for {\em E. coli} by \citet{dres:11}. Two classes of such dipole
swimmers can be distinguished, as shown schematically in
Fig.~\ref{fig:dipole_schematic}. If the swimmer has its ``motor"
in the back, and the passive body drags along the surrounding
fluid in front, the characteristic flow field of a ``pusher"
emerges, see Fig.~\ref{fig:dipole_schematic}(a). Similarly, if the
swimmer has its ``motor" in the front, and the passive body drags
along the surrounding fluid behind, the characteristic flow field
of a ``puller" develops, see Fig.~\ref{fig:dipole_schematic}(b).
It is important to notice that the flow fields of pushers and
pullers look similar, but with opposite flow directions. This has
important consequences for the interactions between swimmers and
of swimmers with walls, as will be explained below.

\begin{figure}
\centering
\includegraphics*[width=0.38\textwidth]{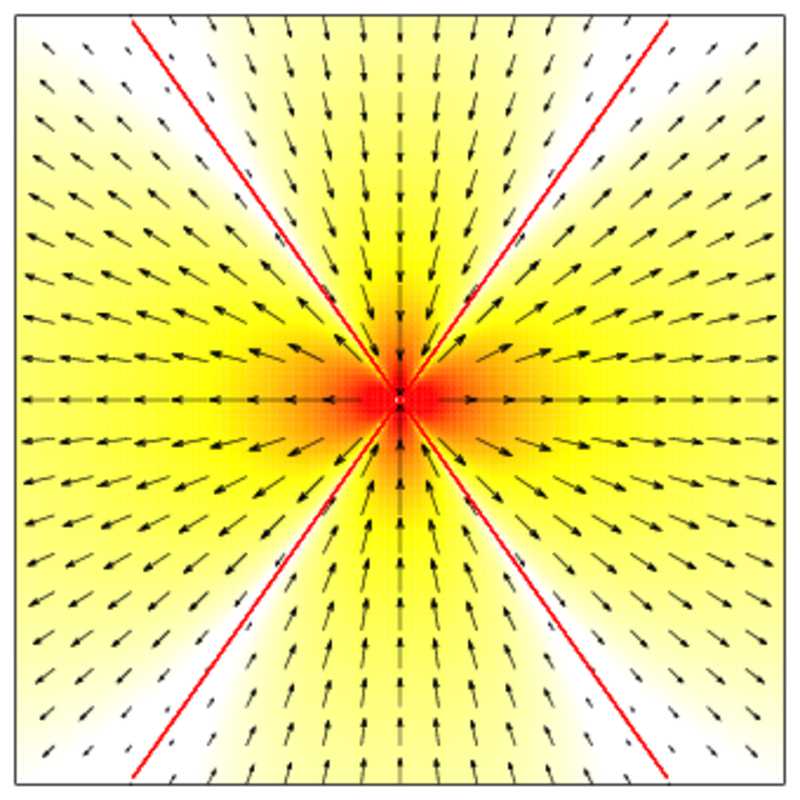}
\caption{
Flow lines in the far-field of a hydrodynamic dipole along the
horizontal direction, as given by Eq.~(\ref{eq:Oseen_dipole}). The separatrices
between the inflow and outflow regions are shown by thick red lines.}
\label{fig:dipole}
\end{figure}

Mathematically, the flow field ${\bm u}_d({\bm r}-{\bm r}_0)$ of a
hydrodynamic force dipole located at ${\bm r}_0$ can be obtained
very easily from the Oseen tensor (\ref{eq:oseen}) by considering two
opposite forces ${\bm f_0 = f_0 \hat {\bm e}}$ of equal magnitude at
${\bm r} = {\bm r}_0 \pm {\bm d}/2$ with ${\bm d =  d \hat {\bm e}}$.
An expansion to leading order in ${\bm d}/|{\bm r}-{\bm r}_0|$  yields
\begin{equation}
\label{eq:Oseen_dipole}
{\bm u}_d({\bm r}) = \frac{P}{8\pi \eta r^3}
\left[ -1 + 3 \frac{({\bm r}\cdot \hat{\bm e})^2}{r^2} \right] {\bm r} ,
\end{equation}
where $P=f_0 d$ is the dipole strength. Note that the flow
field of a force dipole decays as $1/r^2$ from the center of the
dipole, faster than the force monopole or Stokeslet, see
Eq.~(\ref{eq:oseen}). The flow lines of a hydrodynamic dipole
oriented in the $x$-direction are
shown in Fig.~\ref{fig:dipole}. There are two inflow and two
outflow regions in the $xy$-projection, which are separated by the
separatrices $y=\pm \sqrt{2} x$. In three dimensions, the outflow
region is a cone.

The flow field of a hydrodynamic dipole in front of a surface can
be obtained by the image method known from electrostatics.
Considering, for simplicity, a planar wall with slip boundary
conditions. The flow field is then given by
\begin{align}
{\bm v}_{w}({\bm r}-{\bm r}_0) =
   {\bm u}_{d}({\bm r}-{\bf r}_0;{\bm \hat e}) +
   {\bm u}_{d}({\bf r}-{\bm r}_1; \hat{\bm e}')
\end{align}
for a wall at $z=0$, with ${\bm r}_0=(x_0,y_0,z_0)$, ${\bm
r}_1=(x_0,y_0,-z_0)$, where $z_0>0$, and $\hat{\bm e}'$ the mirror
image of $\hat{\bm e}$ with respect to the $z=0$ plane. This
implies that at $z=0$ the velocity field $v_{w,z}$ perpendicular to
the surface vanishes identically, i.e., $v_{w,z}(z=0)\equiv 0$. The dipole
experiences a force near the surface, which is determined by the
hydrodynamic interactions between the dipole and its image. It is
given by the $z$-component of the flow field of the image charge
at the location of the dipole, thus
\begin{align} \label{eq:dipole_interact}
v_{w,z}(z_0) = - \frac{P}{32 \pi \eta z_0^2}
    \left[ 1- 3 (\hat{\bm e}\cdot \hat{\bm e}_z)^2 \right] \ ,
\end{align}
because $(\hat{\bm e}'\cdot \hat{\bm e}_z)^2 = (\hat{\bm e}\cdot
\hat{\bm e}_z)^2$. This result shows that the hydrodynamic force
is attractive to the wall, and that it decays as the dipole flow
field quadratic with the distance from the wall. The exact solution
for a no-slip wall \cite{berk08} yields the same functional
dependence on the angle and the wall distance as
Eq.~(\ref{eq:dipole_interact}), only the numerical prefactor in
Eq.~(\ref{eq:dipole_interact}) is smaller by a factor $2/3$.

\begin{figure}
\includegraphics*[width=0.47\textwidth]{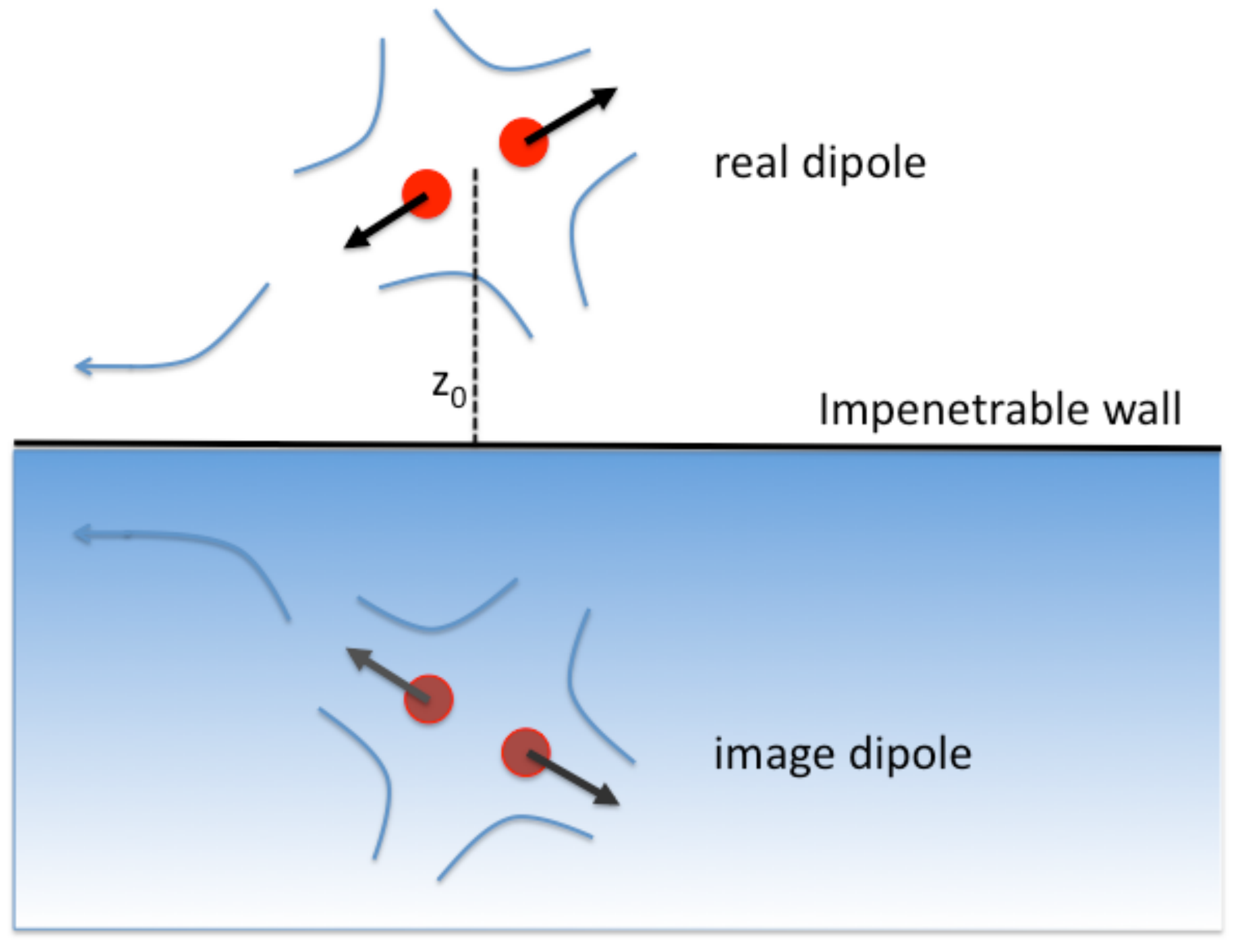}
\caption{Schematic representation of a dipole swimmer (pusher) near
a wall. An image dipole gives the correct boundary conditions
(impenetrability) at the slip wall.
}
\label{fig:dipole_wall}
\end{figure}

\subsection{Fluctuations and Noise}
\label{sec:fluct}

Fluctuations and noise may strongly affect the  motion of microswimmers.
There are two major sources of noise in active systems.
On the one hand, thermal fluctuations are present in both passive and
active systems. On the other hand, active processes themselves can give
rise to fluctuations. For example, in cilia beating, the motor proteins
are sensitive to the availability of ATP molecules, their number attached
to a filament varies in time, and different cilia will typically have
different numbers of active motors.
Furthermore, the smaller a system is, the more
important fluctuations become.

The importance of noise for the dynamics of microswimmers can
be estimated by the P{\'e}clet number $Pe$, which compares advective and
diffusive time scales.
The self-advection time scale is $L/v_0$, where
$v_0$ and $L$ are the typical swimmer velocity and length,
respectively, and the diffusive time scale is $L^2/D_T$, with the
translational diffusion coefficient $D_T$. Hence, a
P{\'e}clet number can by defined by $Pe=v_0 L/D_T$. For motion
in thermal equilibrium and
under the assumption of the swimmer to be spherical, $D_T$ can be related
to the rotational diffusion coefficient $D_R$ via $D_T= D_R L^2/3$.
Thus, the P{\'e}clet number is equivalently given by
\begin{equation} \label{eq:Peclet}
  Pe=\frac{v_0 }{L D_R}.
\end{equation}
For the dynamics of a self-propelled sphere, we consider
the persistent motion of a Brownian particle with independent stochastic
processes for its velocity  and
acceleration. Hence, the equation of motion for translation is
\begin{align} \label{eq:com}
\dot {\bm r} = {\bm v} + \frac{1}{\gamma_T} {\bm \Gamma} ,
\end{align}
which is coupled to changes of the velocity described by
\begin{align} \label{eq:rot}
\dot {\bm v} = - \gamma_R {\bm v} + {\bm \xi} .
\end{align}
Strictly speaking, in this description the magnitude of the velocity is
not preserved, only the relation $\langle {\bm v}^2 \rangle = v_0^2$ applies.
In Eqs.~(\ref{eq:com}) and (\ref{eq:rot}), $\gamma_T$ and $\gamma_R$ are
the translational and rotational friction coefficients, where the latter
is related to the rotational relaxation time $\tau_R$ by
$\tau_R=1/\gamma_R=1/((d-1)D_R)$.  ${\bm \Gamma}$ and ${\bm \xi}$ are
Gaussian and Markovian
stochastic processes, with zero mean and the moments
\begin{align} \label{eq:sto_trans}
\left\langle {\bm \Gamma} (t) \cdot {\bm \Gamma} (t')  \right\rangle & = 2 d \gamma_T^2 D_T  \delta(t-t'), \\ \label{eq:sto_rot}
\left\langle {\bm \xi} (t) \cdot {\bm \xi} (t')  \right\rangle & = 2 (d-1) D_R v_0^2 \delta(t-t')
\end{align}
in $d>1$ dimensions.
Equations (\ref{eq:com}) -- (\ref{eq:sto_rot}) yield the mean square
displacement of the sphere center \cite{uhle:30,risk:89}
\begin{align} \label{eq:msd_trans_rot}
\Delta {\bm r}^2 &= \langle ({\bm r}(t) - {\bm r}(0))^2 \rangle  \nonumber \\
                 &= 2 d D_T t + 2 \tau_R^2 v_0^2
                     \left[  t/\tau_R + \exp(-t/\tau_R)-1 \right],
\end{align}
which implies
$\Delta {\bm r}^2 = 2d D_Tt + v_0^2t^2$ for $t \ll \tau_R$ and
$\Delta {\bm r}^2  = (2dD_T + 2 v_0^2/[(d-1)D_R]t$ for $t\gg \tau_R$.
The latter defines the effective translational diffusion coefficient
\begin{align} \label{eq:D_eff}
D_{T,eff} =   D_T + v_0^2/[d(d-1)D_R] ,
\end{align}
which is much larger than $D_T$ for $v_0 \gg \sqrt{D_TD_R}$. Note that
for a sphere embedded in three-dimensional space, but confined in its
translational motion
to two dimensions, e.g., to a surface, the effective diffusion
coefficient still has the same effective diffusion constant (\ref{eq:D_eff})
with $d=3$, but $\Delta {\bm r}^2 = 4 D_{T,eff} t$. In particular,
these results show that  the P{\'e}clet number (\ref{eq:Peclet}), which
involves the rotational diffusion constant, remains appropriate to
characterize the importance of noise for self-propelled particles.

The mean square displacement of self-propelled Janus-colloids on a surface
has been measured as a function of fuel concentration by \citet{hows07}
as well as by \citet{volpe2011} for various illumination intensities.
The extracted effective diffusion coefficients are well described by
$D_{T,eff}$ of Eq.~(\ref{eq:D_eff}),
indicating that diffusion is dominated by self-propulsion and rotational
diffusion. 
\REV{The relevance of higher moments of the displacement distribution
has been addressed by \citet{zhen:13}. 
In particular, a non-Gaussian character of the  diffusiophoretic motion 
of self-propelled Janus spheres is theoretically predicted and 
experimentally confirmed.}

For active systems, in general, fluctuations have no reason to be
thermal (see Eqs.~(\ref{eq:sto_trans}), (\ref{eq:sto_rot})). While a
Gaussian distribution of the noise is still a reasonable first order
approximation, the  amplitude can be much larger than that due to thermal
fluctuations.
An instructive example is the {\em E. coli} rotary motor.
A motor rotation is typically driven by protons that generate torque by
passing into the cell via the motor \cite{Mora2009}.
However, the flux of protons is not continuous, but rather determined
by discrete events of proton translocation. The stochasticity of this
process creates a ``shot noise''. Modelling the motor in some detail,
\citet{Mora2009} derive an expression for the effective diffusion
coefficient of the flagellar rotation. At high load, the
effective diffusion coefficient is thermal, but at low
loads, diffusion increases and becomes dominated by the shot noise.

Measurements of the rotational
diffusion coefficient of {\em Chlamydomonas} yield
$D_R=0.4 \text{rad}^2/$s \cite{dres:11}. This value  can be compared with
that of a passive sphere of radius $R=5 \mu$m,
which is $D_R={k_BT}/(8\pi\eta R^3) \approx 3\times 10^{-3}\text{rad}^2/$s,
i.e., it is about two orders of magnitude smaller than the value of the active
microorganism.

Similarly, for {\em E.~coli} the values $D_R=0.057 \text{rad}^2/$s
\cite{dres:11} and $D_R \approx 2 \text{rad}^2/$s \cite{sara:12}
have been reported for non-tumbling and tumbling cells, respectively.
A study of paralyzed {\em E. coli} \cite{Tavaddod2011} yields the
rotational diffusion coefficient $D_R=0.032 \text{rad}^2/$s, which is only a
factor two smaller than the value of swimming but non-tumbling cells.
Thus, tumbling events evidently increase the rotational diffusion coefficient by more than an order of
magnitude.

We conclude that noise and fluctuations in microswimmer motion can be much
pronounced compared to that of the dynamics of respective passive objects,
and that activity-based noise can be the dominant contribution to observed phenomena.


\section{Swimming due to Flagellar Motion}
\label{sec:flagellar_swimming}

\subsection{Anisotropic Hydrodynamic Friction of Slender Bodies}
\label{sec:slender_rod}

A microorganism is able to swim forward in a fluid by wiggling or
rotating a flagellum, because the hydrodynamic friction of a long,
slender body in a viscous environment is {\em anisotropic}. This
can be demonstrated easily for a long and thin rod of radius $a$
and length $L$: it experiences less friction when pulled along its
axis than perpendicular to it.

We approximate the rod as a sequence of touching beads of radius $a$. In
general, the equation of motion of the $i$-$th$ bead is given by
\cite{bird:87,wink:07.1}
\begin{align}
\zeta \left(\dot {\bm r}_i - {\bm v}({\bm r}_i) \right) = {\bm F}_i ,
\end{align}
i.e., the frictional force is equal to the applied force. Here,
$\zeta= 6 \pi \eta a$ is Stokes friction coefficient for a sphere
with no-slip boundary conditions moving in a viscous fluid. The
fluid velocity ${\bm v}({\bm r}_i)$ is determined by the motion of all
other beads $j\ne i$ and follows from
Eq.~(\ref{eq:stokes_solution}). The force density on the
fluid originates from the forces of the various beads
\begin{align}
{\bm f}({\bm r}) = \sum_{i} {\bm F}_i \delta ({\bm r}- {\bm r}_i(t)) .
\end{align}
Thus, we obtain \cite{doi86}
\begin{align}
\dot {\bm r}_i(t) = \frac{1}{\zeta} {\bm F}_i + \sum_{j\neq i}
{\mathbf{H}}({\bm r}_i(t)-{\bm r}_j(t)) \cdot {\bm F}_{j} ,
\end{align}
which turns into  \cite{harn:96}
\begin{align} \label{eq:equation_of_motion_continuum}
\dot {\bm r}(s,t) = \frac{1}{3 \pi \eta} {\bm f}(s) + \int
{\mathbf{H}}({\bm r}(s,t)-{\bm r}(s',t)) \cdot {\bm f}(s')  \ ds'
\end{align}
in the continuum limit, where $s$ (with $-L/2 < s < L/2$) is the
contour coordinate along the centerline of the rod, and ${\bm
f}(s)$ the linear force density.

The anisotropic friction coefficients of a rod are defined by the force-velocity relation
\begin{align} \label{eq:anisotropic_friction}
{\bm F} = \zeta_\parallel {\bm v}_\parallel + \zeta_\perp {\bm v}_\perp
\end{align}
for the motion parallel (${\bm v}_\parallel$) and perpendicular
(${\bm v}_\perp$) to the rod axis. Calculations based on
Eq.~(\ref{eq:anisotropic_friction}), with constant friction
coefficients $\zeta_\parallel$ and $\zeta_\perp$, are denoted
``resistive-force theory" \cite{gray55b,ligh:76}.

To calculate $\zeta_\parallel$ and $\zeta_\perp$, it is
easiest to consider the special cases of a rod aligned along the $x$-axis of the reference frame and pulled parallel and
perpendicular to its long axis, respectively, with the constant force ${\bm F} = F \hat {\bm e}$. Since, we consider
a rigid body, the force density is ${\bm f}(s) = F \hat{\bm e}/L$, and the average rod velocity ${\bm v}_{rod}=\int {\dot {\bm r}(s)} ds/L$  becomes
\begin{align} \label{eq:v_rod}
{\bm v}_{rod}  =
   F \left[\frac{ \hat {\bm e}}{3 \pi \eta L} + \frac{( \hat {\bm e} +
          (\hat {\bm e}_x \cdot \hat {\bm  e}) \hat{\bm e}_x)}{4 \pi \eta L^2}
  \int_{2a}^{L}  \, \frac{L-s}{s} \,  ds \right] .
\end{align}
The lower cutoff of the integral
excludes a region of the thickness of the rod and prevents
self-interactions. Because $(\hat {\bm e}_x\cdot \hat{\bm e})
\hat{\bm e}_x=1$ and $0$ for parallel and perpendicular
orientation of the force, respectively, evaluation of the integral yields
\begin{align} \label{eq:friction_anisotropy}
\zeta_\perp = 2 \zeta_\parallel \ , \hspace*{5mm}
\zeta_\perp = \frac{4\pi \eta L}{\ln(L/2a)}
\end{align}
in the asymptotic limit of a long rod \cite{doi86}. It is therefore
easier to pull a long rod along its axis than perpendicular to it by a
factor two. The logarithmic divergence is a result of the
long-range nature of hydrodynamic interactions of different parts
of the rod, which {\em reduce} the friction coefficient compared
to that of a rod of non-interacting beads ($\sim L$). Corrections of the friction coefficients for a  more precise hydrodynamic calculation for a cylinder are provided in \citet{tira:84} and \citet{howa:01}.

\subsection{Swimming Velocity of Beating Flagella and Sperm}
\label{sec:flagellar_velocity}

The result (\ref{eq:friction_anisotropy}) together with
Eq.~(\ref{eq:anisotropic_friction}) can now be used to calculate
the swimming velocity of a sinusoidally beating flagellum. In this
case, the time-dependent shape is given by
\begin{equation} \label{eq:sinusoidal_beat}
y(x,t) = A \sin(kx-\omega t) ,
\end{equation}
where $A$ is the beating amplitude, $\omega$ the frequency, and
$k=2\pi/\lambda$ the wave number with the wave length $\lambda$.
The velocity of a segment of the flagellum at $x$ is then
\begin{equation}
v_y(x,t) = \frac{\partial y}{\partial t} = -A\omega \cos(kx-\omega t) ,
\end{equation}
where geometric nonlinearities are neglected.
With the local tangent vector (not normalized)
\begin{equation}
{\bm t}(x,t) = (1, A k \cos(kx-\omega t), 0)^T ,
\end{equation}
the velocity ${\bm v}(x,t) = (0, v_y(x,t), 0)$ can be decomposed
into ${\bm v}_\parallel = ({\bm v}\cdot {\bm t}){\bm t}/{\bm t}^2$
and ${\bm v}_\perp = {\bm v} - {\bm v}_\parallel$, with
\begin{equation} \label{eq:v_parallel_local}
{\bm v}_\parallel = - \frac{A^2 \omega k \cos^2(kx-\omega t)}
{1+A^2k^2 \cos^2(kx-\omega t)} \, {\bm t} .
\end{equation}
According to Eq.~(\ref{eq:anisotropic_friction}), this generates
the force
\begin{equation} \label{eq:f_flagellum_0}
F_x = (\zeta_\parallel-\zeta_\perp) \frac{1}{L}
\int  \frac{A^2 \omega k \cos^2(kx-\omega t)} {1+A^2k^2 \cos^2(kx-\omega t)}
\ dx
\end{equation}
in the swimming direction, while the force in the perpendicular
direction vanishes when averaged over the whole flagellum. For
small beating amplitudes, Eq.~(\ref{eq:f_flagellum_0}) can easily
be integrated, which yields the average propulsion force
\begin{equation} \label{eq:f_flagellum_1}
F_x = \frac{1}{2} (\zeta_\parallel-\zeta_\perp) A^2 \omega k .
\end{equation}
The swimming velocity then follows from $v_x \simeq
F_x/\zeta_\parallel$ as
\begin{equation} \label{eq:v_flagellum}
v_{flag} = -\frac{1}{2} \left(\frac{\zeta_\perp}{\zeta_\parallel}-1\right)
A^2 \omega k \ .
\end{equation}
This simplified calculation shows several important aspects of flagellar propulsion. First,
swimming is only possible due to the frictional anisotropy, i.e.
 $\zeta_\parallel \neq \zeta_\perp$.
Second, for a travelling wave in the positive $x$-direction, the
flagellum moves in the negative $x$-direction, i.e., movement is
opposite to the direction of the travelling wave. Third, the
swimming velocity increases linearly with the beating frequency
$\omega$ and the wave vector $k$, but quadratically with the
beating amplitude $A$. And finally, the swimming velocity is
independent of the fluid viscosity.

A more refined calculation has been performed by \citet{gray55b},
also employing resistive force theory, to determine the swimming
velocity of sperm. For the sinusoidal beating pattern
(\ref{eq:sinusoidal_beat}) and $\zeta_\bot/\zeta_\parallel=2$,
they find
\begin{align} \label{eq:sperm_velocity}
v_{sperm} & = \frac{1}{2} A^2 \omega k \left[1 + A^2 k^2   \right.  \\
          & \left.  + \sqrt{1+\frac{1}{2}A^2 k^2}
     \frac{3R_h}{L} \left(\ln\left(\frac{kd}{4\pi}\right)-\frac{1}{2} \right) \right]^{-1} .
\end{align}
Here, $L$ is the length of the flagellum and $R_h$ is the radius of
the head. The general conclusions wit respect to Eq.~(\ref{eq:v_flagellum})
remain valid, but additional effects appear. The second term in
the brackets of Eq.~(\ref{eq:sperm_velocity})---its origin is
already recognizable in Eq.~(\ref{eq:v_parallel_local})---arises
from the finite beating amplitude and implies  a {\em saturation}
of the velocity for large beating amplitudes. The last term in the
brackets describes the reduction of velocity due the drag of the
passive head.

\citet{frie10} employed a wave form with increasing
amplitude of the flagellar beat with increasing distance from the head
to describe the beat geometry of bull sperm, and use direct experimental input
for the beat amplitude and frequency. In this way, experimental
trajectories can be reproduced quite accurately by resistive force
theory, when the friction anisotropy is chosen appropriately.
This yields the friction anisotropy 
$\zeta_\bot/\zeta_\parallel=1.81 \pm 0.07$.

The swimming of sperm has also been analyzed by slender-body
theory \cite{hanc:53,ligh:76,john:79} (taking into account the
hydrodynamic interactions of different parts of the deformed flagellum as in
Sec.\ref{sec:slender_rod} for slender rods) by \citet{higd79a}.
Results agree with the resistive-force approach by \citet{gray55b}
within about 10\%.

A higher order solution, taking into account the full hydrodynamics, is
possible for an infinitely long flagellum in {\em two} spatial
dimensions (where hydrodynamics is of longer range than in three
dimensions) --- corresponding to an infinite sheet with a laterally
propagating  wave and transverse oscillations in three dimensions.
Here, the swimming velocity
\begin{equation}
\label{eq:flagellum_velocity_2D}
v_{sperm} = \frac{1}{2} A^2 \omega k \left(1-\frac{19}{16} A^2 k^2 \right)
\end{equation}
has already been obtained by  \citet{tayl51} in his pioneering
work. This result confirms all qualitative features discussed
above, but shows somewhat different numerical coefficients (which
is in part due to the different dimensionality).

The sperm structure or beating pattern is typically not completely
symmetric, but has some chirality. In this case, sperm swim on
helical trajectories \cite{cren89,gg:gomp08a}. In particular, the
helicity of the swimming trajectories is very pronounced for sea
urchin sperm \cite{cren96,kaup03,boeh05}.

\subsection{Propulsion by Helical Flagella}
\label{sec:propulsion_helix}

Resistive force theory \cite{gray55b,ligh:76} as well as slender
body theory \cite{hanc:53,ligh:76,john:79} have been applied to
describe propulsion of rotating helical flagella. Various aspect
of the approaches have been summarized by \citet{laug09} in their
review. Here, we briefly address slender-body results in the light
of recent experiments on macroscopic helices at low Reynolds
numbers \cite{rode:13}.

As shown by \citet{ligh:76}, the velocity of a point at $s$ along the contour of a flagellum of finite thickness can be described as
\begin{align} \label{eq:equation_helix_velocity} {\bm v}(s) & = \frac{1}{4 \pi \eta} {\bm f}_{\perp}(s) \\ \nonumber & + \int {\mathbf{H}}({\bm r}(s)- {\bm r}(s')) \cdot {\bm f}(s') \, \Theta(|{\bm r}(s)-{\bm r}(s')| - \delta) \ ds'
\end{align}
within the far-field approximation (as in Eq.~(\ref{eq:equation_of_motion_continuum})). Here, $\delta = a\sqrt{e}/2$ is the cutoff to avoid self-interactions, $\Theta(x)$ is Heaviside's step function, and ${\bm f}_{\perp}$ is the normal component of the Stokeslet strength (Eq.~\ref{eq:oseen}), i.e., \begin{align} \label{eq:force_stokes_dipol} {\bm f}_{\perp} = \left(\mathbf{I} - \hat{\bm t} \hat{\bm t}\right) \cdot {\bm f} , \end{align} where $\hat{\bm t}$ is the local tangent vector to the filament
and $\mathbf{I}$ the unit matrix \cite{laug09,rode:13}.

A helix oriented along the $z$-axis can be parameterized as
\begin{align} \label{eq:helix_location}
{\bm r}(s) = (R_h \cos \varphi, R_h \sin \varphi, \varphi P/(2 \pi) )^T .
\end{align}
Here, $\varphi = 2 \pi s \cos \vartheta /P$ is the helical phase,
$P$ the pitch of the helix, $\vartheta$ the pitch angle, and $R_h$
the helix radius. From Eq.~(\ref{eq:helix_location}), we obtain
the tangent vector and the force (\ref{eq:force_stokes_dipol}).
Assuming a very long helix $L/P \gg 1$, we can neglect end effects
and approximate the local force density by
\begin{align} \label{eq:force_def}
{\bm f}(s) = \left( - f_{\varphi} \sin \varphi,
f_{\varphi} \cos \varphi, f_z \right)^T
\end{align}
and the local velocity by
\begin{align} \label{eq:velocity_def}
{\bm v}(s) = \left( - \Omega R_h \sin \varphi, \Omega R_h \cos \varphi, v_z \right)^T .
\end{align}
Inserting Eqs.~(\ref{eq:helix_location}), (\ref{eq:force_def}),
and (\ref{eq:velocity_def}) in
Eqs.~(\ref{eq:equation_helix_velocity}) and
(\ref{eq:force_stokes_dipol}), respectively, the translational
velocity $v_z$ and  rotational frequency $\Omega$ can be
represented as
\begin{align} \label{eq:helix_velocity_force}
\left(
\begin{matrix}
v_z \\ \Omega
\end{matrix}
\right)
=
\left( \begin{matrix}
 A_{11} \ A_{12} \\ A_{12} \ A_{22}
\end{matrix}
\right)
\left( \begin{matrix}
 f_z \\ m_z
 \end{matrix}
\right)
\end{align}
by the pulling force density $f_z$ and the moment density $m_z=R_h f_{\varphi}$.
Note that we assume the helix to remain aligned
along the $z$-axis, i.e., other torques are compensated by
additional external forces. Neglecting end effects, the matrix
elements are given by
\begin{align} \nonumber
A_{11} & = \frac{1}{4 \pi \eta}\left( \sin^2 \vartheta + \frac{1}{\sin \vartheta}
\int_{\varphi_0}^{\varphi_L}\left(\frac{1}{\Phi} +
\frac{\varphi^2 \cot^2 \vartheta}{\Phi^3} \right) d \varphi \right) , \\ \nonumber
A_{12} & = \frac{1}{4 \pi \eta R_h}\left( - \frac{1}{2} \sin 2\vartheta  +
 \frac{1}{\sin \vartheta}
\int_{\varphi_0}^{\varphi_L}
\frac{\varphi \sin \varphi \cot \vartheta}{\Phi^3}  d \varphi \right) ,
\\ \label{eq:matrix_elements}
A_{22} & = \frac{1}{4 \pi \eta R^2_h}\left( \cos^2 \vartheta +
\frac{1}{\sin \vartheta}
\int_{\varphi_0}^{\varphi_L} \left( \frac{\cos \varphi}{\Phi} +
\frac{\sin^2 \varphi}{\Phi^3} \right) d \varphi \right) ,
\end{align}
with the abbreviations $\varphi_0= 2 \pi \delta \cos \vartheta/P$,
$\varphi_L = \pi L \cos \vartheta /P$, and $\Phi(\varphi) =
[4 \sin^2 (\varphi/2)+ \varphi^2 \cot^2 \vartheta]^{1/2}$
\cite{rode:13}. Inversion of Eq.~(\ref{eq:helix_velocity_force})
yields the thrust, torque, and drag as function of helix length
and driving frequency $\Omega$ or velocity $v_z$.
In the asymptotic limit $L\to \infty$,  the
thrust $F_T = L f $ obeys \cite{rode:13}
\begin{align}
F_T \sim - \frac{L}{\ln(L\cos \vartheta/R_h)}\Omega ,
\end{align}
i.e., shows a logarithmic dependence on the total helix height
$L\cos\vartheta$. The proportionality factors follow from
Eqs.~(\ref{eq:matrix_elements}). In particular, a logarithmic
dependence is obtained for $A_{11}$, since $\Phi \to \varphi \cot
\vartheta$ as $\varphi \to \varphi_L \to \infty$ ($L\to \infty$).
The other terms converge to finite values. As before, the
logarithm appears due to hydrodynamic interactions.

\begin{figure}
\begin{center}
\includegraphics*[width=0.45\textwidth]{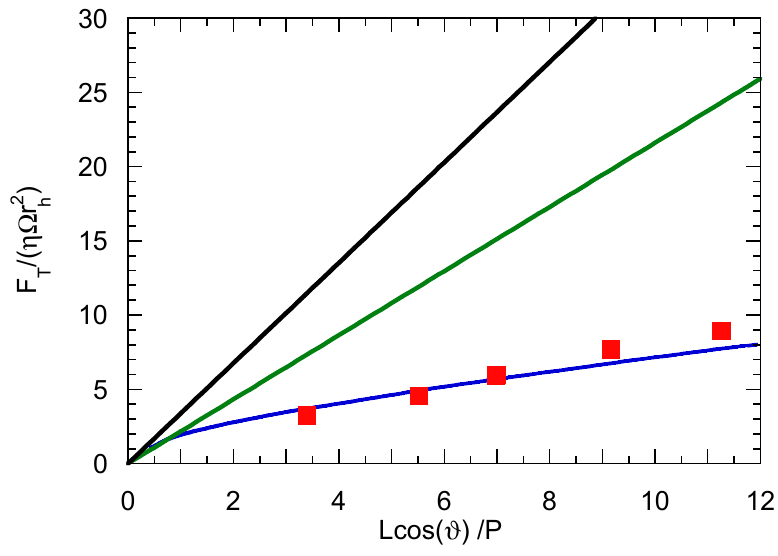}
\caption{Thrust of a helical flagellum as function of its length
according to \citet{rode:13}.
The experimental data (squares) agree well with the slender body
theory by \citet{ligh:76} (blue), the regularized
Stokeslet theory by \citet{cort:05}, and the slender body theory
by \citet{john:80}. The black and green lines are obtained by
resistive force theory according to \citet{ligh:76} and \citet{gray55b},
respectively.
}
\label{fig:helix_thrust}
\end{center}
\end{figure}

As an example, Fig.~\ref{fig:helix_thrust} shows experimental results for the thrust of helical flagella of various lengths obtained in experiments on  macroscopic scale models at low Reynolds numbers by \citet{rode:13}. The data compare very well with the slender body
theories by \citet{ligh:76} and \citet{john:80}, respectively, and
the regularized Stokeslet approach by \citet{cort:05}. The above
asymptotic calculations apply for $L/P> 10^3$. Over the range of
ratios $L/P$ relevant for bacteria ($3 \lesssim L \cos \vartheta
/P \lesssim 11$), there is a quantitative difference between the
full theory and the asymptotic approximation. However, the
qualitative length dependence is well captured.
More results are presented by \citet{rode:13}.

\begin{figure}
\begin{center}
\includegraphics*[width=0.4\textwidth]{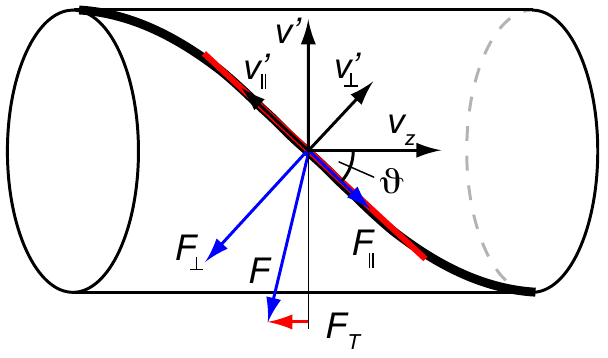}
\caption{Helical segment moving in a viscous fluid. Only half of a
helical pitch is shown. The drag-based thrust force $F_T$ appears by
the motion of the red rodlike segment in the direction $v'$
(see also \citet{laug09}).
}
\label{fig:helix_thrust_sketch}
\end{center}
\end{figure}

In order to determine the swimming properties of a bacterium, we
have to consider both, the flagellum bundle as well as the body.
For simplicity, we adopt resistive force theory to illustrate the
dependence of the swimming velocity on the motor torque rotating a
single flagellum \cite{laug09}. Inversion of
Eq.~(\ref{eq:helix_velocity_force}) yields the force and torque on
a flagellum due to translation and rotation. Considering the limit
of small pitch angles $\vartheta \ll 1$, elimination of rotation
by external torques yields $F_z \approx \zeta_{\parallel} v_z$.
Preventing translation, rotation with the velocity $v'=R_h \Omega$
yields the thrust force $F_z=F_T = (\zeta_{\parallel} -
\zeta_{\perp})\theta v' = (\zeta_{\parallel} - \zeta_{\perp}) R_h
\Omega$ (cf. Fig.~\ref{fig:helix_thrust_sketch}) and the momentum
$M_z= \gamma_{\perp} R_h v' = \gamma_{\perp} R_h^2 \Omega$ with
the friction coefficients (\ref{eq:friction_anisotropy})
\cite{laug09}. Hence, we obtain
\begin{align} \label{eq:helix_thrust}
\left(
\begin{matrix}
F_z \\[5pt] M_z
\end{matrix}
\right)
=
\left( \begin{matrix}
 \zeta_{\parallel}  \ \ \ - (\zeta_{\perp} - \zeta_{\parallel})  -\vartheta R_h \\[5pt]
 -(\zeta_{\perp} - \zeta_{\parallel} )\vartheta R_h \ \ \  \zeta_{\perp} R^2_h
\end{matrix}
\right)
\left( \begin{matrix}
 v_z \\[5pt] \Omega
 \end{matrix}
\right) .
\end{align}
More precise resistive force theory results have been presented by
\citet{chat:06}, \citet{maga:95}, and \citet{purc:97}. Alternatively,
the inverse matrix of Eq.~(\ref{eq:helix_velocity_force}), with the elements
of Eq.~(\ref{eq:matrix_elements}), yields a more
precise description within slender-body theory.

Approximating the cell body by a sphere of radius $R_b$
and  assuming $R_b \ll L$, the frictional body force $F_b$ and
the body torque $M_b$ are
\begin{align} \nonumber
F_b =  - \zeta_b v_z \ , \\
M_b = - \zeta_r^b \Omega_b \ ,
\end{align}
where $\zeta_b=6\pi\eta R_b$ and $\zeta_r^b= 8\pi\eta R_b^3$ are the
translation and rotational friction coefficients. The helix is
driven by a rotary motor with the frequency $\Omega_m$ relative to
the body.
In response, the helix and body rotate
with the frequencies $\Omega$ and $\Omega_b$. These frequencies are
related by $\Omega + \Omega_b = \Omega_m$. Since the whole
bacterium is force and torque free, i.e., $F_z + F_b = 0$ and
$M_z+M_b =0$, its swimming velocity is given by \cite{laug09}
\begin{align} \label{eq:v_bact_swim}
v_z \approx \vartheta \frac{(\zeta_{\perp}
-\zeta_{\parallel}) \zeta_r^b}{\zeta_{\parallel} \zeta_{\perp} R_h} \Omega_m .
\end{align}
The friction coefficient $\zeta_b$ does not appear, since we assume
$\zeta_{\parallel} \gg \zeta_b$ $(L \gg R_b)$. Evidently, swimming
is again---as in the sperm case---only possible due to frictional
anisotropy. Moreover, $v_z$
depends linearly on the body rotational friction coefficient.
Hence, without body, the bacterium could not swim.
Due to the approximation $\vartheta \ll 1$,
$v_z$ depends linearly on the pitch angle. Changing the handedness
of the helix leads to a change of the swimming direction.

We like to mention that a helix driven by
an external torque also swims \cite{ghosh2009}. However, it is not torque free and
therefore is not an autonomes swimmer. Under the same assumptions as above,
the swimming velocity is
$
v_h \approx  \vartheta \Omega [(\zeta_{\perp}-\zeta_{\parallel}) \zeta_r^h]/
                     [\zeta_{\parallel} \zeta_{\perp} R_h] ,
$
very similar to Eq.~(\ref{eq:v_bact_swim}), but now with
the helix frequency $\Omega =M/\zeta_r^h$, determined by the applied
torque $M$, and the overall helix rotational friction coefficient $\zeta_r^h$.


\section{Swimming near Surfaces}
\label{sec:surface}

Surfaces, interfaces, and confinement are ubiquitous in the microswimmer
world. Microswimmers being small, they might be expected to be typically
far away from surfaces.
There are three important points to remember, however. First, many
biological microswimmers regularly encounter surfaces and
confinement, from sperm cells in the reproductive tract to
microorganisms in the soil \cite{Foissner1998,Or2007}.
Second,  microorganisms often rely on the presence of surfaces for their
function and survival; for example, bacteria form biofilms on surfaces for
spreading, to enhance cell-cell exchange and nutrient uptake.
Third, equally important is that modern microfluidic devices can be used to
investigate, control, and manipulate microorganisms in many ways
\cite{Denissenko2012,Kantsler2013}.
In order to make predictions and to interpret and understand experimental
results, it is thus crucial to account for
effects of surfaces and confinement in theoretical models and descriptions.

A generic phenomenon of  microswimmers near
surfaces is an effective surface accumulation.
Already in 1963, Rothschild discovered and quantified an accumulation of
sperm cells near a glass cover slide \cite{roth63}. Other microswimmers
like {\em E.~coli} \cite{berk08} or {\em Chlamydomonas} \cite{Kantsler2013}
also accumulate at walls. Two mechanisms have been suggested to explain this
effect, hydrodynamic interactions and propulsion together with steric
interactions. We will address these two mechanisms in the following sections.

Other surface induced phenomena include rectification of microswimmer
motion by ratchets
\cite{Tailleur2009,Kantsler2013,Berdakin2013}, rotation of microgears
in bacterial suspensions \cite{DiLeonardo2010}, collective surface
adhesion in clusters \cite{wens08} or geometric traps for
microswimmers \cite{Kaiser2012}.

\subsection{Hydrodynamics of Surface Capturing}

The far-field interactions of microswimmers can be understood in
terms of a multipole expansion.
For a force free swimmer, the dominant term is the dipole term, which
distinguishes pushers from pullers, see Sec.~\ref{sec:dipole}.
A microswimmer at a distance $z$ from a no-slip wall, with an
orientation angle $\theta$ between the swimming direction and
the surface normal vector (pointing into the fluid), experiences a
angular velocity \cite{berk08}
\begin{equation}
\label{eq:rotation_wall}
\Omega_r(\theta,z) = -\frac{3P \cos\theta \sin\theta}{64 \pi \eta z^3}
   \left[ 1+ \frac{(\gamma^2-1)}{2(\gamma^2+1)} (1+ \cos^2\theta) \right]
\end{equation}
and a drift velcoity \cite{berk08}
\begin{equation} \label{eq:dipole_wall}
u_z(\theta,z) = - \frac{3P}{64\pi\eta z^2} \, (1-3\cos^2\theta) \ .
\end{equation}
where $P$ is the dipole strength and $\gamma$ the aspect ratio of the
swimmer shape.

Equation (\ref{eq:dipole_wall}) allows several interesting predictions.
First, the hydrodynamic interactions decay slowly as $1/z^2$ with increasing
distance from the surface, as already explained in Sec.~\ref{sec:dipole}.
Second, for pushers (like sperm), the
hydrodynamic interaction is attractive for orientations nearly
parallel to the wall (with $\theta$ near $90^\circ$), but
repulsive for orientations nearly perpendicular to the wall (with
$\theta$ near $0^\circ$); for pullers (like {\em Chlamydomonas}),
the hydrodynamic interaction is
repulsive when they are swimming parallel to the surface.
However, in these considerations, the rotation of the swimmer orientation
due to hydrodynamics interactions has not yet been taken into account.
Equation (\ref{eq:rotation_wall}) shows that for the pusher,
swimming parallel to the surface and
being slowly attracted to it is indeed the stable state.
On the other hand, for pushers the parallel orientation is unstable,
a reorientation toward the surface occurs, and the microswimmer moves
to the surface head-on \cite{berk08,Spagnolie2012}.

\begin{figure}
\includegraphics[width=0.42\textwidth]{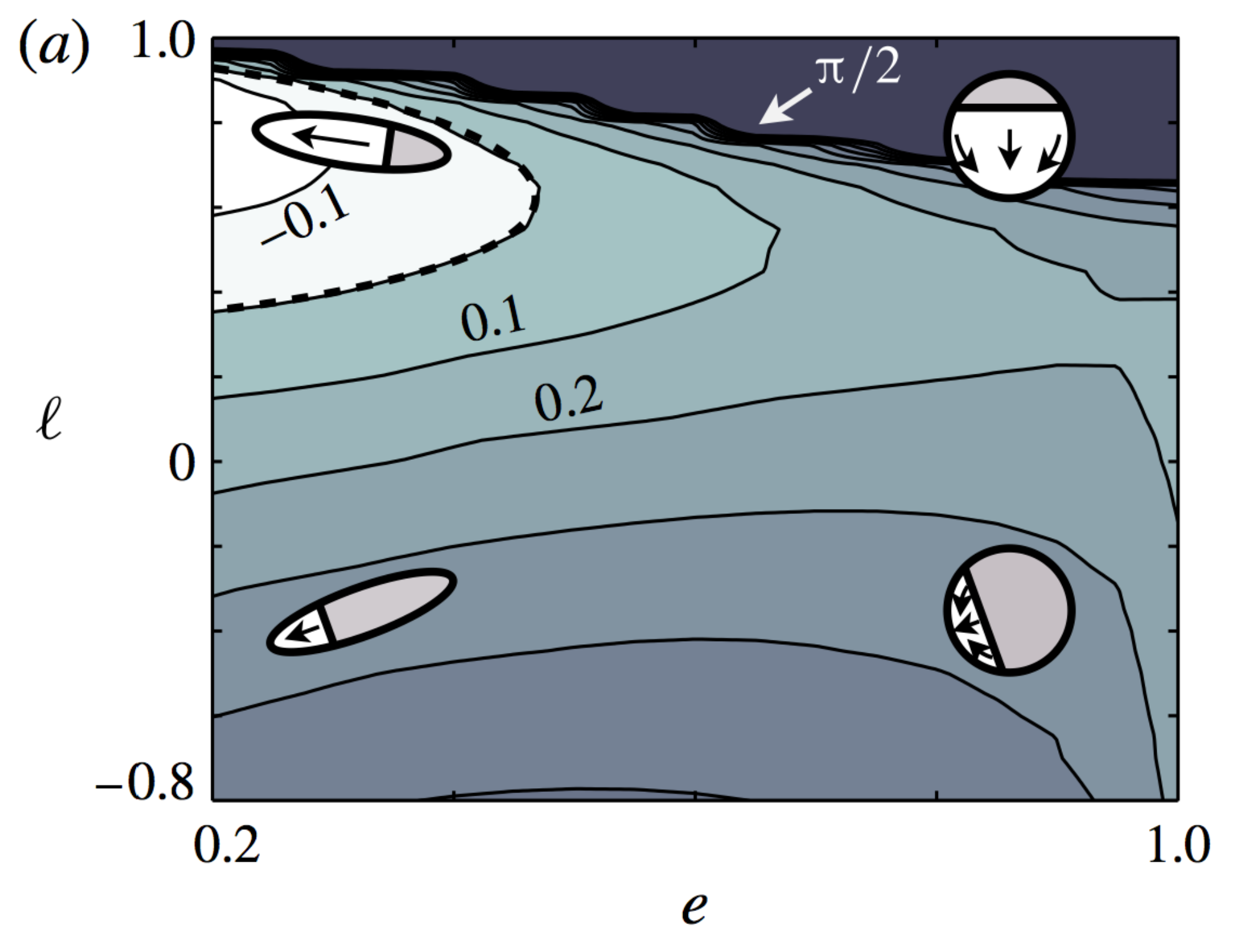}
\caption{Contour plot of stable equilibrium pitching angles of
ellipsoidal microswimmers (pusher), at fixed distance
(equal to the long axis of the ellipsoid)
from a wall, as a function of the inverse aspect
ratio $e$ and the asymmetry parameter $\ell$ for the active part of the
surface, as indicated by the swimmer shapes. Positive pitching angles
are pointing away from the surface. The swimmers are driven by an active
region of the rear end, as indicated by the arrows, where  $(\ell+1)/2$
is the active fraction of length.
From \citet{Spagnolie2012}.
\label{fig:spagnolie}
}
\end{figure}

As both pushers and pullers come closer to the surface, higher orders
in the multipole expansion become important.
For spherical or ellipsoidal squirmers, the importance of
higher order terms has been studied \cite{Spagnolie2012}.
The swimmers
are driven by imposing a surface velocity on part of the particle,
like in the squirmer model (see Sec.~\ref{sec:theoretical_models}).
A boundary-integral formulation of the Stokes equation is used to
generate numerically ``exact" results for comparison. In the
multipole expansion, a general axisymmetric swimmer is described as
a linear combination of fundamental
solutions to the Stokes equations: a Stokeslet dipole, a source dipole,
a Stokeslet quadrupole, and a rotlet dipole.
For spherical and ellipsoidal microswimmers,
the multipole expansion is found to be surprisingly accurate, sometimes down to
surface distances of a tenth of the swimmer length.

Physically, the main result is --- see Fig.~\ref{fig:spagnolie} ---
that microswimmers that are both slender
and sufficiently active (but not completely active, $\ell \neq 1$)
exhibit
pitching equilibria with their noses down toward the wall, while
microswimmers that are not sufficiently slender or not sufficiently active
exhibit pitching equilibria with their noses turned up away from the wall
\cite{Spagnolie2012}.
This has important consequences, because a swimmer which points away from
the wall will usually not remain at the wall, but excape due to is
forward motion.
Furthermore, hydrodynamically caused rotation rates for microswimmers
approaching a wall are found to be typically small, and for large impact
angles not large enough to avoid a collision between the swimmer and the
wall \cite{Spagnolie2012}.  Thus, eventually a more detailed, swimmer-specific
modelling is required.

Ellipsoidal squirmers, with surface
velocities given by a sum of Legendre polynomials (compare
Eq.~(\ref{eq:squirmvel}) for the spherical squirmer), were
studied by \citet{Ishimoto2013}.
A fixed-point analysis yields stable swimming positions for
pushers and pullers over a wide range of aspect ratios and
$B_2$- and $B_3$-amplitudes.
However, when $B_2$ and $B_3$ are too small, no fixed point is found
because hydrodynamic interactions are too weak. Pusher spheres show only
unstable fixed points (unless $B_3$ is very strong), while puller
spheres display stable fixed points with a swimmer orientation toward
the wall. As the aspect ratio increases, puller trajectories at a fixed
distance from the wall become more unstable,
and conversely pushers become stable. Pushers at a fixed point have
an orientation away from the wall.
Some of the unstable fixed points are found to be surrounded by stable
limit cycles, which correspond to swimmers which change their distance from
the wall periodically. For an elongated puller with aspect ratio $a=2$, the
distance from the wall is predicted to vary
between its size and three times its size \citet{Ishimoto2013}.
Furthermore, it is predicted that a change of boundary conditions from no-slip
to slip significantly changes the location and characteristics of
fixed points, and thereby of the swimming behavior near surfaces.

Another important aspect is the competition of hydrodynamic interactions
with rotational diffusion \cite{elge09,dres:11}. When a pusher is deviating
by a (small) angle $\delta \phi$ from parallel alignment with the wall due to
rotational diffusion, then --- in the far-field approximation ---
it takes a time of order $t_r=\delta \phi/\Omega_r$
(see Eq.~(\ref{eq:rotation_wall})) to become aligned again due to
hydrodynamic interactions.  During this time, it can swim a distance
$\Delta z_1= v_0 \sin(\delta\phi) t_r$ away from the wall due to self-propulsion,
but drifts toward the wall by a distance $\Delta z_1 =u_z t_r$ due to
hydrodynamic interactions, see Eq.~(\ref{eq:dipole_wall}). This implies that
for $\delta \phi \gtrsim (r_0/z)^2$, the effective swimmer velocity points
away from the wall, where $r_0$ is the swimmer size and $z$ is the distance
from the wall, and in the time interval $t_r$ travels a distance
$\Delta_z \sim (z^3/r_0^2) \delta \phi$. For distances $z$ a few times $r_0$,
this implies $\Delta_z \sim z$. Thus, for small swimming velocities and
large angular fluctuations, a microswimmer is expected to exhibit also
large fluctuations in its distance from a wall. Here, the importance
of orientational fluctuations can be quantified by the orientational
Peclet number $Pe_\phi(z) = \Omega_r(z)/D_R$.
Of course, very close to a wall, hydrodynamic interactions become more
important, but also the dipole approximation breaks down.

A similar conclusion was reached by \citet{dres:11}, who considered
cell-cell scattering. By estimating the mean-square angular change of
orientation in cell-cell encounters, both due to hydrodynamic interactions
and to rotational diffusion, \citet{dres:11} estimated a hydrodynamic
horizon $r_H$, beyond which hydrodynamic interactions become irrelevant.
For non-tumbling {\em E.~coli}, $r_H$ is found to be comparable to the
length of the cell body, about $3 \mu$m. However, other effects like
flagellar interactions would become important at such short distances.
Of course, rotational diffusion becomes less important with increasing
microswimmer size.

\subsection{Propulsion-Induced Surfaces Accumulation}

Hydrodynamics is not the only mechanism that can explain
accumulation of swimmers at surfaces.
Indeed, it has to be realized that any self-propelled particle in
confinement will eventually encounter a surface. Without a reorientation,
the particle will just stay there.
Therefore, rotational diffusion is required to induce a detachment from
the wall. In order to elucidate this adhesion mechanism,
and how noise drives a self-propelled particle away from a wall,
it is interesting to consider the behavior of ``Brownian" rods ---
in the absence of any hydrodynamic interactions.
In this case, excluded-volume interactions favor parallel
orientation near the wall, while the noise leads to
fluctuations of the rod orientation and thereby an effective
repulsion from the wall. The competition of these two effects
gives rise to an interesting adsorption behavior \cite{elge09,li09}.

\begin{figure}
\centering
\includegraphics[width=0.40\textwidth]{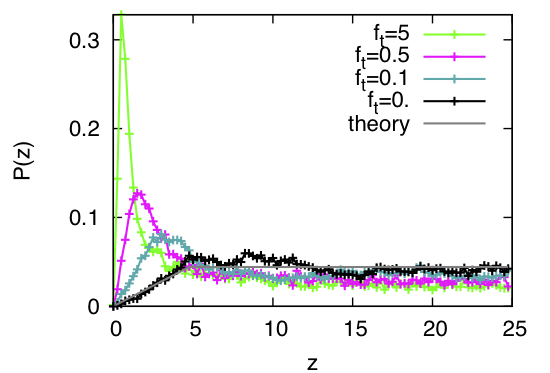}
\includegraphics[width=0.40\textwidth]{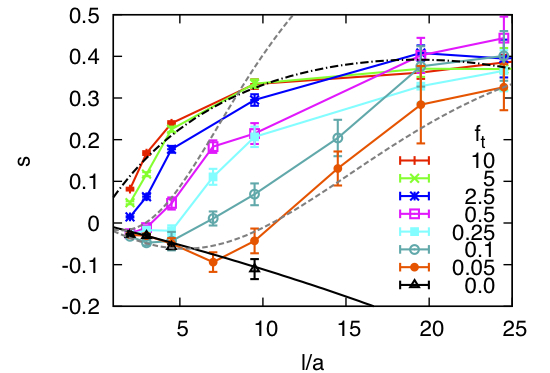}
\caption{
(Top) Probability density $P(z)$ of self-propelled rods as function of
the distance $z$ from the surface, for various propelling forces $f_t$.
The rod length is $l\simeq 10a$ (where $a$ is the rod diameter), the
walls are located at $z=0$ and $z=5 l$.
A solid gray line shows the density profile of passive rods.
(Bottom) Surface excess $s$ as a function of scaled
rod length $l/a$, for various propelling forces $f_t$, as indicated.
The (black) dashed-dotted line is the scaling result  in the
ballistic regime (see Eq.~(\ref{eq:p_ball})),
the (gray) dashed lines are scaling results in the diffusive regime
for $f_t=0.5$ and $f_t=0.05$.
From \citet{elge09}.
\label{fig:rod_excess}
}
\end{figure}

Results of Brownian dynamics simulations \cite{elge09} are shown in
Fig.~\ref{fig:rod_excess}. While passive rods are depleted from
the surface (because their entropy is reduced near the surface due
to restricted orientational fluctuations), active rods show an
increased probability density near the surface, which grows with
increasing propulsion force $f_t$, see Fig.~\ref{fig:rod_excess}(left).
In addition,
the surface accumulation of the rods strongly depends on the rod length $l$.
The surface excess --- the integrated probability density to find
a rod near the surface relative to a uniform bulk density distribution ---
is shown in Fig.~\ref{fig:rod_excess}(right). The results show
(i) that short rods show little or no surface aggregation for any propelling force,
and (ii) that the surface excess initially increases with increasing $f_t$
and $L$, but then saturates and becomes nearly independent of $f_t$
and $L$ for large propulsion forces and rod lengths \cite{elge09}.

The physical mechanism behind this behavior is as follows.
A rod hits the surface at some point in time, because swimming
directions in the bulk are randomly distributed. After contact
with the wall, it gets reoriented parallel to the wall. Then it
moves parallel to the wall, slowly wiggling its trajectory from the
wall again, until it is sufficiently far from the wall that frequent
contacts no longer occur.

The surface accumulation of self-propelled rods can be understood more
quantitatively by exploiting an analogy of the trajectories of rod-like
microswimmers with the conformations of semi-flexible polymers.
The self-propulsion of the rod, combined with rotational diffusion
leads to a trajectory with a persistence length
\begin{equation}
\xi_p \sim v/D_R \sim \eta v l^3/k_BT
\end{equation}
similar to a fluctuating polymer. The key difference is however, that
the path is directed, and thus not forward-backward symmetric like a polymer.
The trajectory of a rod colliding with the wall displays a sharp
kink, impossible for a semi-flexible polymer. Leaving the wall, on the
other hand, happens at a shallow angle, similar to a semi-flexible
polymer attached parallel to the wall at one end.
This difference in possible conformations leads to an effective
attraction of rod-trajectories to the wall, and a repulsion of polymers.
In the end, the polymer analogy predicts very well the scaling of the
surface excess \cite{elge09}.
In particular, in the {\em ballistic regime}, with $\xi_p \gg d$,
the scaling arguments predict the
probability  to find the rod near the wall
\begin{equation} \label{eq:p_ball}
p=l/(l+a_B d) ,
\end{equation}
where $a_B$ is a fitting parameter. Note that in this regime Eq.~(\ref{eq:p_ball})
is independent of the propulsion velocity, as observed in simulations.

This effect of accumulation due to the combination of Brownian motion
and self-propulsion has later been confirmed experimentally for {\em
C.~crecentus} and {\em E.~coli} \cite{li09}.

\begin{figure}
\begin{center}
\includegraphics[width=.42\textwidth]{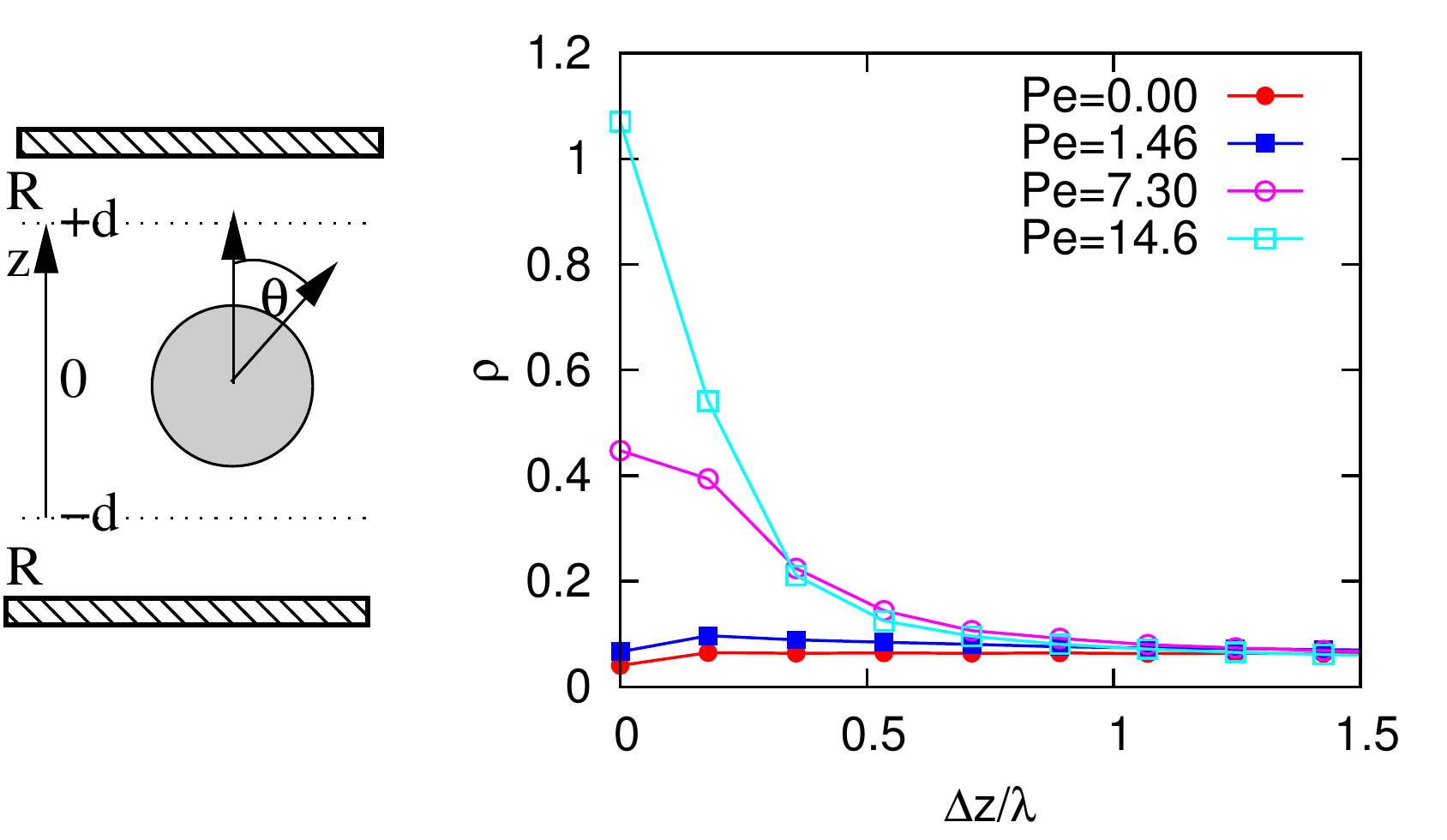}
\caption{
(Left) A self-propelled Brownian sphere of radius $R$ is confined between
two solid walls at $z=\pm (d+R)$. The orientation of the propulsion
direction relative to the $z$-axis, is denoted by $\theta$.
(Right)
Probability density $\rho(\Delta z)$ to find a particle
at a distance $\Delta z$ from the wall. At zero
Peclet number $Pe=v_0/\sqrt{D_rD}$, the probability density is uniform
beyond the short range of the repulsive wall. With increasing Peclet
number particles accumulate near the wall. Results are shown for a system
with wall separation $d/\lambda=15.6$. The curves correspond to a
surface excess of $s=0$, $0.05$, $0.24$, and $0.38$ with increasing
Peclet number, respectively.
From \citet{elge13b}.
\label{fig:schematic}
}
\end{center}
\end{figure}
\begin{figure}
\includegraphics[width=.42\textwidth]{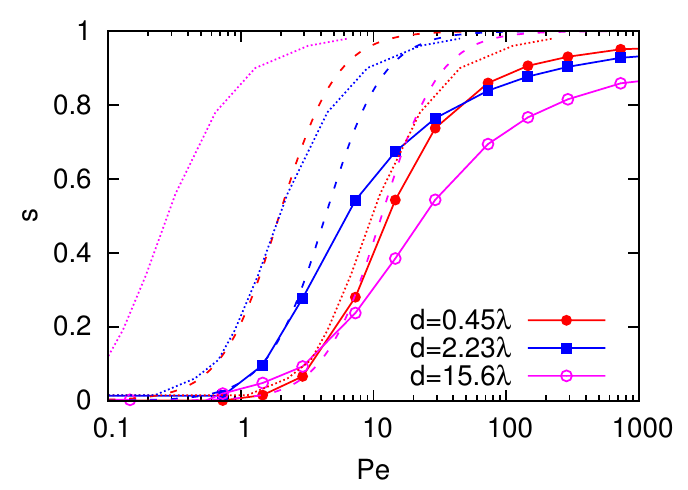}
\caption{
Surface excess $s$ of active Brownian spheres as a function of P{\'e}clet
number $Pe$ for various wall separations $d$, as indicated.
Results from the analytic calculation for very narrow channels (dotted lines)
match well the simulations for
very narrow channels, but fail for wider channels.
The approximation for small P{\'e}clet numbers (dashed lines)
works well for large wall separations. It
overestimates the surface excess for large $Pe$. All analytic
expression have no adjustable parameters.
From \citet{elge13b}.
\label{fig:s}
}
\end{figure}

An even simpler model microswimmer is a self-propelled Brownian sphere.
The essential difference to the self-propelled rod is that the
sphere has no alignment interaction with the wall \cite{elge13b}.
Again, self-propulsion leads to accumulation of particles near the wall. The
simulation results shown in Fig.~\ref{fig:schematic} demonstrate that the
(normalized) probability density $\rho(\Delta z)$ to find a
particle at a distance $\Delta z$ from the wall is strongly peaked close to
the wall for $Pe\gtrsim 5$.
Figure \ref{fig:s} shows the surface excess  $s$ as a function of the P{\'e}clet
number $Pe=v_0/\sqrt{D_R D_T}$ for different channel widths.
Note that $s$ does not
saturate at a value $s_{max}<1$ as is observed for self-propelled
rods above, but approaches unity for large $Pe$ (complete adhesion).

The large degree of symmetry allows for an analytic treatment via the
Fokker-Planck equation
\begin{eqnarray}
\label{eq:main}
\partial_t \rho(z,\theta,t) &=& D_r \frac{1}{\sin(\theta)}
  \partial_\theta \left[ \sin(\theta)\partial_\theta\rho(z,\theta,t) \right] \\
  &-& v_0 \cos(\theta) \partial_z\rho(z,\theta,t)
            + D \partial^2_{z}\rho(z,\theta,t) ,  \nonumber
\end{eqnarray}
where the angle $\theta=0$ corresponds to particles oriented in the
positive $z$-direction.
This equation already demonstrates the main origin of surface
accumulation.
The rotational diffusion is independent of the spatial position, but
particles are driven to one of the chamber walls depending on their
orientation.
Thus particles oriented toward the top, accumulate at the top, those
pointing down, accumulate at the bottom wall. Less particles remain in
the center.
Solutions for small Peclet number and narrow channels are depicted in
Figure \ref{fig:s}, and work well within their respective limits.

\REV{Within a concave confinement (i.e. a ring or an ellipse in two dimensions), 
slow rotational diffusion suppresses surface detachment; 
the particle moves along the confinement wall, until it reaches a location
where its orientation is perpendicular to the wall. This leads to complete 
adsorbtion of self-propelled particles in small enough closed environments, 
and to accumulation in regions of largest wall curvature (because here
the largest reorientation by rotational diffusion is required to remove 
the particle) \cite{fily14b}.}

\begin{figure}
\includegraphics[width=0.40\textwidth]{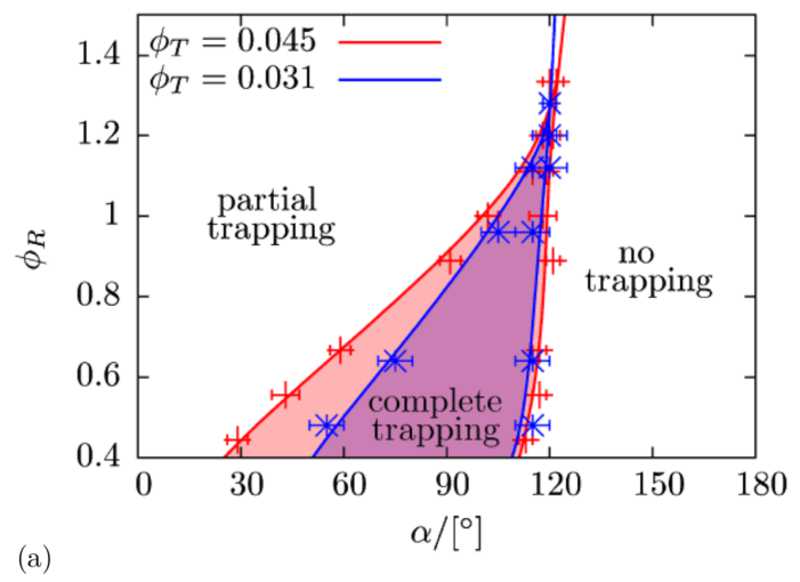}
\includegraphics[width=0.48\textwidth]{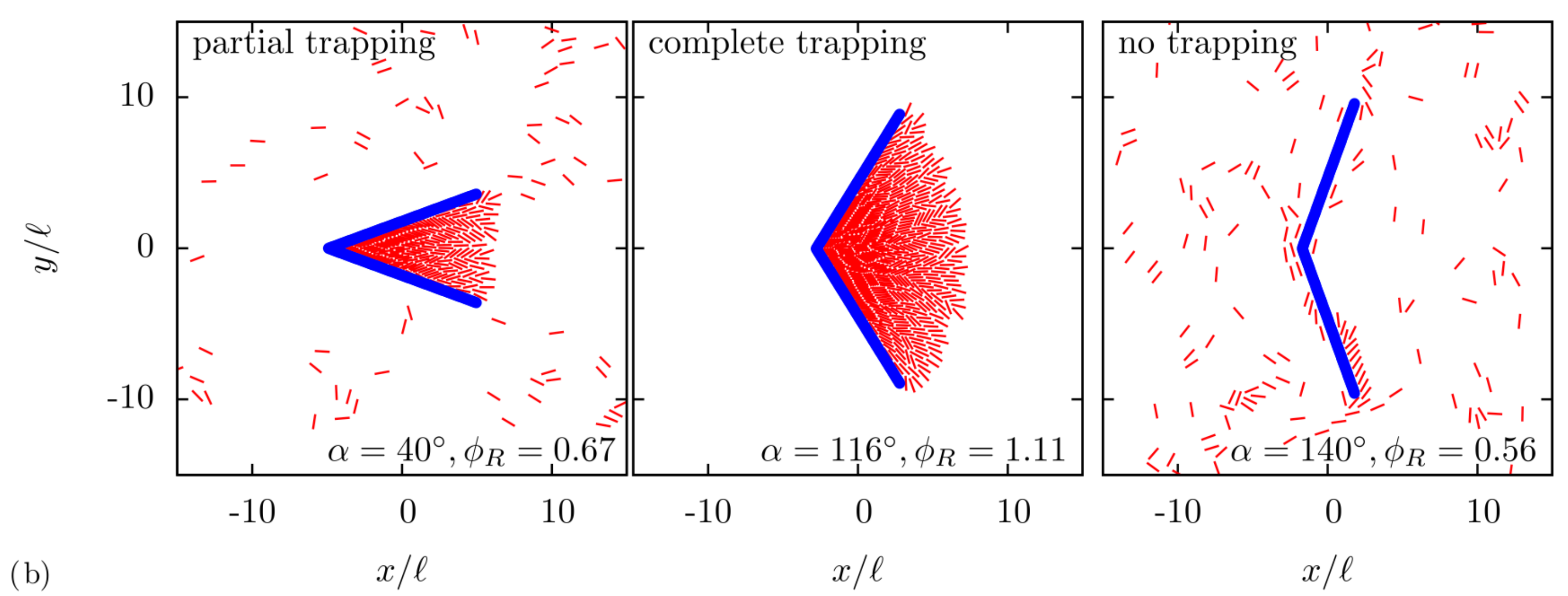}
\caption{
(a) Phase diagram and (b) snapshots from simulations of many self-propelled
rods with funnel-like traps. Active particles can be trapped in
funnels, with an opening angle that would not trap single
particles. Only due to collective effects does a complete-trapping
transition occur at finite densities. From \citet{Kaiser2012}.
\label{fig:kaiser}
}
\end{figure}

These approaches neglect interactions between several
swimmers. However, in experiments, microswimmers often occur at
finite concentration or even dense suspensions. In these cases, collective
effects can play an important role.
For example, in  a system of many self-propelled Brownian rods between
walls in two dimensions, the rods moving along the walls in
opposite directions block each other and lead to the formation
of ``hedgehog-like" clusters \cite{wens08}.
This effect can be exploited to capture active particles.
As shown above, self-propelled rods accumulate already at walls at
infinite dilution.
However, combining the ``hedgehog-effect'' of swimmers blocking each
other in a wedge-like geometry, very strong capture of swimmers can
be achieved, up to the point of complete trapping. In complete
trapping, the swimmers block each other in such a way that none
can escape the trap, even at opening angles where single swimmers
would escape again \cite{Kaiser2012} (see Fig.~\ref{fig:kaiser}).

\REV{Capturing of many particles at interfaces can also be exploited 
to separarte different particle species. Self-propelled discs  or rods of 
different size or velocity in confinement at high densities separate 
with small and fast particles favored at the interface \cite{cost14,yang14b}. 
For soft discs of differenr size and velocity, this segregation 
can be understood by considering the elastic-energy barrier for a
particle to squeeze between two others \cite{yang14b}. For hard rods 
of different velocity, the faster particles push the slower ones away 
from the wall \cite{cost14}. External flows in a channel can be used to 
subsequentially separate the different particles 
into different compartments \cite{cost14}.}

\subsection{Sperm Hydrodynamics near Surfaces}
\label{sec:sperm_hydro_wall}

Sperm frequently encounter situations of swimming in confined
geometry, for example in the female reproduction tract \cite{fauc06}.
The swimming behavior of sperm near walls and surfaces
has been studied for sperm of bull \cite{roth63,frie10},
human \cite{Winet1984}, sea urchin \cite{coss03,kaup03,boeh05},
mouse, rat, and chinchilla \cite{wool03}.
All experiments reveal circular or curvilinear trajectories
close to the surface.
Such trajectories are relevant for the movement of sperm on the
epithelial layer that lines the oviduct, and when sperm reach the
surface of the much larger egg. For example, many fish eggs on their
surface possess a small orifice, called pylum, that must be reached
by the sperm for successful fertilization.

For sea urchin sperm, portions of the tail were observed to be outside
of the focus plane of the microscope \cite{coss03},
suggesting an out-of-plane component of the beating pattern. Mouse
sperm, which are characterized by a strongly curved midpiece, adhere
to the wall only when the ``left" side of the head faces the glass
surface \cite{wool03}. Chinchilla sperm undergo a rolling motion as they
move along the surface, thereby touching the wall with different parts
of their head \cite{wool03}. Finally, the beating pattern of bull
sperm seems to be nearly planar parallel to the wall \cite{frie10}.

Several explanations have been proposed to account for the capture
of sperm near a surface \cite{roth63,coss03,wool03}.
In his pioneering study of bull sperm
near surfaces, Rothschild concluded that hydrodynamic interactions
are the most likely origin of this effect \cite{roth63}.
For rodent sperm, two mechanisms have been proposed \cite{wool03}.
For sperm
that exhibit a three-dimensional beating pattern and display a rolling
motion as they progress (like chinchilla sperm), it was argued that the
conical shape of the flagellar envelope establishes a thrust toward
the surface.
Alternatively, for sperm that exhibit a two-dimensional
beating pattern (like mouse sperm), the discoidal shape of the
sperm head, which is slightly tilted with respect to the plane of the
flagellar beat, may act as a hydrofoil \cite{wool03}.

\begin{figure}
\centering
\includegraphics[width=0.45\textwidth]{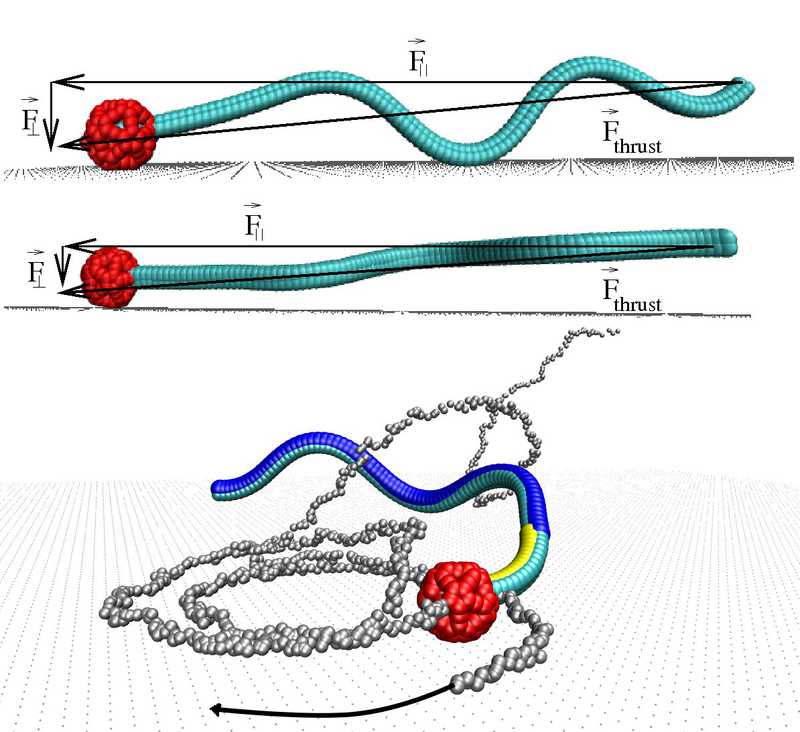}
\caption{
Sperm model used in mesoscale hydrodynamics simulations of
sperm adhesion.
(a) Sketch of forces responsible for the adhesion of symmetric sperm
at a surface.
Top: Without hydrodynamic interactions, the beating-plane of adhering
sperm is oriented perpendicular to the surface.
Bottom: With hydrodynamic interactions, the beating-plane is oriented
parallel to the surface.
(b) Sperm with large preferred curvature $c_0^{(m)}$ of the midpiece
at a surface
(with $c_0^{(m)}L_m=1$, where $L_m$ is the midpiece length).
The head touches the wall and blocks rolling. The beating
plane is approximately perpendicular to the surface.
From \citet{elge10}.}
\label{fig:adhesion}
\end{figure}

Since sperm is a pusher, far-field hydrodynamics predicts a parallel
orientation with the wall and an effective hydrodynamic attraction.
When the microswimmer comes closer to the wall, so that the dipole
approximation is no longer justified, mesoscale simulations
(see, e.g. reviews by \citet{kapr08} and \citet{gg:gomp09a})
can be employed to elucidate the physical mechanism of attraction
to the wall \cite{elge10}. Alternatively, sperm motion near a wall
can be studied by a numerical solution of the Navier-Stokes equations
\cite{smit09}.

\begin{figure}
\centering
\includegraphics[width=0.40\textwidth]{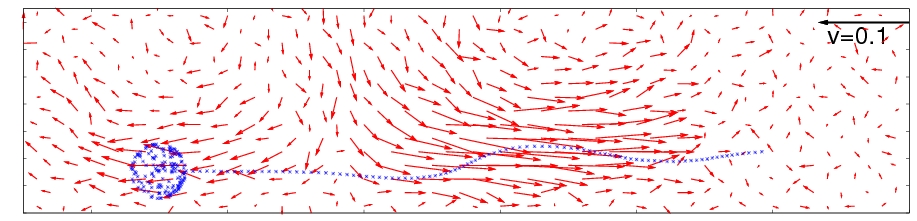}
\includegraphics[width=0.40\textwidth]{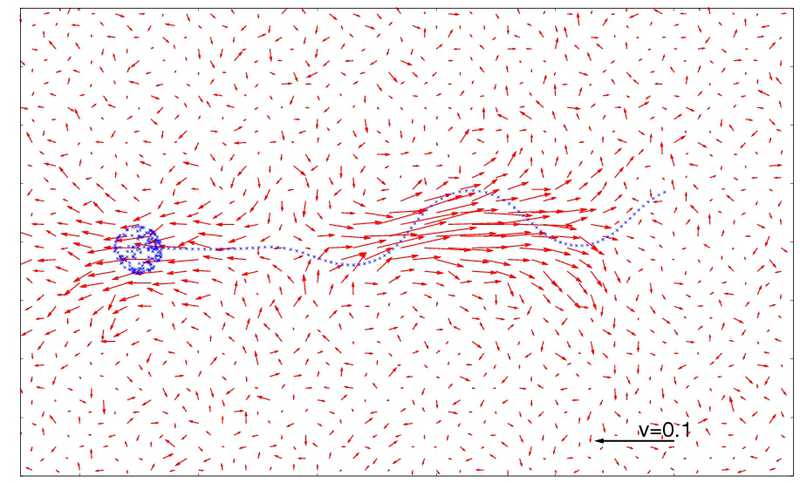}
\caption{
Averaged flow field in the vicinity of a symmetric sperm cell
adhering to a wall. (a) Plane perpendicular to the wall, and (b)
plane parallel to the wall, with both planes containing the average
sperm shape. A snapshot of a sperm is superimposed.
The flow field generated by the beating tail is directed away from
the sperm along their swimming direction and toward the sperm along
its side. From \citet{elge10}.
}
\label{fig:flow_field}
\end{figure}

The sperm model used in the simulations of \citet{elge10}
is shown in Fig.~\ref{fig:adhesion}. The main qualitative result is
that sperm swims very close to the wall.
Fig.~\ref{fig:flow_field} shows
the flow field of a symmetric, non-chiral sperm near a surface, as
obtained from the simulation.
The influx of fluid in the midpiece region, which is characteristic for
dipole swimmers in the bulk, becomes very asymmetric near
the surface: the flow onto the midpiece from above relative to flow from
below is greatly enhanced due to the presence of the wall
(Fig.~\ref{fig:flow_field}a). This imbalance of fluxes generates
an attraction to the wall in the midpiece region, in
qualitative agreement with the predictions by a force-dipole
approximation for large distances from the wall \cite{berk08}.
The flow field near the end of the tail also has a
component toward the wall (Fig.~\ref{fig:flow_field}a);
this component is responsible for
a hydrodynamic repulsion of the tail from the wall,
which induces a tilt of the sperm axis toward the surface
(Fig.~\ref{fig:adhesion}). Furthermore, due to the no-slip boundary
conditions, the flow in the plane parallel to the wall is screened
(Fig.~\ref{fig:flow_field}b). The far-field approximation predicts
the decay $r^{-3}$ for large distances \cite{dres:11,Spagnolie2012}.

Thus, the simulations \cite{elge10} reveal that
(i) the beating plane of sperm gets oriented parallel to the wall
(in agreement with experimental observations \cite{frie10}),
(ii) the sperm develops a small tilt angle toward the wall (compare
Fig.~\ref{fig:adhesion}a, which enhances the attraction compared to
the pure dipole force, and (iii) the elongated shape of sperm
contributes to the adhesion effect.

In contrast, on the basis of the numerical solution of the Navier-Stokes
equations, \citet{smit09} predict that sperm swims nearly parallel to
the wall, but with a distance comparable to the sperm length, and with
a small angle {\em away} from the wall.

\begin{figure}
\centering
\includegraphics*[width=0.8\columnwidth]{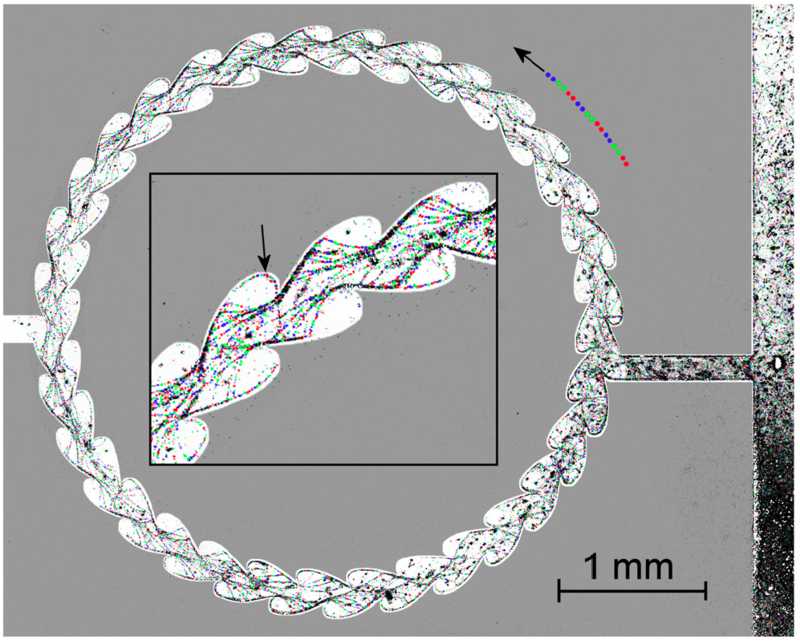}
\caption{A microfluidic device, micro-fabricated in PDMS, which
presents a ``one-way street'' for swimming sperm . In this
anisotropic channels sperm swim preferentially in a \REV{counter-clockwise}
fashion. Center: A magnification showing how sperm traveling in the
wrong direction are forced to turn.
The colors represent the time sequences red-green-blue.
From \citet{Denissenko2012}.
\label{fig:denissenko}
}
\end{figure}

The trend of sperm to follow boundaries can than be exploited
to create ``one-way channels'', recently realized for human sperm
swimming in micro-fabricated channels \cite{Denissenko2012}, see
Fig.~\ref{fig:denissenko}.  The main idea here is to give the
channel walls a ``clover-leaf" structure, where sperm following the walls are
redirected by $180^0$, then leave the wall at a sharp corner.
In the same experiments, it was found that  sperm indeed seem to
swim at interfaces with finite angle toward the surface
confirming the predictions of \citet{elge10}, but at  variance
with the results of \citet{smit09}.

Chiral sperm also adhere to surfaces. In this case, sperm swim
on helical trajectories in the bulk \cite{gray55b,cren89,cren96}, and in
circles at a wall, see Fig.~\ref{fig:adhesion}b, in good agreement with the
experiments mentioned above. The radius of the swimming circles
decreases with increasing chirality. Weakly chiral sperm display
a rolling motion, while strongly chiral sperm do not roll \cite{elge10}.
This effect can be understood from the steric hindrance of a strongly
asymmetric shape.

\subsection{Bacteria Swimming at Surfaces}
\label{sec:bacteria_swimming_surface}

The swimming behavior of bacteria close to surfaces differs from
the ``run-and-tumble'' motion in free solution. Experiments
show that individual {\em E.~coli} bacteria swim in clockwise,
circular trajectories near planar glass surfaces
\cite{berg:90,frym:95,dilu:05,hill:07,kaya:12,mola14}. In the following,
individual cells at surfaces are considered, which could be swimmer or
swarmer cells.

\begin{figure}
\centering
\includegraphics*[width=0.8\columnwidth]{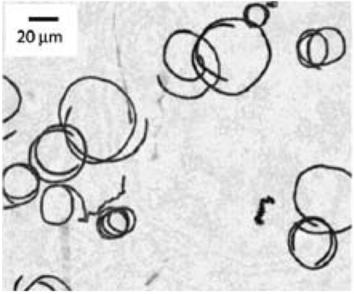}
\caption{Superimposed phase-contrast video microscopy images of
non-tumbling {\em E.~coli}
bacteria (HCB437) swimming in circular trajectories near a glass surface.
From \citet{laug:06}.
\label{fig:bacteria_circles}
}
\end{figure}

The direction of circular motion at a wall depends on the boundary
conditions. Since the cell is force and torque free, the body and
flagella bundle rotation imply a clockwise rotational motion for a
no-slip boundary condition \REV{\cite{laug:06,li:08,leme:10,lope:14}},
see Fig.~\ref{fig:bacteria_circles}.
Hydrodynamics gives rise to an increased drag of the fluid
confined between the cell and the surface  and a torque appears on
the swimmer, which turns the cell in clockwise direction for
no-slip boundaries. In contrast, as shown by \citet{leme:10} and \citet{lope:14}, slip boundary
conditions allow for counter-clockwise circular motion. As a consequence of the preferred
circular swimming direction for no-slip boundary conditions, {\em
E.~coli} swim preferentially on the right-hand side of a
microchannel, when it is confined to the bottom surface \cite{dilu:05}.

\begin{figure}
\includegraphics*[width=0.76\columnwidth]{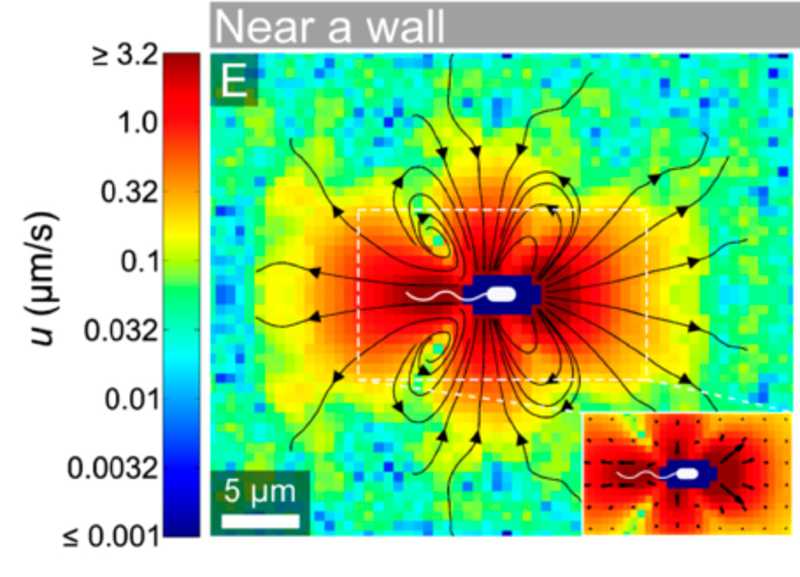}\\
\hspace*{2.1cm}
\includegraphics*[width=0.70\columnwidth]{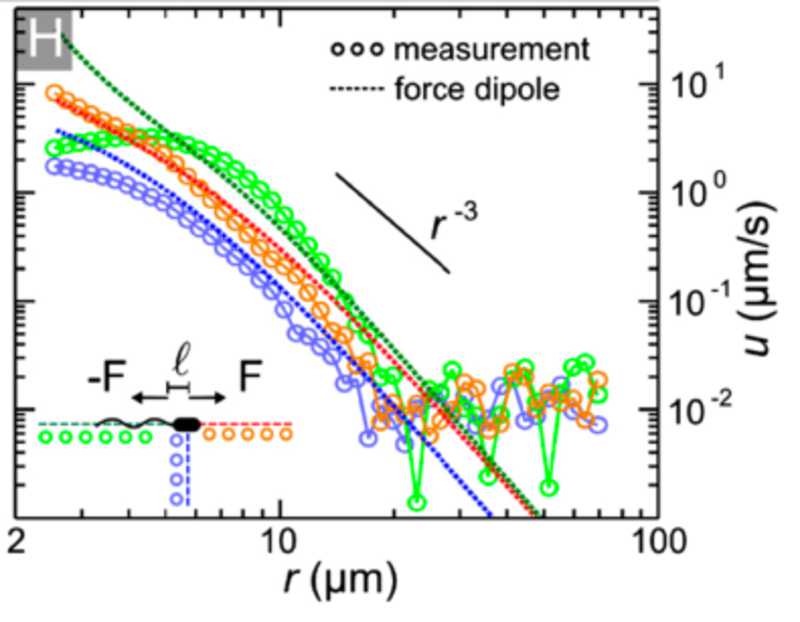}
\caption{(a) Experimentally determined flow field of an {\em E.~coli}
near a wall.
(b) The distance dependence of the flow strength as a function of
distance from the microswimmer follows the theoretically predicted
power of a dipole swimmer, which decays like $r^{-2}$ in the bulk,
see Eq.~(\ref{eq:Oseen_dipole}), but like $r^{-3}$ near a surface.
From \citet{dres:11}.
\label{fig:bacteria_wall_flowfield}
}
\end{figure}

Recent experimental studies revealed the flow field near a swimming
{\em E.~coli} bacterium, in particular next to a surface
\cite{dres:11}, see Fig.~\ref{fig:bacteria_wall_flowfield}. In this case,
the flow field can be well described
by a simple force dipole model, as discussed in Sec.~\ref{sec:dipole}.
Moreover, these studies support the argument that hydrodynamic effects
contribute to the observed long residence times of bacteria close
to no-slip surfaces \cite{berk08}.

In contrast, the experiments of \citet{dres:11} suggest that
cell-surface collisions determine the cell behavior next to surfaces,
as proposed by \citet{li09}, \citet{elge09}, and \citet{hern:09},
rather than long-range hydrodynamic interactions
\cite{berk08}. To arrive at the more complete picture of the
cell-surface interactions, more precise measurements and more
detailed theoretical considerations are necessary.


\section{Synchronization}
\label{sec:synchronization}

Synchronization of motion is a common phenomenon in nonlinear
many-particle systems, and thus appears in a broad range of physical,
biological, engineering, and social systems \cite{piko:02,stro:04}. The
phenomenon appears at all length scales from atoms to macroscopic
bodies. For microswimmers, synchronization is fundamental for
coordinated cyclic motion of cilia and flagella. The synchronous beating of the two flagella
of {\em Chlamydomonas} causes straight
swimming, while asynchronous beating implies  tumbling motion
\cite{qian:09,dres10a,guas10,gold:11,laug:12,benn:13}. The helical
flagella of bacteria, like {\em E. coli}, synchronize their
rotational motion during bundling
\cite{kim:03,kim:04,reic:05,reig:12,reig:13}. Multi-ciliated and
multi-flagellated microorganisms such as unicellular {\em Paramecia}
\cite{knig:54} or {\em Volovoxes} \cite{brum:12} exhibit
metachronal waves (MCW) \cite{sle_book.2}. Here, synchronization
is essential for microswimmer motility. Furthermore, coordinated
flagellar motion plays a major role in eukaryotes \cite{stoc:09,poli:09},
where they transport fluid in the respiratory system in form of
cilia \cite{afze:76}, are involved in cellular communications
\cite{wang:06}, and even determine the morphological left-right
asymmetry in the embryo \cite{cart:04}.

As early as 1951, \citet{tayl51} suggested that
hydrodynamic interactions lead to synchronization of nearby
swimming spermatozoa. Since then, the hydrodynamic interactions of
active systems at low Reynolds numbers has become a subject of major
interest. A recent review by \citet{gole:11} addresses a
number of important aspects of hydrodynamical induced synchronized
motions. Here, we will present and discuss the most important
aspects and address recent developments.

Synchronization is not easily achieved for systems governed by
low-Reynolds number hydrodynamics and thus described by
the Stokes equation (\ref{eq:stokes}). The presence of kinematic
reversibility of this equation combined with swimmer symmetries
may prevent synchronization \cite{kim:04,reic:05,elfr09,thee:13}.
To overcome this fundamental limitation of life at low Reynolds
numbers and to generate a time-irreversible dynamics, various alternatives
have been suggested. This comprises inclusion of additional
degrees of freedom such as system flexibility
\cite{kim:04,reic:05,qian:09,nied:08,uchi:12,reig:12,reig:13},
or specific, non-reversible driving forces
\cite{uchi:11,uchi:12,benn:13,benn:13.1}.
In addition, specific system designs combined with hydrodynamic
interactions lead to synchronization, as has been shown for models of
undulating sheets \cite{elfr09,elfr:11} and for the flagellar beat of
{\em Chlamydomonas} \cite{benn:13,benn:13.1,geye13}. For the
latter, it has been shown that synchronization of flagella beating can
be achieved even without hydrodynamic interactions \cite{frie:12,polo:13}.
Alternatively, a more general
linear unsteady Stokes equation can be adopted to describe the fluid properties
\cite{laug:11,thee:13}. Here, the time-reversal symmetry of the
fluid-dynamical equation is broken by inertial terms, see
Eq.~(\ref{eq:linear_n-s}).

\subsection{Basic Concepts}
\label{sec:synchrobasic}

Minimalistic models are useful to shed light on the mechanisms
of hydrodynamic synchronization. An example is the rigid rotor
model of \citet{lenz:06}, where each rotor possesses a
single degree of freedom only. Hence, it is particularly useful
for analytical studies of fluid mediated interactions between several
rotors. Adopting this model, we consider two beads of radius $a$
moving along circles of radius $R$. The circles are centered at
${\bm r}_{i}^0= (-1)^i {\bm d}/2 $ ($i=1,2$), where ${\bm d} = d
\hat{\bm e}_x$ and $d$ is the center-to-center distance; both
beads are confined in the $xy$-plane. The trajectories of the bead
centers can be expressed as
\begin{align} \label{trajectory}
{\bm r}_i(t)= {\bm r}_i^0+ R \hat{\bm e}_{ri}(t) ,
\end{align}
with the radial unit vectors $\hat {\bm e}_{ri} = (\cos
\varphi_i(t), \sin \varphi_i(t),0 )^T$, in terms of the phase
angles $\varphi_i(t)$. The driving forces
\begin{align} \label{driving_force}
{\bm F}_i^d(t) = F_i (\varphi_i) \hat{\bm t}_i(t)
\end{align}
point along the tangents $\hat {\bm t}_i(t)=(-\sin
\varphi_i(t),\cos \varphi_i(t),0)^T$ of the trajectories. The
equations of motion of the particles are given by
\begin{align} \label{equation_motion}
\dot {\bm r}_i  = \frac{1}{\zeta} {\bm F}_i + {\bm v}({\bm r}_i)=
\frac{1}{\zeta} {\bm F}_i + \sum_{j\ne i}
\mathbf{H}_{ij} \cdot {\bm F}_j
\end{align}
within the Stokes description (\ref{eq:stokes}) of the fluid, with
the hydrodynamic tensor $\mathbf{H}_{ij} = \mathbf{H} ({\bm
r}_i-{\bm r}_j)$ (\ref{eq:oseen}), and the total force ${\bm F}_i$
on a particle.  Since the flow field ${\bm v}({\bm r}_i)$
at the position of particle $i$, induced by the motion of other
particles, is typically not aligned with the tangent vector
$\hat{\bm t}_i$, a constraining force ${\bm F}_i^c$ is necessary
to enforce a circular trajectory (${\bm F}_i={\bm F}_i^d + {\bm
F}_i^c$). However, within the far-field approximation of the
hydrodynamic tensor, the constraining force yields corrections of
the order $O((a/d)^2)$ to the flow field ${\bm v}({\bm r}_i)$.
Hence, for $a/d \ll 1$, we can neglect such contributions and the
equations of motion of the phase angles are (${\bm F}\cdot
\hat{\bm t}= {\bm F}_i^d \cdot \hat {\bm t} = F_i$)
\begin{align}
\dot \varphi_i(t) = \omega_i + \frac{1}{R} \sum_{j\ne i} F_j \hat{\bm t}_i(t)
\cdot \mathbf{H}_{ij} \cdot  \hat{\bm t}_j(t)  ,
\end{align}
with the frequencies $\omega_i=F_i/(6 \pi \eta a R)$. As pointed
out by~\citet{lenz:06}, \citet{gole:11}, \citet{uchi:11}, and
\citet{thee:13}, for constant forces $F_1=F_2=F$, the symmetry
of the Oseen tensor
$\mathbf{H}({\bm r}) = \mathbf{H}(-{\bm r})$ implies a constant
phase difference $\Delta \varphi = \varphi_2(t)-\varphi_1(t)$ --- so
no synchronization occurs.

The same conclusion has been reached by \citet{kim:04} and \citet{reic:05}
for rigid rotating helices. An extra
degree of freedom can be introduced by adding some flexibility in
the model. The flexibility can arise from different origins, such
as intrinsic flexibility of a filament or flexibility of the
anchoring point. Indeed, as shown by \citet{flor:05}, \citet{reic:05},
\citet{jans:11}, and \citet{reig:12}, flexible filaments exhibit
synchronized motion.

As far as rotors are concerned, various models for elastic
deformation have been investigated. \citet{nied:08} allowed for
radial fluctuations of a harmonically bound particle.
\citet{qian:09} considered a sphere attached to a rigid rod with
its other end confined in a harmonic potential.
\citet{uchi:12} investigated a bead driven by an optical tweezer,
whose focus is dragged along a prescribed trajectory. We will
adopt the latter approach to illustrate the consequences of
elastic deformations on the synchronization of rotors.
The position of the bead $i$ is given by
${\bm r}_i(t)= {\bm r}_i^0+ R \hat{\bm e}_{ri}(t) + {\bm u}_i(t)$ for a
circular trajectory of the tweezer focus $R \hat{\bm e}_{ri}(t)$
and the displacement ${\bm u}_i$ in the potential.
With the approximation of the tweezer field by a harmonic potential,
the equations of motion (\ref{equation_motion}) again apply,
with the force balance ${\bm F}_i + k {\bm u}_i =0$, where $k$
is the trap stiffness. For an isolated rotor, i.e,
in the absence of fluid flow due to other rotors, ${\bm v}({\bm
r})=0$, and in the stationary state the force ${\bm F}_i$ obeys
the relation
\begin{align}
{\bm F}_i = \zeta \omega_i R \left(\hat{\bm t}_i - \frac{1}{kR}
\frac {d {\bm F}_i}{d \varphi_i} \right) ,
\end{align}
with $\omega_i = \dot \varphi_i$. For a constant driving force,
this yields the relation $\omega_i = F_i/(\zeta R)$ between the
force and intrinsic frequency $\omega_i$. Multiplication of
Eq.~(\ref{equation_motion}) by $\hat{\bm e}_{ri}$ yields
\begin{align}
{\bm u}_i\cdot \hat{\bm e}_{ri} = \frac{\zeta}{k} \hat{\bm e}_{ri}
\cdot {\bm v}({\bm r}_i)
\end{align}
to $O(k^{-1})$. Similarly, multiplication of
Eq.~(\ref{equation_motion}) by $\hat{\bm t}_i$ yields the equation
of motion for the angles $\varphi_i$
\begin{align}
\nonumber \dot \varphi_i = \ & \omega_i \left( 1 -
    \frac{\zeta}{kR} \sum_j \hat{\bm e}_{ri} \cdot \mathbf{H}_{ij}
                  \cdot \hat{\bm t}_j F_j \right) \\
   & + \zeta \sum_j \omega_j \hat{\bm t}_i \cdot \mathbf{H}_{ij} \cdot
       \left[\hat{\bm t}_j + \frac{F_j}{kR} \hat{\bm e}_{rj} \right]
\end{align}
to $O(k^{-1})$. By assuming circular trajectories, the far field
approximation for the Oseen tensor, and equal forces $F_1=F_2=F$,
we finally obtain the equation of motion for the phase difference
$\Delta \varphi = \varphi_2 - \varphi_1$
\begin{align}
\frac{d \Delta \varphi}{dt} =
     - \frac{9 a}{2 d} \frac{\omega F}{k R} \sin \Delta \varphi .
\end{align}
For $\Delta \varphi \ll 1$, we find an exponentially decaying phase
difference $\Delta \varphi \sim \exp(- t/\tau_s)$, with the
characteristic synchronization time \cite{qian:09,uchi:12}
\begin{align} \label{synch_time_pot}
\tau_s = T \frac{k d R}{9 \pi a F} .
\end{align}
The two beads synchronize their rotational motion on a time scale,
which is proportional to the trap stiffness $k$. In the limit $k
\to \infty$, i.e., the rotation on the circle discussed above, the
synchronization time diverges. The same linear dependence
on the constant of the harmonic potential has been obtained by
\citet{nied:08}.

In addition, \citet{qian:09} considered dumbbell-like rotors with
harmonically bound centers as models for symmetric paddles. Here,
also a linear dependence of the synchronization time on $k$
is obtained, but with the much stronger dependence
$\tau_s \sim k d^5/(a^3 RF)$ on the rotor distance.

Flexibility of rotors is not a necessary requirement for synchronization
\cite{uchi:11,uchi:12,thee:13}. Alternatively, synchronization can
be achieved for rigid trajectories by a particular, phase-angle
dependent driving forces and/or for trajectories of certain
non-circular shapes \cite{uchi:11,uchi:12}.
For cilia or the flagella of {\em Chlamydomonas}, significant
force modulations can be expected due to the asymmetry between
power and recovery strokes.
\citet{uchi:11} and \citet{uchi:12} analyzed the necessary conditions for
in-phase synchronization for phase-dependent driving forces ${\bm
F}_i^d=F(\varphi_i) \hat{\bm t}_i$, where now $\hat{\bm t}_i =
{\bm r}_i' /|{\bm r}_i'|$ with the abbreviation ${\bm r}_i' = d
{\bm r}_i/d \varphi_i$. For circular trajectories,  the equations
of motion (\ref{equation_motion}) yield
\begin{align}
\frac{d \Delta \varphi}{d t} = \omega_2 - \omega_1 + \frac{1}{R}
\left[F (\varphi_1) - F (\varphi_2) \right] \hat{\bm t}_1
\cdot \mathbf{H}_{12} \cdot \hat{\bm t}_2
\end{align}
for the phase difference. For small phase differences $\Delta \varphi$,
linearization leads to
\begin{align}
\frac{d \Delta \varphi}{d t} = \left(\omega'(\varphi_1)
- \frac{1}{R} F'(\varphi_1)  \hat{\bm t}(\varphi_1)
\cdot \mathbf{H}({\bm d}) \cdot \hat{\bm t}(\varphi_1) \right)
      \Delta \varphi \ .
\end{align}
Integration of the ratio
$\dot{\Delta \varphi}/\Delta \varphi $ over one time period $T$
yields the cycle-averaged characteristic time \cite{uchi:11,uchi:12}
\begin{align} \label{synch_time_force_int}
\tau_s^{-1} = - \frac{2}{T} \int_0^{2 \pi} \frac{d \ln F}{d \varphi}
\hat{\bm t}(\varphi)
\cdot \mathbf{H}({\bm d}) \cdot \hat {\bm t}(\varphi) \ d \varphi
\end{align}
in the limit $a/d \ll 1$. \citet{uchi:12} analyzed and identified
force profiles which lead to synchronization. Here, we mention and
consider the relation
\begin{align}
F(\varphi) = F[1-A \sin(2 \varphi)]
\end{align}
only, where $A$ ($0< A <1$) is a constant. Up to $O(A^2)$, integration
results in
\begin{align} \label{synch_time_force}
\tau_s = \frac{2d}{3\pi a A} T ,
\end{align}
i.e., a characteristic synchronization time which depends on the rotor
distance $d$ and the bead diameter $a$ only as far as the rotor geometry is concerned.

It is interesting to note that in this description, synchronization
is caused by the tensorial character of the hydrodynamic
interactions. The contribution of the diagonal part of the Oseen
vanishes by the integration (\ref{synch_time_force_int})
\cite{uchi:11,uchi:12}. In contrast, for the harmonically bound
beads, all parts of the hydrodynamic tensor contribute to the time
$\tau_s$.

More details of the phase-angle dependent driving forces and
different trajectory shapes are discussed by \citet{uchi:12}. In
particular, the combined effect of harmonically bound beads,
driven by phase-angle dependent forces are addressed. The
characteristic decay rate $\tau_s^{-1}$ of the combined effects is
the sum of the contributions of the individual contributions
Eq.~(\ref{synch_time_pot}) and (\ref{synch_time_force}).

\citet{kota:13} recently presented experimental
results for the synchronization of two colloids, which are driven
by feedback-controlled optical tweezers. Both, the elasticity of the
trajectory and a phase-angle dependent driving force was implemented,
as discussed above. The colloids exhibit strong synchronization
within a few cycles, even in the presence of noise, consistent with
the considerations described above.

The relevance of the various contributions depends on the
actual parameters, e.g., the stiffness of the radial (harmonic)
potential and amplitude of the force modulations along the
trajectory. These values have been chosen by \citet{kota:13} such
that both contributions are important. As pointed out by
\citet{uchi:12}, however,  for small disturbance of a trajectory
by hydrodynamic interactions, compared to the size of a
trajectory, flexibility has only a weak effect in establishing
synchronization. Although this may apply to certain systems only,
it reflects a shift in paradigm from a flexibility dominated
synchronization mechanism to driving-force governed processes.

A different hydrodynamic route was adopted by \citet{thee:13},
who started from the linearized Navier-Stokes equations
(\ref{eq:linear_n-s}). In this case, the hydrodynamic tensor
is time dependent. This leads to the equations of motion
\begin{align}
\dot \varphi_i = \omega + \frac{F}{R} \sum_{j \ne i}
\int_0^t \hat{\bm t}_i(t) \cdot \mathbf{H}({\bm r}_i(t)-{\bm r}_j(t'), t-t')
\cdot \hat {\bm t}_j(t') dt'
\end{align}
for the phase angles of circular rotors driven by the constant
force $F_1=F_2=F$, with the intrinsic frequency $\omega = F/(\zeta
R)$. Here, the whole time history contributes to the dynamical
behavior. In the limit $a/d \ll 1$ and for times $t \gg \tau_{\nu}
= d^2/\nu$, an approximate expression for the integral can be
derived, which becomes
\begin{align} \frac{d \Delta \varphi}{d t} = -
\frac{2 \sqrt{\pi} a \nu}{d^3 T} \int_0^t \left( \frac{d^2}{\nu
t'}\right)^{3/2} \sin(\omega t') \ dt' \ \Delta \varphi(t)
\end{align}
for small phase differences. In the long-time limit, the phase
difference decays exponentially with the characteristic time
\cite{thee:13}
\begin{align} \label{eq:sync_time_retard}
\tau_s= \frac{d}{4 \pi^{3/2} a} \sqrt{\frac{T}{\tau_{\nu}}} T .
\end{align}
Compared to the other mechanisms described above, $\tau_s$
depends on the square-root of the ratio $T/\tau_{\nu}$ of the rotation
period $T$ and the shear-wave propagation time $\tau_\nu$. This
ratio can be identified with
the oscillatory Reynolds number $Re_T$ (\ref{eq:n-sdimless}). Hence,
the synchronization time (\ref{eq:sync_time_retard}) is determined
by the oscillatory Reynolds number, but with a square-root
dependence.

For typical parameters of {\em E. coli} bacteria, the
time-dependent hydrodynamic correlations have a weak effect on
synchronization only. Studies of the combined effect of
phase-angle dependent driving forces and time-dependent
hydrodynamic interactions yield a faster synchronization dynamics
than the individual mechanisms only.

\begin{figure}[t]
\begin{center}
\includegraphics[width=0.45\textwidth]{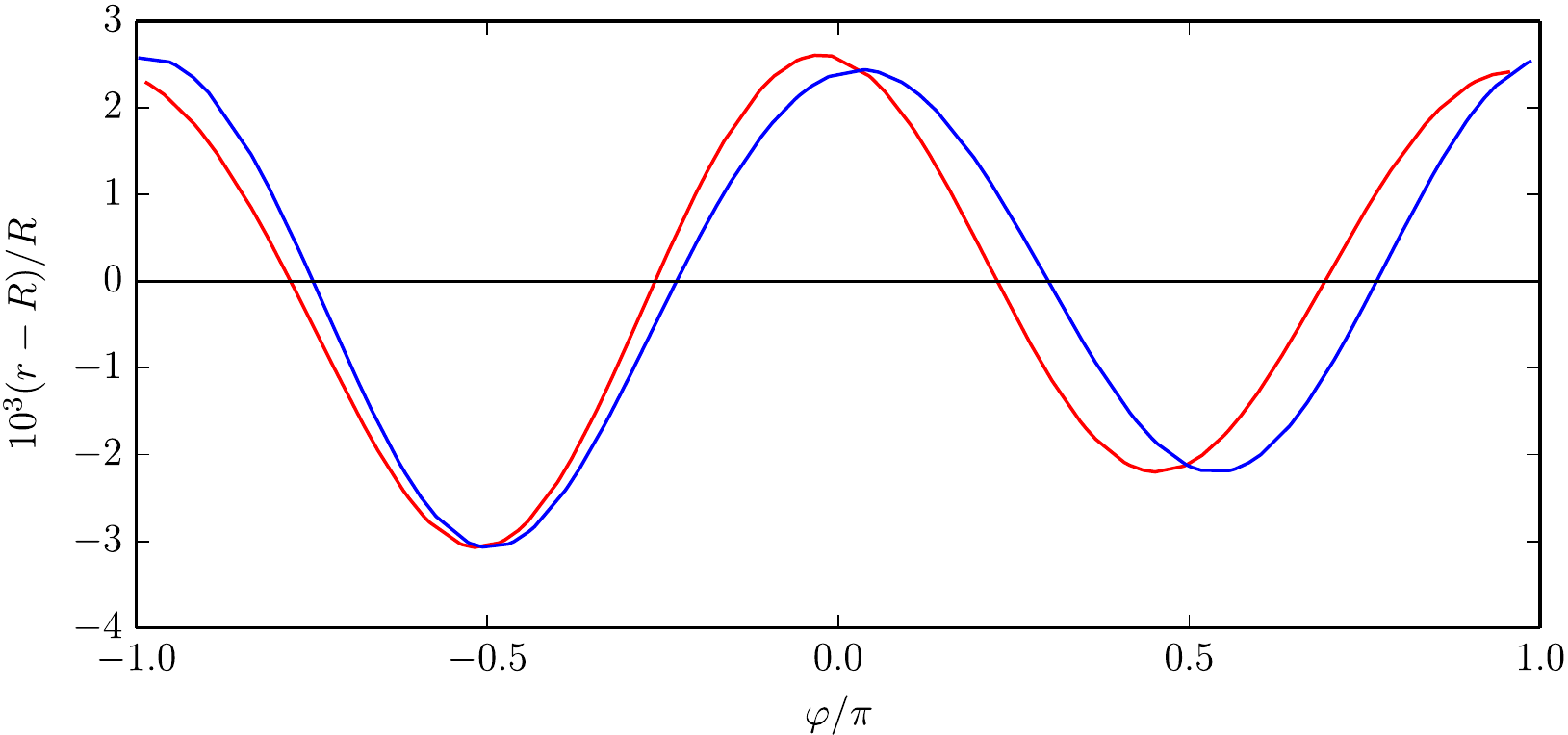} \\
\includegraphics[width=0.2\textwidth]{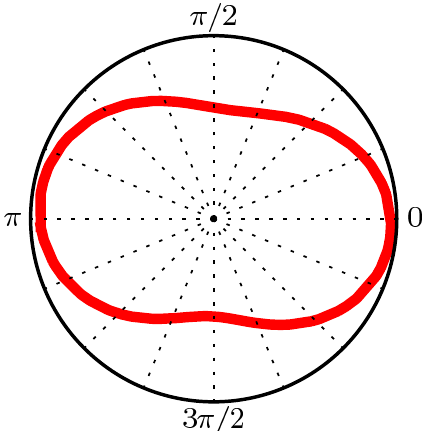} \hspace*{3mm}
\includegraphics[width=0.2\textwidth]{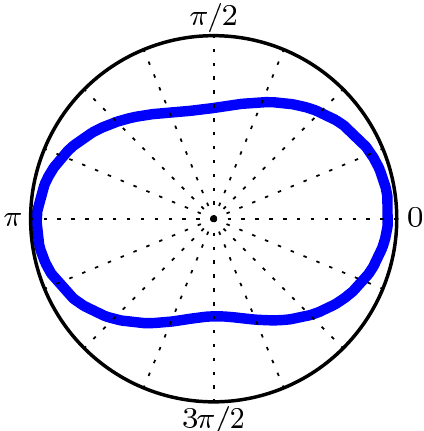}
\end{center}
\caption{Deviation of the radial distance $r$ from a circle of radius
$R$ for beads of two hydrodynamically coupled rotors in the
vicinity of a surface at $z=0$. The trajectory is confined in a plane
perpendicular to the surface. The phase angle $\varphi=0$
corresponds to the unit vector $\hat{\bm e}_r=(1,0,0)^T$. In the polar representations
(bottom), the deviation $(r-R)/R$ is multiplied by 100 for better visibility.
} \label{fig:metachrona_angle}
\end{figure}

Various other concepts have been put forward to study
synchronization of two rotors and, in particular, the appearance
of a phase lag as a prerequisite for the formation of metachronal
waves. For the latter, in addition, hydrodynamic interactions with
a surface are taken into account by employing the Blake tensor
\cite{blak71a} for hydrodynamic interactions near a no-slip wall.
Among the first to study this phenomenon were \citet{lenz:06},
\citet{vilf:06}, and \citet{nied:08}.
As discussed above (Eq.~(\ref{synch_time_pot})), synchronization is
obtained for beads confined in radially harmonic potentials and
equal driving forces.  A phase-locked motion has been found for
different driving forces by \citet{nied:08}. In contrast,
\citet{brum:12} found a stable phase lag even for equal driving
forces. They considered the equation of motion of a bead
\begin{align}
\label{eq:Blake_synchro}
\dot {\bm r}_i = {\boldsymbol{\zeta}}^{-1} \cdot {\bm F}_i + \sum_{j\ne i}
\hat{\mathbf{H}}_{ij} \cdot {\bm F}_j ,
\end{align}
where ${\boldsymbol{\zeta}} = \zeta[\mathbf{I} + 9 a(\mathbf{I} +
\hat{\bm e}_z \hat{\bm e}_z)/(16z) ]$ is the friction tensor and
$\hat{\mathbf{H}}_{ij}$ is Blake's tensor, which take  into
account the no-slip boundary condition at the wall at $z=0$.
The force is given by
${\bm F}_i = -k (|{\bm r}_i|- R) \hat{\bm e}_{ri} + F \hat{\bm t}_i$.
As pointed out by \citet{brum:12}, a numerical solution of the
equation yields a stable phase lag $\Delta \varphi \neq 0$.
A numerical solution of Eq.~(\ref{eq:Blake_synchro}) is shown
in Fig.~\ref{fig:metachrona_angle}
for bead trajectories perpendicular to the surface. Due to the
boundary condition, the flow properties are different for the motion
toward and away from the surface. This shifts the
phase difference from perfect synchronization to a finite phase lag
\cite{brum:12}. Interestingly, $\Delta \varphi$ is
independent of the initial condition and the strength $k$ of the
potential. Phase-locking seems to be a consequence of nonlinear
interactions in the system, because \citet{nied:08} found
synchronization rather then phase-locking for their approximate
description.
As indicated in Fig.~\ref{fig:metachrona_angle}, the trajectory is
ellipse-like, with the long axis along the $x$-axis and the
short axis along the $z$-axis (perpendicular to surface), with a
somewhat smaller deviation $|r-R|/R$ on the side further away from
the surface than close to it.

\subsection{Experimental Results: Microrotors and Colloidal Oscillators}

As an experimental realization of driven rotors, chiral propellers
have been designed, which are driven by radiation pressure \cite{dile:12}.
In the light of the discussions of Sec.~\ref{sec:synchrobasic}, the
symmetry-breaking mechanism is provided by the flexibility of the optical
trap, which is used to provide stable alignment and the driving
torque. As shown by \citet{dile:12}, two such rotors synchronize
their rotational dynamics by hydrodynamic interactions. However,
the coupling is very weak, and requires a fine-tuning of the relative
torque within a resolution of $10^{-3}$. Even then, phase slips
occur, where one of the rotors moves faster than the other until a
stable phase-locked state may appear again.

In contrast, strong static correlations are obtained
through hydrodynamic interactions between colloidal particles in
rotating energy landscapes \cite{koum:13}. Thereby, the particles are
driven periodically around a circle by rotating energy
landscapes with a variable number of minima. In the
co-moving, rotating frame, a colloid experiences a tilted periodic
potential due to the hydrodynamic interaction with the
other colloids.
This enhances the probability to overcome the barriers
and allows the different colloids to synchronize their rotational
motion.

Another strategy to synchronize the dynamics of beads has been
proposed by \citet{guer:97,guer:98}. It is based on the difference
between the power stroke and the recovery stroke of cilia, or, more
generally, the presence of two phases with broken symmetry. The latter
concept has been applied to colloidal systems by \citet{lago:03},
\citet{kota:10}, \citet{bruo:11}, \citet{loic:12}, and \citet{lher:12}.
Here, the periodic motion of colloids is controlled by a
configuration-dependent external input.
Thereby, various properties can be switched from one state to
another upon a particular geometry being assumed by the
colloids. Examples are different effective drags \cite{lago:03},
or harmonic potentials with different equilibrium positions
\cite{kota:10}. The latter can be realized by optical traps.
These strategies lead to a synchronized motion by
hydrodynamic interactions
\cite{lago:03,kota:10,bruo:11,loic:12,lher:12,woll:11}.

\subsection{Synchronization of {\it Chlamydomonas} Beating}

{\em Chlamydomonas} with its two beating flagella
(see Sec.~\ref{sec:bio-swimmers}) has become a model system for experimental
studies of synchronization \cite{polo:13}. When the flagella beat synchronously,
the alga swims along a straight path, while dephasing leads to reorientation
in a run-and-tumble-like manner \cite{gold:09,poli:09,laug:12}.

\begin{figure}[t]
\begin{center}
\includegraphics[width=0.46\textwidth]{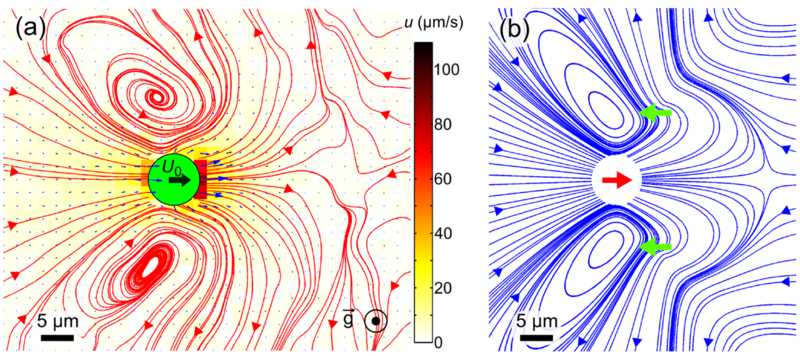}
\end{center}
\caption{Time- and azimuthally-averaged flow field of {\em Chlamydomonas}.
(a) Streamlines computed from the experimentally measured velocity vectors (blue).
(b) Streamlines of the azimuthally-averaged flow of an assumed three-Stokeslet
model: flagellar thrust is distributed between two Stokeslets located at the
approximate flagellar position (green arrows), whose sum balances drag on the
cell body (central red arrow). From \citet{dres10a}.
\label{fig:chlamydomonas_flow_field}
}
\end{figure}

The measured flow field of a swimming {\em Chlamydomonas} \cite{dres10a,guas10}
is well described by three Stokeslets (see Fig.~\ref{fig:chlamydomonas_flow_field}).
This finding has  stimulated a more detailed theoretical modeling of
{\em Chlamydomonas} by three spheres --- two spheres driven on circular orbits,
which are mimicking the flagellar beat
(see Sec.~\ref{sec:synchrobasic}), are linked by a frictionless scaffold to a
third sphere representing the cell body \cite{frie:12,polo:13,benn:13,benn:13.1}.

The extra translational and overall rotational degrees of freedom,
compared to the two degrees of freedom of the rotors of the systems of
Sec.~\ref{sec:synchrobasic}, combined with non-linearities, give rise to
additional features.
As shown by \citet{frie:12,polo:13,benn:13.1}, force and momentum balance
predominantly couple the phases of the rotors via translation and rotation of
the cell body. This coupling suggest the possibility of
flagella synchronization by local hydrodynamic friction even in
the absence of hydrodynamic interactions. The efficiency of synchronization
naturally depends on the geometry. It is weaker than hydrodynamic coupling,
when the cell body is much larger than the distance between the neighboring
flagella or cilia \cite{uchi:12}.

The three-sphere model exhibits a very rich dynamical swimming
behavior, when a phase-angle dependent driving force and noise is
added. Without hydrodynamic interactions, the cell just
moves forward and backward and there is no net motion. However, in
the presence of hydrodynamic interactions, which vary in strength
during a beating cycle, the broken symmetry between the power and recovery
stroke leads to a net propulsion \cite{benn:13.1}.
Interestingly, a run-and-tumbling motion is obtained in the presence
of noise. The threshold-like run-and-tumble behavior displayed by
bacteria like {\em E. coli}
and believed to be controlled by a sophisticated biochemical
feedback mechanism, emerges here naturally from nonlinearities in the
propulsion mechanism and could be triggered even by white noise.
Moreover, the same model could describe the experimentally observed antiphase
beating pattern of a {\em Chlamydomonas} mutant \cite{benn:13.1,lept:13}.

\begin{figure}[t]
\begin{center}
\includegraphics[width=0.5\textwidth]{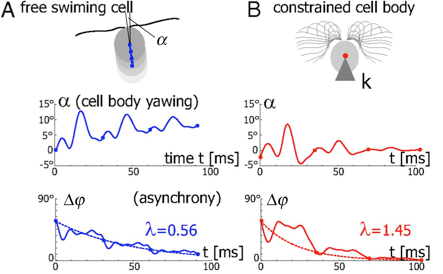}
\end{center}
\caption{Flagella synchronization by cell-body yawing.
(A) For a free-swimming cell, the cell body exhibits a yawing motion characterized
by the angle $\alpha(t)$, when the two flagella beat initially asynchronously. The
flagellar phase difference $\Delta \varphi$ between the left and right phase
(see Sec.~\ref{sec:synchrobasic}) decreases with time approximately exponentially,
$\exp(-\lambda t/T)$, where $\lambda (\approx 0.5-0.6)$ defines a dimensionless
synchronization strength.  Dots indicate the completion of a full beat cycle of
the left flagellum.
(B) Trapped cell body to prevent translation. Cell-body yawing  is restricted by
an elastic restoring torque that acts at the cell body center. The two flagella
synchronize with a synchronization strength $\lambda$ that can become even larger
than in the case of a free swimming cell.  For very large clamping stiffness, the
cell body cannot move and the synchronization strength attenuates to a basal value
$\lambda\approx 0.03$, which arises solely from direct hydrodynamic interactions
between the two flagella. From \citet{geye13}.
} \label{fig:chlamydomonas_synch}
\end{figure}

An experimental and theoretical study of the effect of the hydrodynamic coupling
on the synchronization of the flagella beating is presented by \citet{geye13}.
The central mechanism of an efficient synchronization is the adaptation of the
flagellar beat velocity to environmental conditions --- the flagella beat
speeds up or slows down in response to the exerted hydrodynamic friction forces.
A perturbation of the synchronized beating leads to a yawing motion of the cell,
which is reminiscent to rocking of the cell body
(see Fig.~\ref{fig:chlamydomonas_synch}) \cite{geye13}. This rotational motion
confers different hydrodynamic forces on the
flagella, with the consequence that one of them beats faster and the other slows
down. The coupling between cell-body rotation and flagellar beating speed leads
to a rapid reestablishment of a synchronized motion.

The studies by \citet{geye13} suggest that cell rocking is an important mechanism for
flagella synchronization. However, synchronized beating was also observed for
{\em Chlamydomonas} cells with their bodies restrained by a micropipette
\cite{ruef:98,poli:09,gold:09}. A rather fast synchronization has been obtained,
faster than pure hydrodynamic interactions would suggest \cite{geye13}. Rotational
motion with a small amplitude of a few degrees only, which may result from either
a residual rotational compliance of the clamped cell or an elastic anchorage of the flagellar
pair, provides a possible mechanism for rapid synchronization, which is analogous
to synchronization by cell-body rocking in free-swimming cells
(see Fig.~\ref{fig:chlamydomonas_synch}).

These results lead to the conclusion that hydrodynamic interactions are an integral
part of the synchronization mechanism of flagella beating. However, additional degrees
of freedom of a specific system may enhance synchronization. In this context, we
like to point out that purely hydrodynamic synchronization of the motion of two
flagella has recently been documented by \citet{brum:14}.

\subsection{Synchronization of Rotating Bacterial Flagella and
Bundle Formation}

Synchronization of the rotational motion of helical flagella of
bacteria is essential for bundle formation and hence their
swimming motion. Theoretical \cite{reic:05} and
simulation \cite{kim:04} studies show that rigid helices rotated by
stationary motors do not synchronize their rotational motion.
However, addition of certain flexibility, e.g., for helices with
their ends confined in harmonic traps, combined with hydrodynamic
interactions lead to a synchronized rotation even for separated
and mechanically not interacting helices \cite{reic:05} as already
suggested by macroscopic-scale model experiments \cite{macn:77,kim:03}.

\begin{figure}
\includegraphics*[width=8cm]{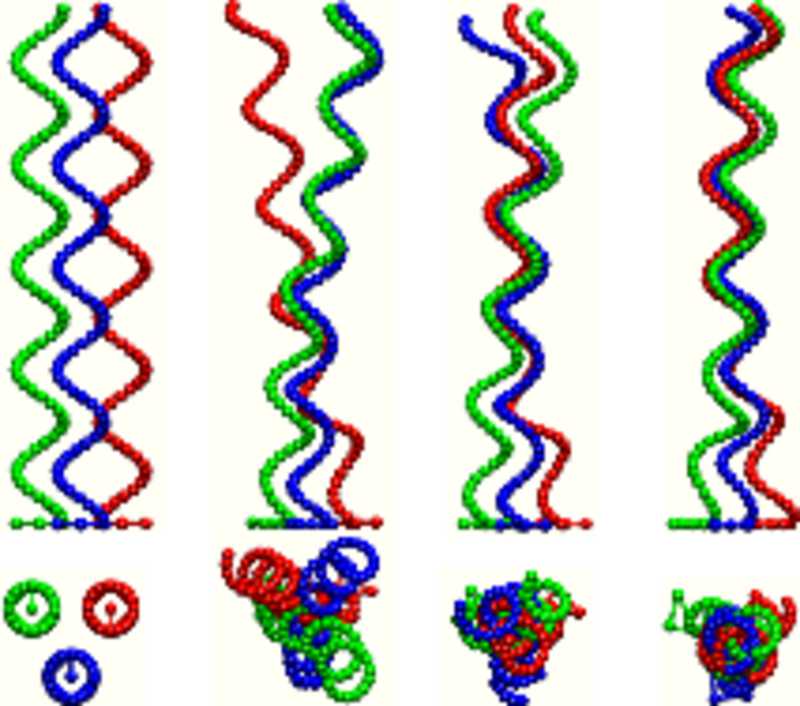}
\caption{\label{fig:helices_3} Snapshots, side views (top)
and top views (bottom), of
various stages of the bundling process for $d/R_h=3.5$. (i)
Initial state, the red helix is out of phase. (ii) The helices
synchronized their rotation and start to bundle. (iii) Parts of the
helices are bundled. (iv) Final, bundled state.
From Ref.~\cite{reig:12}.}
\end{figure}

A more detailed picture of bundle formation has been achieved by
computer simulations of model flagella
\cite{flor:05,jans:11,reig:12,reig:13}. In particular, the role of
fluid dynamics has been elucidated. Considering two parallel
helices composed of mass points (see Fig.~\ref{fig:helices_3})
with bending and torsional elastic energy, and using a
mesoscale hydrodynamics simulation technique \cite{kapr08,gg:gomp09a},
\citet{reig:12} calculated the hydrodynamic forces on each helix due to their
rotation motion. The simulations reveal a zero force along the radial distance,
consistent with the results for macroscopic helices of \citet{kim:04}.
However, there are large transverse forces in opposite directions. The
rotation of a helix creates a flow field, which tries to drag the other
helix in the transverse direction.  A ``tipping'' momentum has been
determined by \citet{kim:04}, which expresses the same effect.
Hence, there is no simple attraction between helices by hydrodynamic
interactions, as has been speculated. Bundling is rather induced
by the common flow field created by the rotation of the individual
helices. This expresses the importance of the rotational flow
field as suggested by \citet{flor:05}.

\begin{figure}
\includegraphics*[width=8cm]{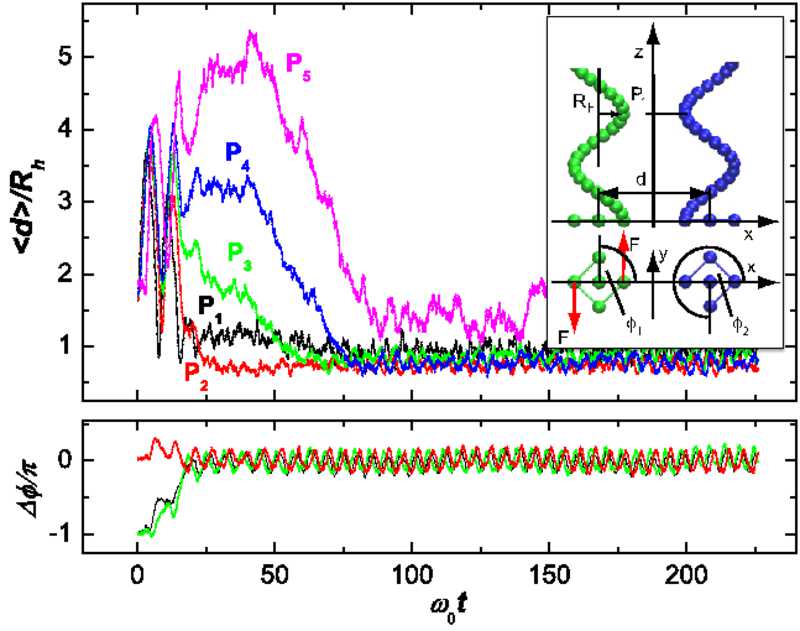}
\caption{\label{fig:helix_sync} Phase angle difference $\Delta \phi$ (bottom) and
average distances $\langle d \rangle$ (top)
between equivalent points $P_i=iP$ along the
helix contour as a function of time for the separation $d/R_h = 2.5$.
The three helices have the initial phase differences $\Delta \phi_{12} = - \pi$
(black), $\Delta \phi_{13} = 0 $ (red), and $\Delta \phi_{12} = - \pi$ (green). From Ref.~\cite{reig:12}.    }
\end{figure}

Releasing the constraints on the orientation provides insight into the full
bundling process \cite{jans:11,reig:12}, as illustrated in
Fig.~\ref{fig:helices_3}.  Starting from an aligned initial state,
the tangential hydrodynamic forces cause a tilt of the individual
helices, which brings them in closer contact near their fixed ends
and simultaneously separates their free ends. Such a spatial approach was
already assumed to be necessary by \citet{macn:77}. The whole
mechanism is quantitatively reflected in the time dependence of the helix
separations shown in Fig.~\ref{fig:helix_sync}, where the mean distance at
$P_2$ rapidly approaches its stationary-state
value, while $P_5$ increases initially and only slowly reaches its
stationary-state value.  Naturally, the details depend on the separation $d$
between the anchoring points \cite{reig:12}. In the stationary state,
a compact bundle is formed, where the helices are wrapped around each other.
The stationary-state distances are assumed in sequence from $P_1$ to
$P_5$, which implies that bundling occurs from the anchoring plane
to the tail.
Figure~\ref{fig:helix_sync} shows that, starting from an initial phase
mismatch, the helices first synchronize their rotational motion
and then subsequently form a bundle. Thereby, synchronization by
hydrodynamic interactions is achieved within a few rotations. The
simulations predict a synchronization time proportional to $d^2/(R_h^2\omega_0)$
and a bunding time proportional to $d/(R_h\omega_0)$, where $\omega_0$ is the
helix rotation frequency \cite{reig:12}. The scaling of the synchronization
time can be understood by estimating the time needed for the helices to come
close to each other in the rotational flow field $v(r)\sim R_h^2/(\omega_0 r)$
generated by the other helices,
\begin{equation}
\omega_0 t_{syn} \sim \omega_0 \int_0^{\pi} d\phi \ \  d/v(d) \sim (d/R_h)^2 .
\end{equation}

Bundling is a rich phenomenon, with a multitude of bundling
states depending on the elastic properties of a flagellum.
For example, it has been shown by \citet{jans:11} that for certain flagellum flexibilities
and depending on the initial condition, either rather tight bundles,
with the flagella in mechanical contact, or loose bundles, with
intertwined, non-touching flagella can be found.

\begin{figure}[t]
\includegraphics*[width=8cm]{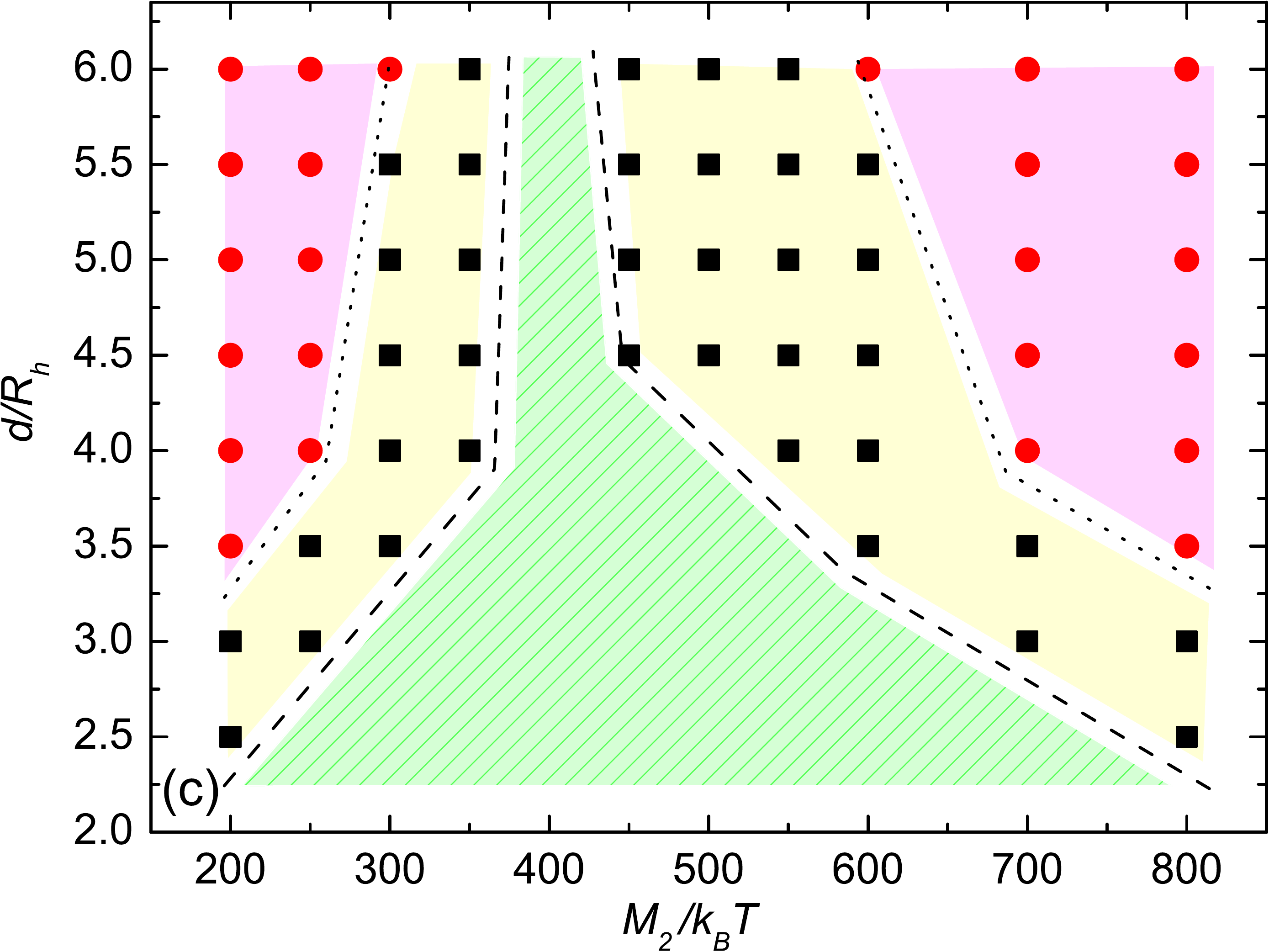}
\caption{\label{fig:phase_diagram}
Phase diagram indicating
stable bundles (green), intermittent slippage (yellow), and drifting
states (red) of a system of three helices
exposed to different torques and for various helix-base separations $d$.
Two helices ($1,3$) experience the torque
$M_1=M_3=400 k_BT$, whereas the torque $M_2$ is varied
\cite{reig:13}.
}
\end{figure}

This raises the question of the stability of a flagellar bundle. There are
various sources, which give rise to fluctuations and variations in the
torque of a flagellum, such as an intrinsic noise in the motor torque,
or the inequality of motors \cite{chen:00,xing:06}.
Variations may even be induced on purpose, as for bacteria such as
{\em R. lupini}, which control the motor torque to induce tumbling
\cite{scha:02}. Simulations of three flagella, in which one helix is
driven by a different torque than the other two, yield the following
qualitative classification of the bundling dynamics (see
Fig.~\ref{fig:phase_diagram}) \cite{reig:13}. At small torque
differences, the bundle remains stable with a phase lag between
the various flagella. For very large torque differences, the
bundle disintegrates and the flagella rotate asynchronous and
independently, i.e., the phase differences of neighboring helices are
drifting. In between, there is an intermittent regime, where phase
slippage occurs, i.e., the synchronized rotational motion is
interrupted by events, where the flagellum with the larger torque
leaves the bundle, rotates faster, and rejoins the bundle. The
time interval between individual slippages decreases with increasing
torque difference and ultimately drifting is obtained. This is
quantitatively shown in Fig.~\ref{fig:phase_diagram}.  It is important to
note that bundle formation is rather robust over a wide range of torques and
separations, so that bundles are able to sustain considerable torque
differences. Even for distances as large as $d/R_h =4$, phase
locking occurs for torque differences as large as $\Delta M/M_1
\approx 1/3$. Both, the high bundle stability and the asymmetry of the
phase diagram with respect to the reference torque $M_{1,3}/k_BT =400$
is a consequence of the flagellar flexibility, which allows the helices to
partially wrap around each other --- which happens differently for smaller and
larger torques $M_2/k_BT$ due to the chiral structure and the different
hydrodynamic interactions. This can be
compared with the synchronization of rigid, three-arm colloidal
micro-rotators discussed by \citet{dile:12}, which can tolerate
only very small torque differences on the order of $\Delta M/M_1 \approx
10^{-4}$ at much smaller distances of $d/R_h =2.0$ to $2.5$. The
largely enhanced bundle stability is due to the flexibility of the
bacterial flagella;
thus, intertwisting stabilizes a bundle and ensures a coordinated
and synchronized motion, which cannot be provided solely by
hydrodynamic interactions.

\subsection{Synchronization of Sperm and Flagella}
\label{sec:sperm_synchro}

When two sperm swim close to each other, at distances smaller than
the sperm length, then the dipole approximation does not apply and
the full hydrodynamic interactions between two time-dependent
flagellar shapes have to be taken into account. An particularly
interesting aspect of this interaction is that the flagellar beats
of the two sperm can now affect each other. What typically happens
is phase locking, {\em i.e.}, the two flagella adjust such as to
beat in synchrony.

This synchronization of the beat of swimming sperm has been studied by
mesoscale simulations in two dimensions \cite{gg:gomp08l}.
Two sperm, S1 and S2, are placed inside a fluid, initially with straight
and parallel tails at a small distance with touching heads.
They start to beat at $t=0$ with fixed frequency $\omega$, but with
different initial phases $\varphi_1$ and $\varphi_2$. Here, the beat
is driven by the local time-dependent preferred curvature
$c_0(s,t) = A \sin(\omega t - ks + \varphi)$ at position $s$ along the
flagellum.
The flagellar shapes and the
relative positions can adjust due to hydrodynamic interactions.

\begin{figure}
\centering
\includegraphics[scale=0.25]{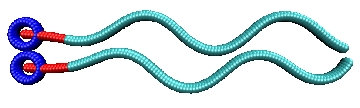}
\includegraphics[scale=0.25]{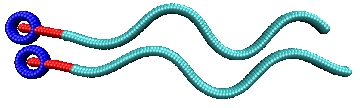}
\includegraphics[scale=0.25]{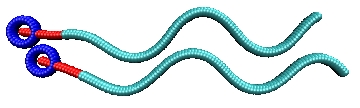}
\includegraphics[scale=0.25]{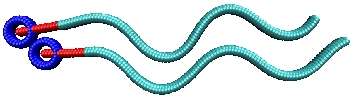}
\includegraphics[scale=0.25]{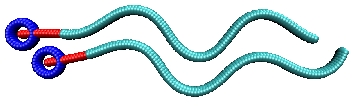}
\includegraphics[scale=0.25]{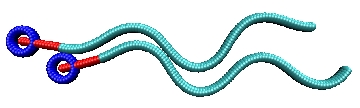}
\includegraphics[scale=0.25]{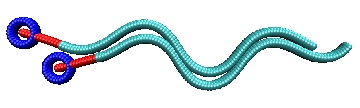}
\caption{Snapshots of two sperm in two dimensions, with phases $\varphi_1$
(upper), $\varphi_2$ (lower), and phase difference
$\Delta\varphi=\varphi_2-\varphi_1=0.5\pi$.
(a) $t\omega=\pi/3$ initial position; (b) $t\omega=4\pi/3$; (c) $t\omega=7\pi/3$;
(d) $t\omega=10\pi/3$; (e) $t\omega=13\pi/3$; (f) $t\omega=61\pi/3$;
(g) $t\omega=601\pi/3$ (from top left to bottom).
From (a) to (e), the synchronization process takes place. The tails are
already beating in phase in (e). From (e) to (g), two synchronized sperm form
a tight pair due to hydrodynamic attraction.
From \citet{gg:gomp08l}.
}
\label{fig:snapshots}
\end{figure}

In the dynamical behavior
of these hydrodynamically interacting sperm, two effects can be
distinguished, a short time ``synchronization" and a longer time
``attraction" \cite{gg:gomp08l}.
If the initial phase difference $\Delta\varphi=\varphi_2-\varphi_1$
is not too large,
``synchronization" is accomplished within a few
beats. This process is illustrated by snapshots at different simulation times in
Fig.~\ref{fig:snapshots}.
The synchronization time depends on the phase difference, and varies
from about two beats for $\Delta\varphi=0.5\pi$ (see Fig.~\ref{fig:snapshots})
to about five beats for $\Delta\varphi=\pi$.
A difference in swimming velocities adjusts the relative positions of
the sperm. After a rapid transition, the velocities of two cells
become identical once their flagella beat in phase.

\begin{figure}
\centering
\includegraphics[width=0.45\textwidth]{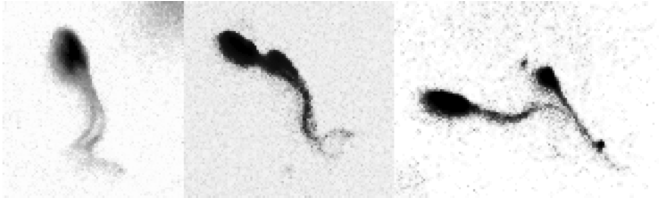}
\caption{Experimental snapshots of two human sperm swimming together at
different times. (Left) Two sperm with
initially well synchronized tails and very small phase difference;
(Middle) the sperm are still swimming together and are well
synchronized after 4 seconds; a phase difference has developed;
(Right) the sperm begin to depart after 7 seconds.
From \citet{gg:gomp08l}.
Pictures courtesy of Luis Alvarez, Luru Dai and U. Benjamin Kaupp
(Forschungszentrum caesar, Bonn).}
\label{fig:exp_human_sperm}
\end{figure}

The synchronization and attraction of sperm has also been observed
experimentally \cite{gg:gomp08l,wool09}.
Figure~\ref{fig:exp_human_sperm} shows two human sperm with
synchronized beats, which swim together for a while, but then
depart again due to somewhat different beat frequencies and thus
different swimming velocities. On the other hand, for bull sperm,
persistent synchrony of the flagellar beats
has been observed only when the heads are tightly coupled
mechanically \cite{wool09}.

\begin{figure}
\centering
\includegraphics[width=0.45\textwidth]{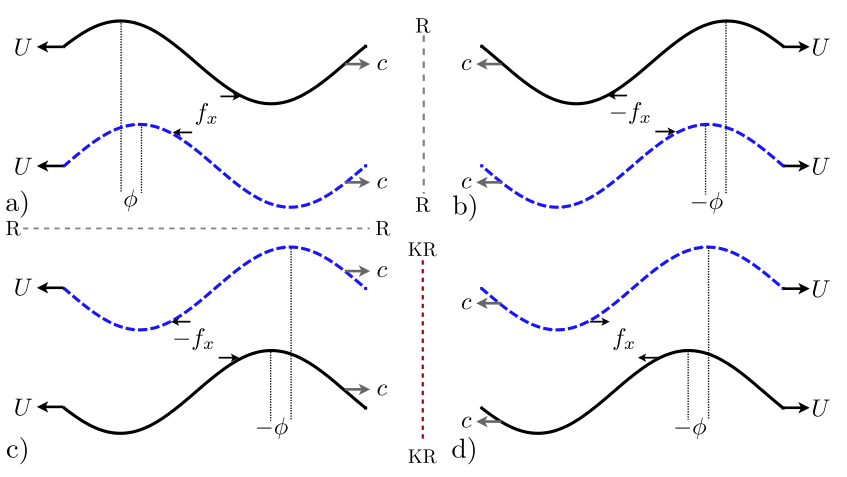}
\caption{Swimmers with reflection symmetries
by the horizontal and vertical axes cannot phase lock: If a relative
force exists in (a), we obtain the forces in (b) and (c) by
reflection symmetries (R planes). Applying kinematic reversibility
to (c) (KR plane) leads to a force in (d) which is minus the one
in (b), indicating that both must be zero.
From \citet{elfr09}.}
\label{fig:synchro_symm}
\end{figure}

Symmetry arguments show that synchronization is {\em not}
possible for two co-swimming flagella with a prescribed,
{\em reflection-symmetric} wave form \cite{elfr09}, described
in the spirit of Taylor \cite{tayl51} by two infinite parallel
two-dimensional sheets with propagating lateral waves with the
same wave vector ${\bm k}$ but a phase shift $\phi$, as
illustrated in Fig.~\ref{fig:synchro_symm} --- and thus corresponding
to the
two-dimensional system described above, except that the flagella are
infinitely long. Suppose that the force
$f_x$ between the swimmers acts to decrease the phase
difference $\phi$, as shown in Fig.~\ref{fig:synchro_symm}a.
This setup can be reflected at the vertical plane to obtain
Fig.~\ref{fig:synchro_symm}b, or first at the horizontal plane
and then by kinematic reversal to obtain Fig.~\ref{fig:synchro_symm}d.
Obviously, Figs.~\ref{fig:synchro_symm}b and \ref{fig:synchro_symm}d
describe exactly the same physical situation. However, the force
$f_x$ is reducing the phase difference in Fig.~\ref{fig:synchro_symm}b
while it is increasing $\phi$ in Fig.~\ref{fig:synchro_symm}d,
indicating that $f_x\equiv 0$. Thus, flagella with pure sine waves
cannot synchronize.

On the other hand, wave forms which are not
front-back symmetric, such as wave forms of sperm with increasing
amplitude of the flagellar beat with increasing distance from the head
\cite{frie10}, or flagella which can respond elastically to hydrodynamic
forces \cite{gg:gomp08l,llop13} (compare Fig.~\ref{fig:snapshots}), this
symmetry argument does not apply, and non-zero forces can appear.
A more detailed calculation by \citet{elfr09} --- based on the
lubrication approximation valid for small distances between the
oscillating sheets --- shows that the time evolution of the phase
difference between co-swimming cells depends on the nature of the
geometrical asymmetry,
and that microorganisms can
phase-lock into conformations which either minimize (in-phase mode) or
maximize (peristaltic mode) energy dissipation.
However, the relative arrangement of the two sperm also plays an
important role \cite{llop13}.  For two sperm with parallel beating
planes, even the sign of the interaction can
change depending on whether the beating planes are co-planar or are
stacked on top of each other.

\subsection{Cilia Synchronization}
\label{subsec:cilia}

Motile cilia are abundant in eucaryotic microswimmers to generate
propulsion. From {\em Paramecium} over {\em Volvox}, cilia
are used from unicellular to multicellular organisms to propel fluid
across their surface \cite{afze:76}. In higher organisms and humans, cilia
are not only involved in moving mucus in the lungs, but also in embryonic
development, e.g. in breaking the left-right symmetry
\cite{cart:04}, and in cell signaling \cite{wang:06}.
Already in the 1960s \cite{sle_book.2}, it was
observed that arrays of cilia beat neither randomly nor synchronously,
but in a wave pattern called metachronal wave (MCW).

\begin{figure}[htbp]
\includegraphics[width=0.48\textwidth]{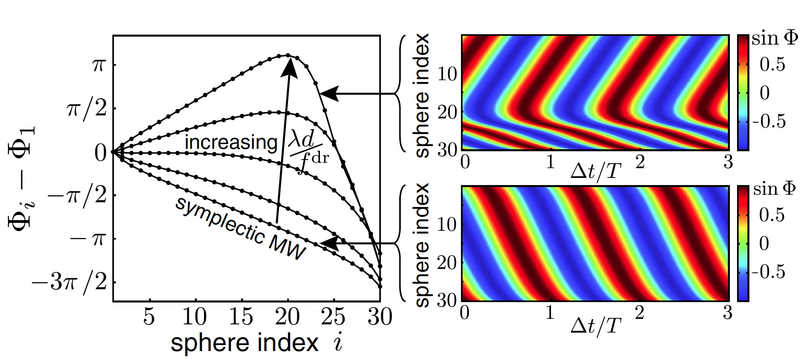}
\caption{
Phase profile of an array of 30 spheres
after 1200 beats. Results are shown for different driving forces
$f_{dr}$, spring constants $\lambda$, and distances from the surface $d$.
From \citet{brum:12}.
\label{fig:brum}
}
\end{figure}

Several theoretical models have been proposed to explain
the hydrodynamic origin of this phenomenon.
One approach is to focus on highly simplified systems similar to
those presented in Sec.~\ref{sec:synchrobasic}. Either the
rotating spheres are placed near a no-slip wall
\cite{vilf:06,nied:08,uchi:10,gole:11,brum:12}, or spheres oscillate on a
line with different hydrodynamic radii in the two directions of
motion \cite{lago:03,woll:11}. One-dimensional chains of such
simplified cilia, can show metachronal waves under special
conditions \cite{lago:03,woll:11,brum:12}. For {\em Volvox}, such a
rotating-sphere model has been compared in detail with the
experimental situation \cite{brum:12}.
By choosing the right orbit parameters, the experimentally
observed metachronal wave pattern can be reproduced. Hence, the
simple theoretical approach can help to understand the origin of the
nontrivial synchronization with finite phase lag. As it turns out,
the presence of a surface already suffices to create a  non-zero
phase lag (see Sec.~\ref{sec:synchrobasic}).

\begin{figure*}[htbp]
\includegraphics[width=0.92\textwidth]{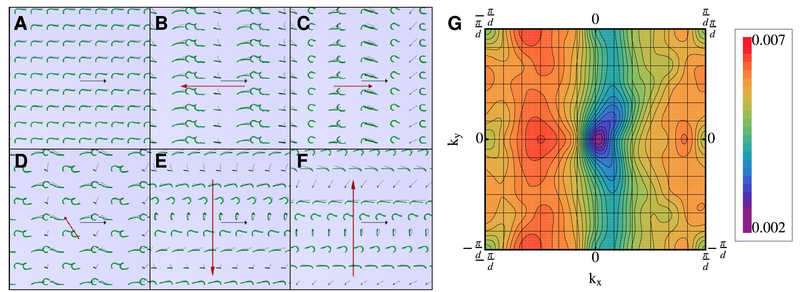}
\caption{
Optimal solutions at fixed wave vectors for inter-ciliary distance
$d = 1.0 L$, $N=20$, where $L$ is the cilium length.
(A-F) Optimal solutions for wave vectors
(A) $(k_x,k_y) = (0,0)$, (B) $(-\pi/(2d),0)$, (C) $(5\pi/(6d),0)$,
(D) $(-2\pi/(3d),\pi/d)$, (E) $(0,-\pi/(3d))$, and (F) $(0,\pi/3d)$.
The blue arrow ($x$-axis) denotes the direction of
pumping and the red arrow the wave length and direction of
metachronal-wave propagation. (G) Efficiency $\epsilon_c'$
(red color represents high efficiency) as
a function of the wave vector $(k_x,k_y)$. The maximum efficiency
in this case is achieved for $k = (\pi/(2d), 0)$, and antiplectic
waves are generally more efficient than symplectic ones. The
synchronous solution $(k_x,k_y)=(0,0)$ represents the global minimum
of efficiency.  From \citet{Osterman2011}.
\label{fig:osterman}
}
\end{figure*}

A second approach uses efficiency arguments.
The idea is that the cilia beat has been optimized during evolution
to attain maximum efficiency \cite{Osterman2011,Eloy2012}.
Here, efficiency is defined as minimal power
needed to actuate the cilium along a certain path to generate
the fluid transport with a velocity $v$. By simple dimensional
analysis \cite{Eloy2012,Osterman2011}, or by comparison with a
reference flow driven by a constant force density parallel to a
wall in a slab of thickness $L$ (the cilium length) \cite{Elgeti2013},
it has been shown that the  efficiency should scale as
\begin{equation}
  \label{eq:efficiency}
  \epsilon=\frac{Q^2}{P}.
\end{equation}
where $Q$ is the average volumetric flux and $P$ the average power
consumption. This approach yields a well defined cilia stroke
\cite{Osterman2011,Eloy2012} that looks very similar to observed cilia
strokes, see Fig.~\ref{fig:ciliumbeat}.
This mechanism  can be understood as follows.
In the fast power stroke, the cilium is nearly fully extended, to reach as
far out from the wall as possible, where the fluid flow is least restricted
by the presence of the no-slip wall; in contrast, in the slow recovery
stroke, the cilium curves and bends sideways to move closely to the
wall to generate as little backflow as possible, and at the same time not
to bend too much to avoid damage of the microtubule structure.
This efficiency argument can be taken one step further to provide a
potential explanation of metachronal waves.
If all cilia beat synchronously, the whole fluid body flows back and
forth with every stroke, which implies a high energy dissipation.
Thus the synchronous beat is highly inefficient.
By numerically calculating the dissipation for different wave vectors,
\citet{Osterman2011} predict antiplectic metachronal
waves, see Fig.~\ref{fig:osterman}. However, it is not obvious
whether global efficiency optimization can be a criterion for the
evolution of a collective cilia beat in a system, which should perform
well under a variety of environmental conditions.

\begin{figure*}[htbp]
\includegraphics[width=0.95\textwidth]{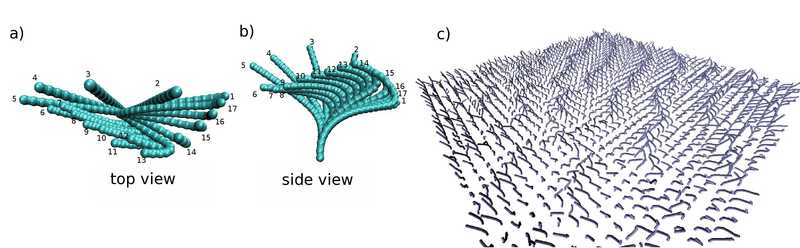}
\caption{
(a) Side and (b) top view of the beat pattern of the
computational cilia model. Subsequent conformations are equally spaced
in time. The simulated beat pattern is, for example, similar to the
beat pattern of rabbit tracheal cilia in culture medium \cite{Sanderson1981}.
The fast, planar power stroke (frames 1 to 5) continues until a positive
curvature threshold in the lower part of the cilium is reached.
The cilium then switches to a slow, out-of-plane recovery stroke
(frames 6 to 17), which ends when a negative curvature threshold is exceeded.
(c) Simulation snapshot of an array of $40\times40$ beating cilia.
Cilia are placed on a square lattice, with lattice constant $d_c$.
The metachronal wave can easily recognized by the lines of fully extended
cilia during the power stroke. From \citet{Elgeti2013}.
\label{fig:configs}
}
\end{figure*}

\begin{figure}[htbp]
\includegraphics[width=0.49\textwidth]{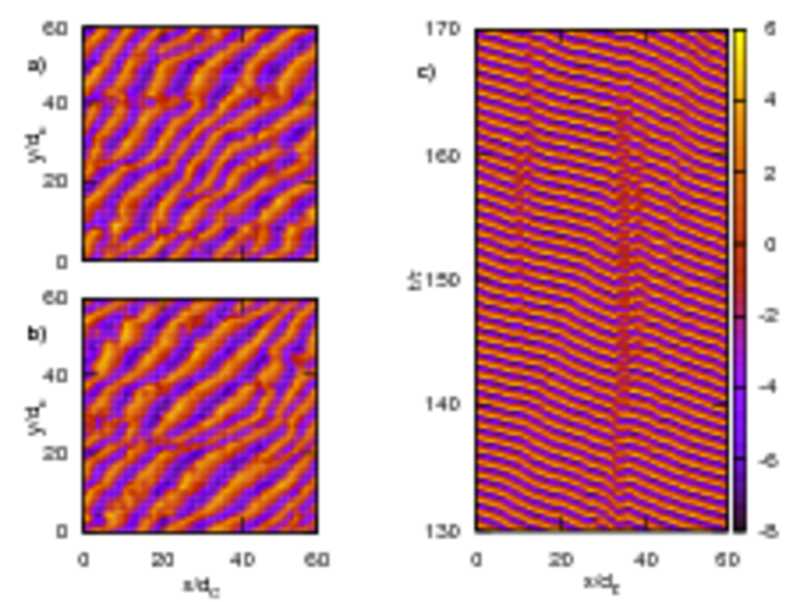}
\caption{
(a),(b) Phase-field representations of a metachronal wave, for an
array of 60 × 60 cilia, at two times separated by about 40 beats. The color
denotes the projected displacement of the tip of a cilium from its base in the
direction of the power stroke. (c) Time dependence of a selected line of cilia
along the x axis. As a function of time, defects in the metachronal wave
pattern appear and disappear. From \citet{Elgeti2013}.
\label{fig:config_def}
}
\end{figure}

A third approach is to model cilia as semiflexible filaments with a
pre-defined beating mode, and to allow for self-organized metachronal
waves by hydrodynamic interactions. This approach has been followed
for one-dimensional chains of cilia \cite{guer:97,guer99}, for
small two-dimensional patches of cilia \cite{guer01}, and in a
mean-field approach \cite{guir07}.
\citet{Elgeti2013} were able to extend this type of
approach recently to much larger two-dimensional arrays (up to $60\times 60$
cilia) in a three-dimensional fluid with noise.
The beat pattern of a single cilium looks like that of a cilium of
paramecium. The beat pattern of an individual cilium can react to
the surrounding fluid flow, because the model only imposes
time-dependent curvature forces, and employs geometric thresholds
for the switch  between power and recovery stroke, and {\em vice versa}
\cite{lind10b}. Here, metachronal waves emerge autonomously
despite the presence of strong noise (see Fig.~\ref{fig:configs}).
Furthermore, this approach allows the study of the appearance of defects,
transport efficiency, and wave vectors.
In particular, it predicts a large increase in efficiency and
propulsion velocity due to metachronal coordination --- without the
assumption that the system has evolved to optimal
efficiency. This efficiency gain is due to the rectification of fluid
flow across a ciliated surface, which avoids the oscillatory back-and-forth
motion of a perfectly synchonized beat with large viscous energy loss.
This result indicates that the efficiency gain by metachronal
coordination is a rather universal feature, and does not require
an evolutionary optimization strategy. \citet{Elgeti2013} could also
demonstrate the presence of defects in the wave
pattern of MCWs and characterize some of their properties.
It seems that defects assume a form similar to dislocations
in two-dimensional crystals (see Fig.~\ref{fig:config_def}).
However, the defects do not travel with the wave, but remain stationary
while the wave passes over them. More detailed studies are needed to fully
elucidate the defect dynamics in MCWs.

\section{Collective and Cooperative Motion}

Microswimmers hardly ever swim alone.
In nature, the density of microswimmers can reach astonishing densities.
Sperm cells are released by the millions to compete in the run for the egg.
Coordinated motion is exploited, for example, by spermatozoa of the wood
mouse {\em Apodemus sylvaticusi}, which self-assemble into unique
train-like aggregates of hundreds or thousands of cells and thereby
significantly increased sperm motility
\cite{sivi:84,haya96,moor95,haya98,imml07,moor02}.
Bacteria grow by dividing and invading their surroundings together.
Artificial microswimmers will only be able to deliver useful quantities
of pharmaceuticals or modify material properties when present in
large numbers.

In assemblies of motile microorganisms, cooperativity reaches a new
level as they exhibit highly organized movements with remarkable
large-scale patterns such as networks, complex vortices, or swarms 
\REV{\cite{hein:72,kear:10,wens12,gach:14}}. Such patters are typically
displayed by bacteria confined to two dimensions, e.g., {\em E. coli}
or {\em Bacillus subtilis} involving hundreds to billions of cells
\cite{hars:94,liu:00,avro:04,darn:10}. Flagella are an essential
ingredient in swarming of biological cells, as is evident for, e.g.,
{\em E.~coli}, which produce more flagella and, in addition,
elongate and become multinucleate
\cite{darn:10,cope:09,kais:07,dani:04,hars:03}, which underlines
the complexity of the interactions in such assemblies.

The full characterization of the complex dynamical behavior requires
an understanding of the underlying (physical) cooperative mechanism
on various levels, starting from the interactions of individual cells,
fluid-mediated interactions, up to the generic
principles of the formation of large-scale patterns.

\subsection{Hydrodynamic Interactions between Microswimmers}
\label{sec:attract}

\begin{figure}
\centering
\includegraphics[width=0.45\textwidth]{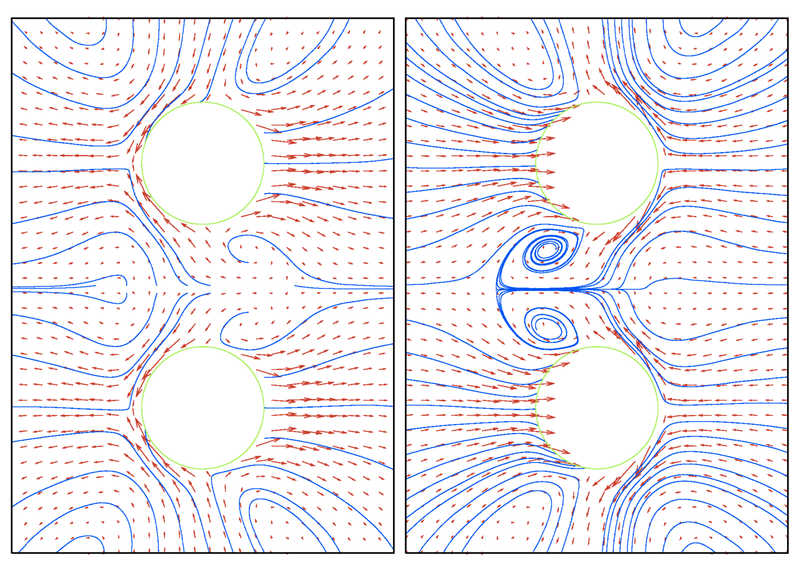}
\caption{
Velocity fields for fixed parallel pairs of squirmers,
for (a) pusher (with $B_2/B_1=-3$) and (b) puller (with $B_2/B_1=+3$),
with P{\'e}clet number ${\rm Pe}=1155$. For definition of mode
amplitudes $B_1$ and $B_2$ see Eq.~\ref{eq:squirmvel} in Sec.~\ref{sec:intro}.
Swimmers move to the right. Streamlines serve as a guide to the eye.
Only a fraction of the simulation box is shown.
(Due to the finite resolution of the measured velocity field,
some streamlines end on the squirmers' surfaces.)
From \citet{gg:gomp10l}.
\label{fig:squirmer_int}}
\end{figure}

Similar as the interactions of microswimmers with a surface,
the interaction of two microswimmers at long distances is determined by
their dipole flow fields.
The dipole approximation predicts that the interactions of
microswimmers depends on their relative orientation, and that
pushers and pullers in equivalent positions and orientations
have interactions equal in magnitude and opposite in sign, because
their dipole strengths $P$ have opposite signs, see Sec.~\ref{sec:dipole}.
This behavior can be
understood easily by considering the flow fields of two parallel-swimming
pushers or pullers, where pushers attract and pullers repel each other.
This effect is also present for shorter distances
between the swimmers; the results of mesoscale
hydrodynamics simulations of two squirmers \cite{gg:gomp10l} shown in
Fig.~\ref{fig:squirmer_int} demonstrate that for {\em pushers}, the
fast backward flow velocity in the rear part extracts fluid from
the gap between the swimmers, and thereby induces attraction
(Fig.~\ref{fig:squirmer_int}a); in contrast, for {\em pullers},
the fast backward flow velocity in the front part injects fluid
into the gap, and thereby induces repulsion (Fig.~\ref{fig:squirmer_int}b).

\begin{figure}
\centering
\includegraphics[width=0.36\textwidth]{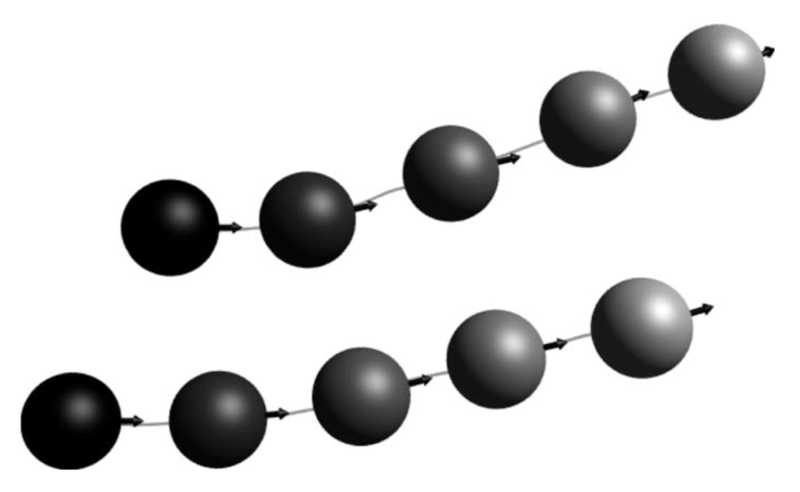}
\caption{
Trajectories of a pair of initially parallel squirmers (pushers)
with P{\'e}clet number $Pe = 1155$. Snapshots are shown at fixed time intervals
$\Delta t v / \sigma = 1.67$. The initial offset is
$\Delta z= 2\sigma$ (perpendicular to direction of motion)
and $\Delta x = \sigma$ (parallel), where $\sigma$ is the squirmer diameter.
The instantaneous squirmer orientations are indicated by the arrows.
From \citet{gg:gomp10l}.
\label{fig:squirmer_trajectory}}
\end{figure}

However, the hydrodynamic interaction between microswimmers is much more
complex, because they will usually meet at different relative positions and
orientations, and possibly different relative phases of their internal
propulsion mechanisms. This gives rise to a rich behavior of attraction,
repulsion, or entrainment depending on these parameters
\cite{alex08,ishi06,gg:gomp10l}. An example is shown in
Fig.~\ref{fig:squirmer_trajectory}, which shows the trajectories of two
initially parallel oriented pushers, which have a small offset in the
forward direction. As a result of this offset, the two squirmers do not
attract each other, but rather change their direction of motion together
toward the pusher in front. Pullers show the opposite trend and change their
direction of motion toward the puller in the back.

\subsection{Generic Model of Flocking}
\label{sec:Vicsek}

Collective behavior of active bodies is frequently found
in microscopic systems such as bacteria
\cite{BenJacob2000,Sokolov2007,Peruani2012,chen2012prl,Gachelin2013}
and synthetic microswimmers \cite{ibel09,theu:12,palacci2013,butt:13,wang13},
but also in macroscopic systems
such as flocks of birds and schools of fish \cite{vics12,cavagna2010pnas}.
Despite the very different propulsion mechanisms and interactions in
these systems, they all favor alignment of neighboring bodies, thus leading
to similar forms of collective behavior.

Therefore, it is natural to look for a model, which is able to capture
the generic collective properties of all the various systems of
self-propelled particles and organisms.
Such a model was proposed in a pioneering work by \citet{Vicsek1995}.
In this model, now often called the ``Vicsek model", $N$ polar point
particles move in space with constant magnitude of velocity $v_0$. The
dynamics proceeds in two steps, a streaming step of duration $\Delta t$,
in which particles move ballistically, and a interaction step, in which
particles align their velocity direction with the average direction of
motion of their neighbors. In two spatial dimensions, this implies the
dynamics for the position ${\bm r}_i$ and velocity ${\bm v}_i$ of particle
$i$,
\begin{eqnarray}
{\bm r}_i(t+\Delta t) &= {\bm r}_i(t) + {\bm v}_i(t) \Delta t ,\\
\theta_i(t+\Delta t) &= \langle \theta(t)\rangle_\sigma + \Delta \theta ,
\end{eqnarray}
where $\theta_i$ is the angle between ${\bm v}_i$ and the $x$-axis of
a Cartesian coordinate system, $\langle \theta(t)\rangle_\sigma$ is
the average orientation of all particles within a circle of diameter
$\sigma$. and $\Delta \theta$ is a random noise uniformly distributed in
the interval $[-\eta/2, +\eta/2]$. The essential control parameters of the
Vicsek model are the particle density $\rho$, the noise strength $\eta$,
and the propulsion velocity $v_0$ in units of $\sigma/\Delta t$.

A numerical investigation of this model by \citet{Vicsek1995} shows a phase
transiton with increasing density or decreasing noise strength from a random
isotropic phase to an aligned phase, in which particles move collectively
in a spontaneously selected direction.
This observation of a non-equilibrium phase transition in a system of
self-propelled point particles has led to numerous analytical
\cite{Toner1995,Simha2002,Ramaswamy2003,Toner2005,Peruani2008,
Baskaran2009,Bertin2009,gole09}
as well as computational
\cite{redn:13,Szabo2006,Gregoire2004,Huepe2004,Dorsogna2006,Aldana2007,
Chate2008,Ginelli2010} studies.
A review is provided by \citet{vics12}.

\subsection{Self-Propelled Rods}
\label{sec:SPR}

When we consider a system of self-propelled particles beyond the phenomenological
description of interactions as in the Vicsek model \cite{Vicsek1995}, the simplest
physical interaction which leads to alignment is volume exclusion of
rod-like microswimmers and self-propelled rods.  Here polar interactions, with
alignment in the swimming directions, have to be distinguished from
nematic interactions, with aligment independent of the direction of motion.
Of particular interest are experiments with elongated self-propelled
particles on the microscopic scale in two dimensions, such as motility
assays where actin filaments are propelled on a carpet of myosin motor
proteins \cite{Harada1987,Schaller10}, microtubules propelled by
surface-bound dyneins \cite{Sumino12}, and microswimmers that are
attracted to surfaces (as described in Sec.~\ref{sec:surface}).

Self-propelled rods (SPRs) in two dimensions
are often modelled as linear chains of (overlapping) beads
\cite{wens08,Yang2010,Abkenar2013}.
Both, models with strict excluded-volume interactions and with a
finite overlap energy, which allows rods to cross, have been employed.
In the latter case, rod crossing is included to mimic in a two-dimensional
simulation the possible escape of rods into the third dimension when two
rods collide (like actin filaments or microtubules in a motility
assay) \cite{Abkenar2013}.

In such models, self-propulsion of rods leads to enhanced aggregation and
cluster formation, as well as to various kinds of ordered phases
in Brownian Dynamics simulations
\cite{peru06,Kraikivski2006,wens08,Mccandlish2012,wens12,Yang2010,Abkenar2013},
as well as in theoretical approaches based on the Smoluchowski equation
\cite{Baskaran2008b,Baskaran2008,Peruani2010} and on continuum models in
the ``hydrodynamic limit" \cite{Baskaran2008b,Baskaran2008}.

\begin{figure}
\centering
\includegraphics[width=0.23\textwidth]{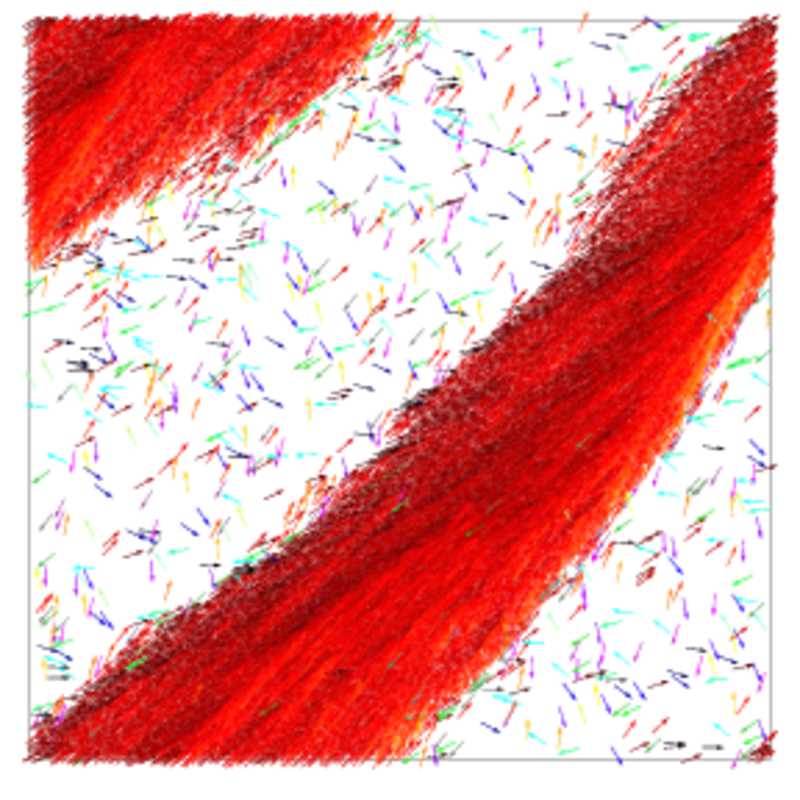}
\includegraphics[width=0.23\textwidth]{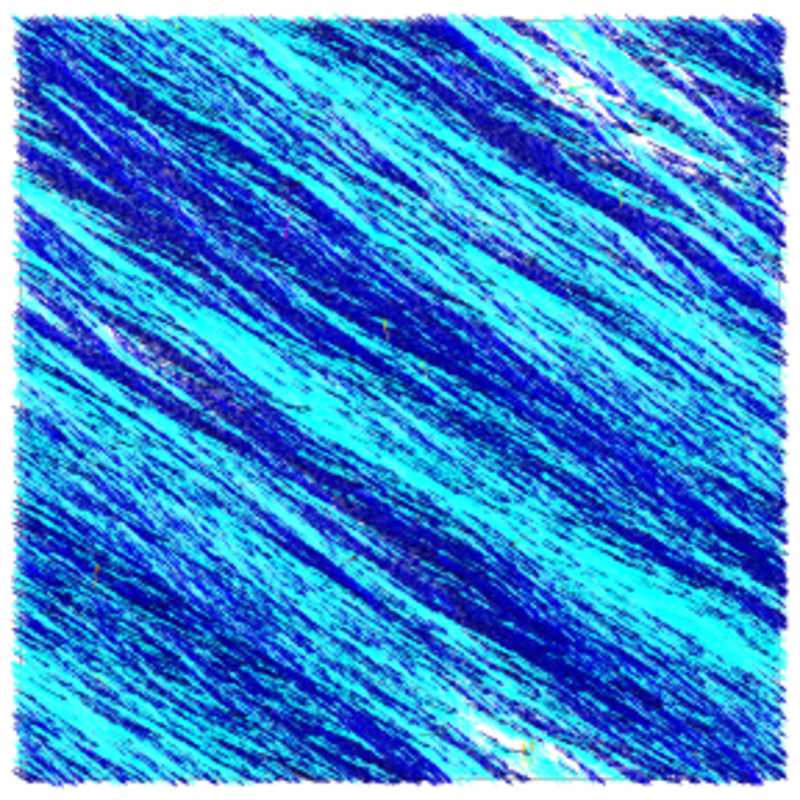}
\caption{Simulation snapshots of a system of self-propelled soft rods
(with aspect ratio 18) display (left) a giant polar cluster coexisting with
a very dilute phase of single swimmers at intermediate density
$\rho L_{rod}^2=10.2$ and P{\'e}clet number $Pe=25$, and (right) a
phase of nematic aligment, in which rods moving in opposite directions
self-organize in lanes at high rod density $\rho L_{rod}^2=25.5$ and
P{\'e}clet number $Pe=75$.
Colors indicate rod orientation.
From \citet{Abkenar2013}.
\label{fig:SPR_phases}
}
\end{figure}

\begin{figure}
\centering
\includegraphics[width=0.48\textwidth]{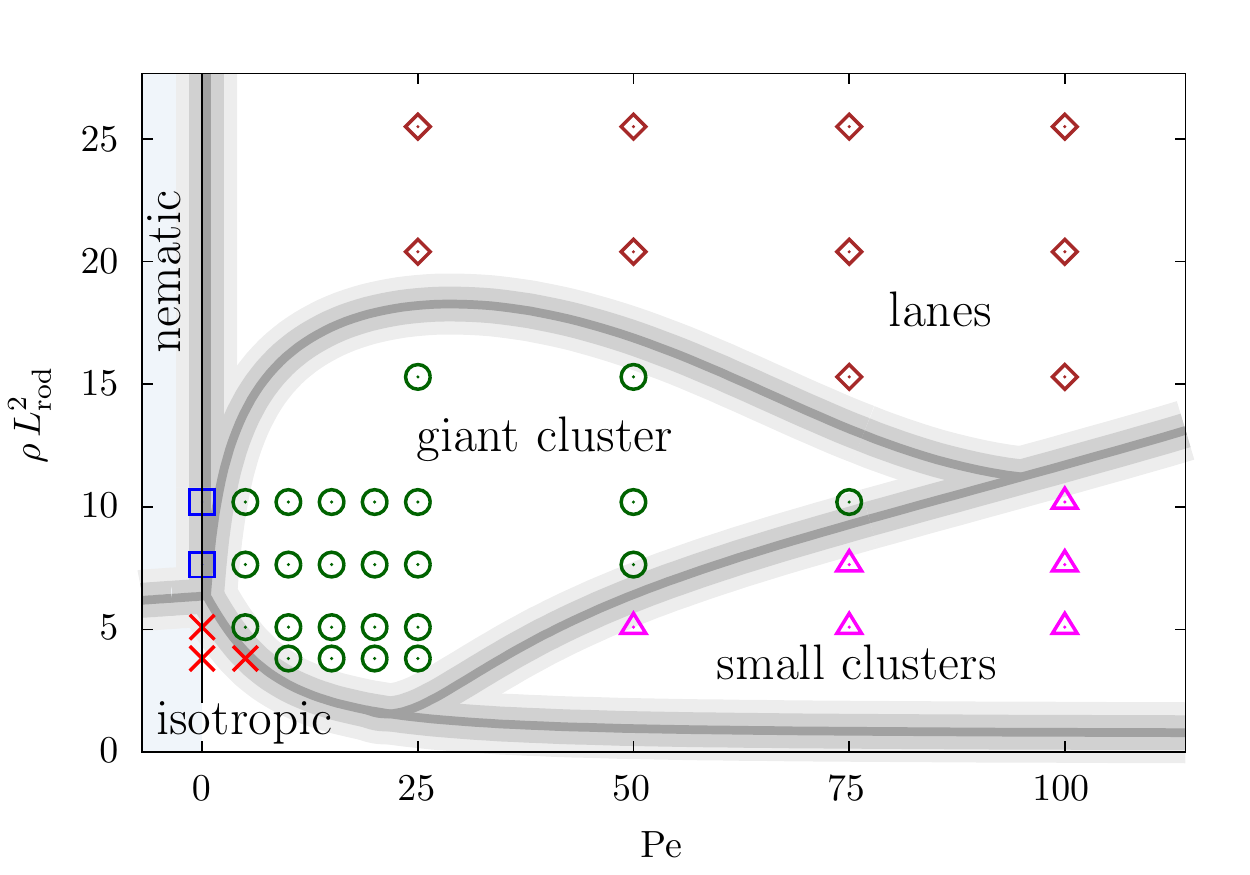}
\caption{Phase diagram for self-propelled rods (with aspect ratio 18) as a
function of density $\rho$ and P{\'e}clet number $Pe$. The energy barrier
is $E=1.5\,k_BT$; the gray lines are guides to the eye. The region $Pe<0$
has no physical meaning; it only indicates the presence of isotropic and
nematic states for passive rods (with $Pe=0$).
From \citet{Abkenar2013}.
\label{fig:SPR_PD_noise}
}
\end{figure}

A few simulation snapshots of a two-dimensional system,
with different rod number densities $\rho$, propulsion velocities $v_0$,
and noise levels are shown in Fig.~\ref{fig:SPR_phases}, and a
corresponding phase diagram in Fig.~\ref{fig:SPR_PD_noise}.
Here, the importance of propulsion compared to noise is characterized by the
P{\'e}clet number $Pe=L_{rod}v_0/D_\parallel$, where $L_{rod}$ is the
rod length, $v_0$ the swimming velocity of an isolated swimmer, and
$D_\parallel$ the diffusion coefficient in the direction of the instantaneous
rod orientation.
The phase diagram of Fig.~\ref{fig:SPR_PD_noise} describes a system in which
rods interact with a ``soft" interaction potential, so that the rods are
penetrable at high P{\'e}clet numbers \cite{Abkenar2013}.
Disordered states, motile clusters, nematic phases, and lane formation are
observed. In particular, Fig.~\ref{fig:SPR_phases} displays a phase with
a single polar cluster, which is formed at intermediate rod density and
not too large P{\'e}clet numbers, and a phase with both nematic
order and a lane structure where rods moving in opposite directions,
which is formed at high rod densities \cite{Mccandlish2012,Abkenar2013}.
Near the transition from a disordered state of small- and intermediate-size
clusters to an ordered state, the cluster-size distribution obeys a
power-law decay, $P(n) \sim n^{-\gamma}$, with an exponent $\gamma\simeq 2$
\cite{Huepe2004,peru06,Huepe2008,Yang2010,Abkenar2013}. This
prediction has been compared in detail with experimental results for
the clustering of myxobacteria on surfaces (for a mutant which shows
polar motion), and very good agreement has been found \cite{Peruani2012}.

The simulation results for systems of self-propelled rods can also be compared
with the results of simulations and analytical calculations for the
Vicsek model and continuum hydrodynamic models \cite{pesh12}. In the
Vicsek and continuum models, polar and nematic interactions have to be
distinguished. For polar interactions, moving
density waves are predicted \cite{Bertin2009,Ginelli2010}, while for nematic
interactions, high-density bands with rods moving parallel to the band in
both directions are expected \cite{pesh12}. Because rods in the simulations
are constructed as chains of beads, the dominant contribution of the
interaction is nematic, but there is also a polar component (i.e.
a somewhat larger probability for parallel than anti-parallel alignment
after collisions) due to an effective friction resulting from the (weakly)
corrugated interaction potential \cite{Abkenar2013}.
However, neither with polar nor with nematic interactions do
Vicsek and continuum hydrodynamic models predict the
polar bands with rod orientation parallel to the band, as shown in
Fig.~\ref{fig:SPR_phases}a.
Thus, clearly more work is needed to elucidate the origin of the differences
in the results of models of rod-like particles and with anisotropic
interaction potentials and of effective continuum descriptions.

\begin{figure}
\centering
\includegraphics[width=0.48\textwidth]{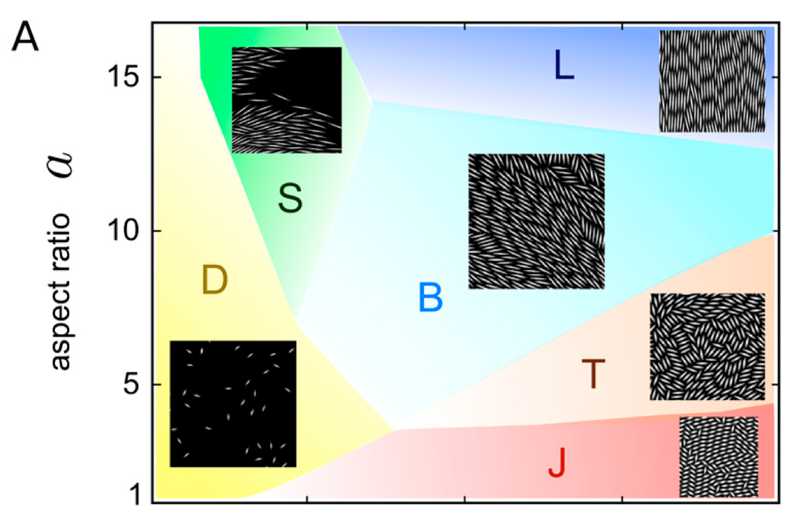}
\includegraphics[width=0.45\textwidth]{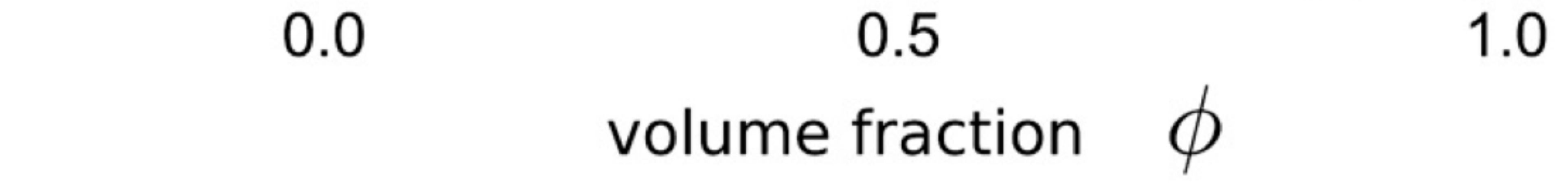}
\caption{Phase diagram of self-propelled rods as a function of aspect ratio
$a$ and volume fraction $\phi$, as obtained from simulations in the absence
of noise. Snapshots indicate the
phases $D$-dilute, $J$-jamming, $S$-swarming, $B$-bionematic,
$T$-turbulence, and $L$-laning.
From \citet{wens12}.
\label{fig:PD_rods_nonoise}
}
\end{figure}

Self-propelled hard rods show also interesting structures in the
absence of noise. A phase diagram as a function of aspect ratio $a$ and
volume fraction $\phi$ as obtained from simulations \cite{wens12} is shown in
Fig.~\ref{fig:PD_rods_nonoise}. At small volume fractions, a dilute phase
of single rods is observed for all aspect ratios. For higher volume fractions,
several phases of densely packed rods are found, which differ in their
internal structure. With increasing aspect ratio, this is a jammed phase
of very short rods, a tubulent phase for intermediate aspect ratios,
local nematic alignment for somewhat larger aspects
ratios, and a swarming phase --- which is reminiscent
of the giant cluster of noisy rods in Figs.~\ref{fig:SPR_phases}(a) and
\ref{fig:SPR_PD_noise} ---, and finally a
laning phase for long rods with aspect ratio $a \gtrsim 12$
--- again similar to the corresponding phase of noisy
rods in Figs.~\ref{fig:SPR_phases}(b) and \ref{fig:SPR_PD_noise}.

\begin{figure}
\centering
\includegraphics[width=0.48\textwidth]{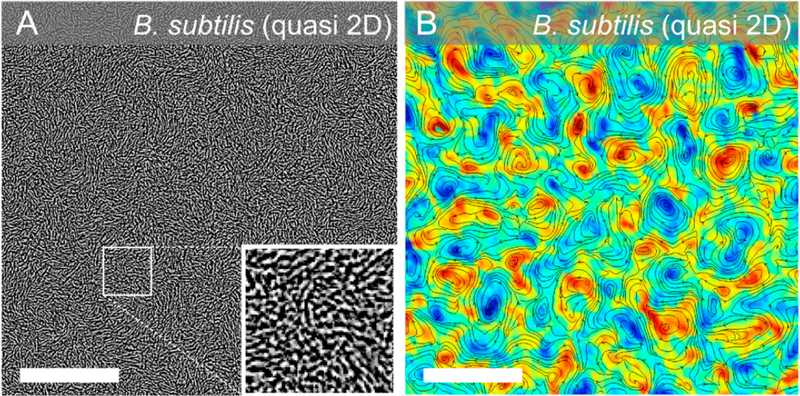}
\caption{Snapshots of (A) bacteria arrangements and (B) the extracted vorticity
field in the turbulent state of a dense bacterial population on a surface.
From \citet{wens12}.
}
\label{fig:bact_turbulence}
\end{figure}

A phase of particular interest is here the ``turbulent" phase.
This phase is denoted turbulent, because it velocity field displays
the typical random swirls and vortices, which are characteristic for
high-Reynolds-number turbulence. Such a phase has indeed been observed
in dense populations of {\em Bacillus subtilis}, both at surfaces and in
three-dimensional bulk suspensions
\cite{Mendelson1999,Sokolov2007,cisn11,wens12}, see
Fig.~\ref{fig:bact_turbulence}a.
This phase has also been studied in some detail theoretically and
in simulations \cite{aran07,wolg08,wens12,dunk13}.
The structure and dynamics of the turbulent phase can be characterized
by velocity distributions, velocity increment distributions, and spatial
correlation functions (or equivalently structure functions). The results
of such an analysis is a typical vortex size $R_v$, see
Fig.~\ref{fig:bact_turbulence}b, which in two dimensions
is about three times the bacterial length $L$. A comparison with
high-Reynolds-number turbulence can be made by calculating the
energy spectrum $E(k)$ as a function of the wave vector $k$, which
is closely related to the Fourier transform of the spatial velocity
correlation function
$\langle {\bm v}(t,{\bm r})\cdot {\bm v}(t,{\bm r + R}) \rangle$.
The main difference between classical and bacterial
turbulence is that the energy input occurs in the former case on
large length scales, but in the latter case on small
scales of the bacterium size $L$.  This has important consequences
for the behavior of $E(k)$. In classical turbulence,
the famous Kolmogorov-Kraichnan scaling \cite{krai80} predicts
in three dimensions an energy-inertial downward cascade with
$E(k) \sim k^{-5/3}$. In two dimensions, there can be both
an energy-inertial
upward cascade with $E(k) \sim k^{-5/3}$ and an enstrophy-transfer
downward cascade with $E(k) \sim k^{-3}$. For bacterial turbulence,
similar power-law regimes are found, but the exponents are different.
Here, $E(k) \sim k^{+5/3}$ for small $k$, and $E(k) \sim k^{-8/3}$ for
large $k$ \cite{wens12}. Thus, self-sustained bacterial turbulence
shares some properties with classical turbulence on small scales, but
differs on large scales \cite{wens12}.

\subsection{Active Brownian Spheres}

The collective phenomena of Sec.~\ref{sec:SPR} above and
Secs.~\ref{sec:cluster_sperm} and \ref{sec:swarming_bacteria} below
are affected by the anisotropic shape and specific interactions between
individual cells. These specificities give rise to particular phenomena but
may be masked by more general underlying organization principles of
microswimmers. Indeed, studies of Brownian, spherical self-propelled
colloidal particles without any alignment rule reveal a rich structural
and dynamical collective behavior
\cite{dese:10,theu:12,bial:12,buttinoni2012,fily:12,butt:13,palacci2013,redn:13,sten:14,wyso:14,fily:14}.

Experiments using either spherical gold/platinum Janus particles \cite{theu:12}
and polymer spheres with an encapsulated hematite cube \cite{palacci2013}
in a hydrogen peroxide solution,  or carbon-coated colloidal Janus particles
dissolved in a near-critical mixture of water and lutidine
\cite{butt:13} and heated by laser light, yield well-ordered crystalline
aggregates at low densities and a phase-transition into large clusters and a
dilute gas phase at higher densities in two dimensions. Thereby, the clusters
are highly dynamic. Colloids adsorb and evaporate from a cluster, and clusters
continuously merge and dissociate \cite{theu:12,butt:13}. The mean cluster
size itself depends linearly on the swimming speed \cite{butt:13}. Large
clusters exhibit a well-ordered, crystalline internal structure, which
dynamically changes mainly due to movements of dislocations.

In order to elucidate the mechanism governing the aggregation of the active
colloids, computer simulations have been performed of a minimal model.
Every colloid is described as a point-particle, which is propelled with
constant velocity along a body-fixed direction. In addition, it is exposed
to a random, white-noise force and the colloid-colloid interaction force,
e.g., arising from a Yukawa-like potential. The orientation of the colloid
performs a random walk with the respective rotational diffusion coefficient.
Hence, no alignment forces or hydrodynamic interactions are taken into account
\cite{bial:12,bial:13,fily:12,butt:13,palacci2013,redn:13,sten:14,wyso:14,fily:14}.
These simulations indeed yield cluster formation and phase separation
solely by propulsion and excluded-volume interactions.

For two-dimensional systems, simulation results are in qualitative agreement
with experiments. For monodisperse spheres, clusters are formed with
pronounced crystalline order. For polydisperse spheres \cite{fily:14}, the
high-density phase remains fluid-like, with very little collective dynamics.

\begin{figure}
\centering
\includegraphics[width=0.4\textwidth]{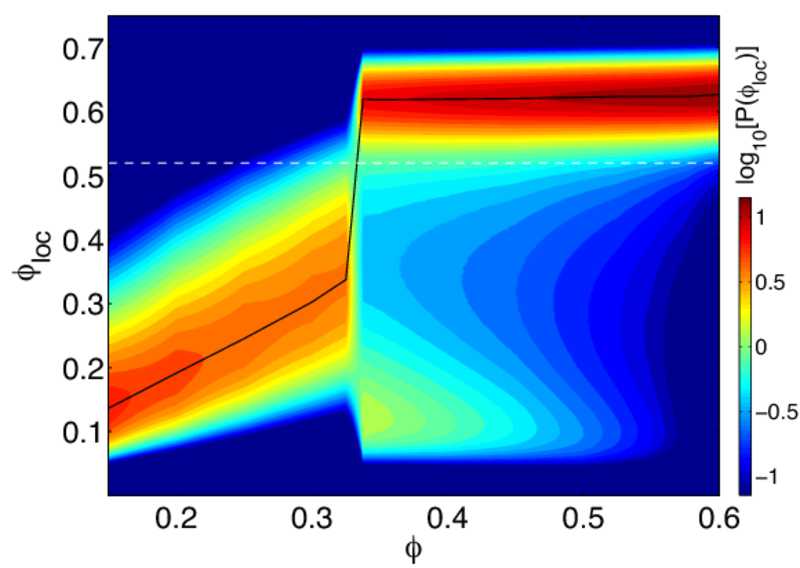}
\caption{Probability distribution $P(\phi_{l})$ of the local packing fraction
$\phi_{loc}$ as a function of the global packing fraction $\phi$ for $Pe=272$.
The most probable $\phi_{loc}$ is indicated by the solid black line.
Redrawn from \citet{wyso:14}.
}
\label{fig:prob_dens_abp}
\end{figure}

\begin{figure}
\centering
\includegraphics[width=0.47\textwidth]{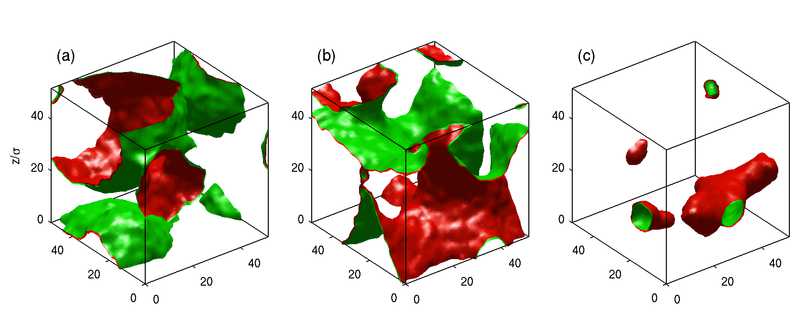}
\caption{Snapshots of the gas-liquid interfaces of
systems (a,b) just above the clustering transition ($\phi=0.3375$ and $\phi=0.4$) and (c) at a
high density ($\phi=0.6$). The red isodensity surfaces corresponds to $\phi_{l}=0.54$ and
the green to $\phi_{l}=0.5$.
 From \citet{wyso:14}.
}
\label{fig:isodensity_abp}
\end{figure}

Simulations reveal a far richer behavior of selfpropelled spheres in
three-dimensions \cite{wyso:14,sten:14}. As for two-dimensional systems, the
system phase separates into a dilute gas-like and dense fluid-like phase above
a critical density at a given swimming velocity, or more precisely,
P\'{e}clet number $Pe = v_0/(\sigma D_r)$, where
$v_0$ is the swimming velocity, $\sigma$ the colloid diameter, and $D_r$
the rotational diffusion coefficient. Figure \ref{fig:prob_dens_abp}
presents the probability distribution $P(\phi_{l})$ of the local packing
fraction $\phi_{l}$ as a function of the average density $\phi$, where
$\phi_{l}$ is obtained by Voronoi construction \cite{rycroft2009chaos}.
Below a critical volume fraction $\phi_c \approx 0.3375$, the system is
essentially homogeneous and $P(\phi_{l})$ is unimodal. While approaching
$\phi_c$ with increasing $\phi$, $P(\phi_{l})$ broadens (by formation of
transient clusters) and becomes bimodal above $\phi_c$. The appearing
structures
are illustrated in Fig.~\ref{fig:isodensity_abp}. At low overall density,
transient spherical clusters are formed, which turn into a bicontinous structure,
whose surfaces is reminiscent of the Schwarz P surface, and finally at high
$\Phi$, gas-phase droplets float in a dense matrix \cite{wyso:14,sten:14}.

\begin{figure}
\centering
\includegraphics[width=0.42\textwidth]{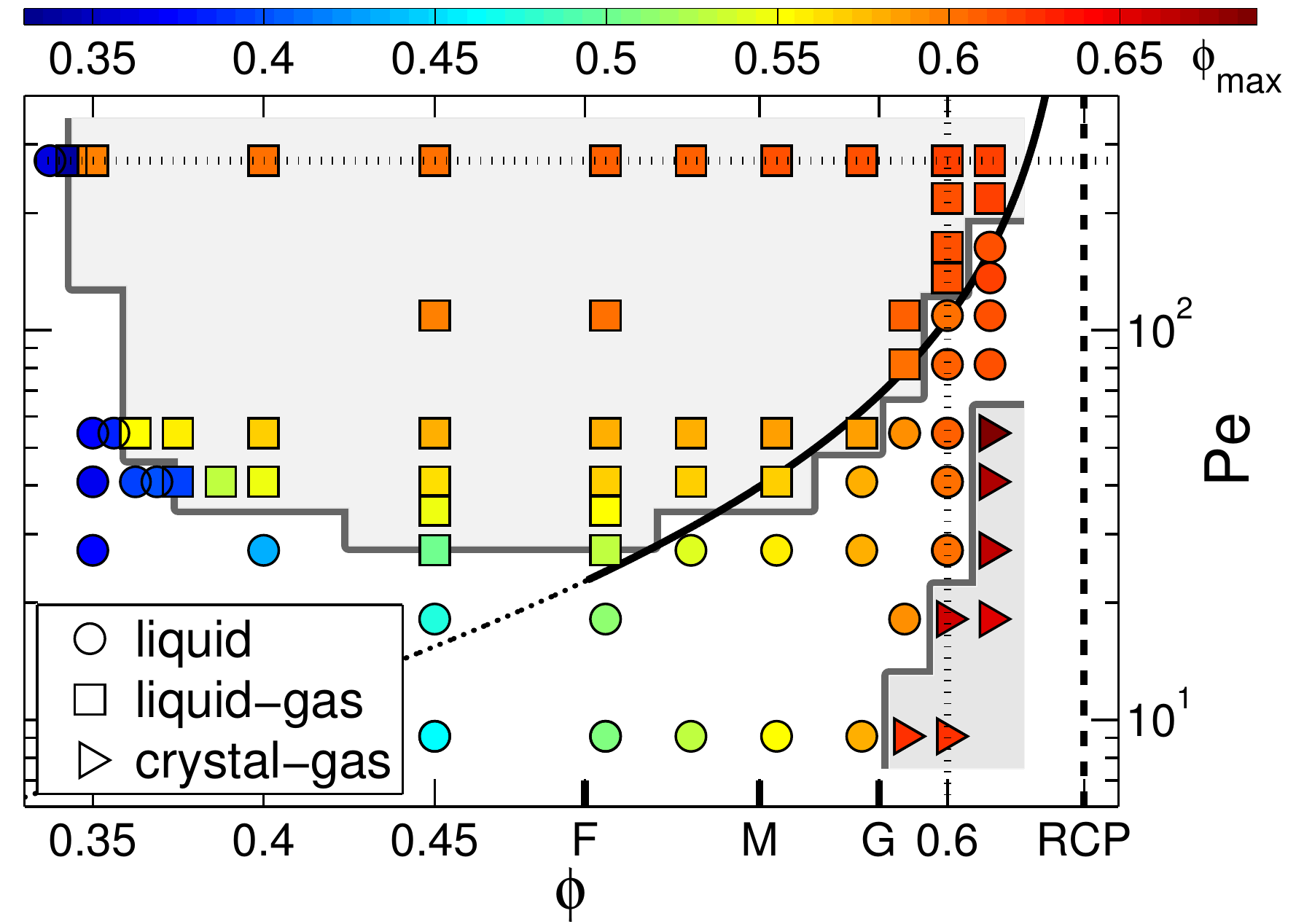}
\caption{Phase diagram of an suspension of active Brownian particles.
Symbols denote the homogenous liquid phase ($\circ$),
the gas-liquid coexistence ($\Box$) and the crystal-gas coexistence
($\triangleright$).
The equilibrium transition points of hard-spheres for freezing ($\phi_F=0.494$),
melting ($\phi_M=0.545$), glass-transition point ($\phi_G\approx0.58$), and
random close packing ($\phi_{RCP}\approx0.64$), are indicated
by F, M, G, RCP, respectively. The most probable $\phi_{l}$ is color-coded, i.e.,
in the two-phase region one can read off the density of the dense liquid phase $\phi_{liq}$.
The solid line marks $Pe_c(\phi)$ which is proportional to Eq.~(\ref{e:pressure}).
From \citet{wyso:14}.
}
\label{fig:phase_diagram_abp}
\end{figure}

The phase diagram for various P\'{e}clet numbers and densities is displayed in
Fig.~\ref{fig:phase_diagram_abp} \cite{wyso:14}. Similar results were obtained by
\citet{sten:14}. The gray line indicates
the spinodal line separating the two-phase region from the one-phase region.
Binodal lines are predicted by \citet{sten:14}. The existence of the spinodal
(and binodal) lines in the phase diagram are in analogy with equilibrium phase
diagrams, despite the non-equilibrium character of the present phase separation \cite{sten:14}.
\REV{\citet{spec:14} derived an effective Cahn-Hilliard equation for 
repulsive active Brownian particles on large length and time scales, 
which underlines the similarities in the phase behavior between passive 
and active systems.}

The phase separation at the higher densities can be understood as follows.
The pressure of a hard sphere fluid increases with $\phi$ and diverges at random close packing for the
metastable branch as 
\begin{align}\label{e:pressure}
p_{HS}=-\frac{6k_BT}{\pi\sigma^3}\phi^2\frac{\mathrm{d}}{\mathrm{d}\phi}
\ln{\left\{\left[\left(\phi_{RCP}/\phi\right)^{1/3}-1\right]^3\right\}},
\end{align}
according to free volume theory \cite{kamien2007prl}.
Self-propelled particles at low $\phi$ easily overcome this
pressure, coagulate due to their slow-down during collisions
(overdamped dynamics) and hence form clusters. The density within
the cluster, $\phi_{liq}$, adjusts such that $p_{HS}(\phi_{liq})$ balances
the active pressure $p_a$. An initially homogenous system can only
phase separate if the active pressure
$p_a\sim\gamma_tV_0/\sigma^2$ exceeds $p_{HS}(\phi)$, which leads to a
critical line $Pe_c(\phi)$ (spinodal) between the liquid-gas coexistence
and the homogenous liquid phase at high $\phi$,
(cf. Fig.~\ref{fig:phase_diagram_abp}); a similar argument has been put forward
by \citet{fily:14} in the context of soft disks.

Remarkably, the local packing fraction in the liquid phase can assume rather
large values, deep within the glassy region
($\phi_G\approx0.58\leq\phi_{liq}\leq\phi_{RCP}\approx0.64$) of
passive hard spheres \cite{brambilla2009prl,kamien2007prl,pusey1986nature}.
Nevertheless, particles remain mobile and no long-range crystalline order is
detected. This indicates a activity-induced shift of the glass-transition
point toward higher concentrations, as also discussed by \citet{ni:13},
\citet{fily:14}, and \citet{bert:13}.

\begin{figure}
\centering
\includegraphics[width=0.35\textwidth]{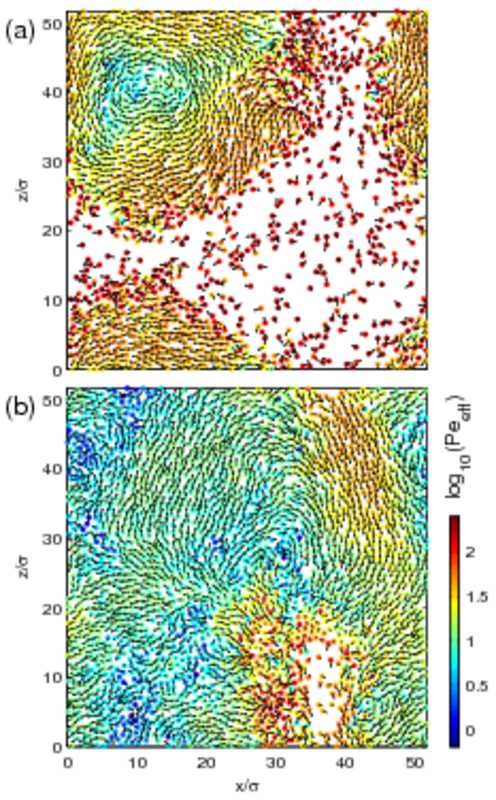}
\caption{Collective motion of active Brownian spheres, in the steady state
at $Pe=272$. Snapshots (slices of thickness $\sigma$) of a system
(a) just above the clustering transition ($\phi=0.3375$)
and (b) at a high concentration ($\phi=0.6$).
Arrows indicate the direction of the displacements over a
short lag time. The magnitude is color-coded
and is expressed by the effective P\'eclet number $Pe_{eff}$.
From \citet{wyso:14}.
}
\label{fig:dynamics_abp}
\end{figure}

Another remarkable aspect is the intriguing dynamics within the liquid phase
of the phase-separated system (cf. Fig.~\ref{fig:dynamics_abp}). Large-scale
coherent displacement patterns emerge, with amplitudes and directions strongly
varying spatially. In addition, transient swirl- and jet-like structures appear
frequently. This is in contrast to two-dimensional systems, which exhibit
far less coherent motions.
Figure~\ref{fig:dynamics_abp}(a) shows a swirl spanning the whole
cluster while Fig.~\ref{fig:dynamics_abp}(b) displays a large mobile
region moving between the gas-phase regions of the system. Thereby,
the fluid density is above the glass transition density
$\phi_{liq}\approx0.62>\phi_G\approx0.58$. Spatio-temporal correlation
functions for the colloid displacements indicate the existence of a universal
finite-size scaling function, which suggests that the system becomes scale
invariant with large-scale correlated fluid flow pattern in the asymptotic
limit of large systems \cite{wyso:14}. A similar behavior has been discussed
in other active systems, however, with a notable polar alignment mechanisms,
such as starling flocks \cite{cavagna2010pnas} and motile bacteria
colonies \cite{chen2012prl}.

Various analytical calculations have been performed to achieve a
theoretical understanding of the generic principles underneath the propulsion-induced phase transitions
\cite{sten:13,sten:14,fily:12,fily:14,bial:13,cate:13,tail:08,witt:14}. Adopting an
effective continuum theory with a density-dependent effective propulsion speed,
\citet{bial:13} predicted an instability region in the density driving-speed
diagram above a lower minimal velocity. A similar approach using the free
energy, yields a dynamical equation for the colloid packing fraction.
With suitable parameters, this equation qualitatively reproduces the
colloid density distributions (see Fig.~\ref{fig:isodensity_abp}) found
in simulations for two- and three-dimensional systems \cite{sten:14}.
\citet{fily:14} predicted the spinodal lines of their soft,
polydisperse active colloids in two dimensions. The success of this
approach suggest a remarkable analogy between non-equilibrium and
equilibrium phase transitions \cite{sten:14}.

\begin{figure}
\centering
\includegraphics[width=0.48\textwidth]{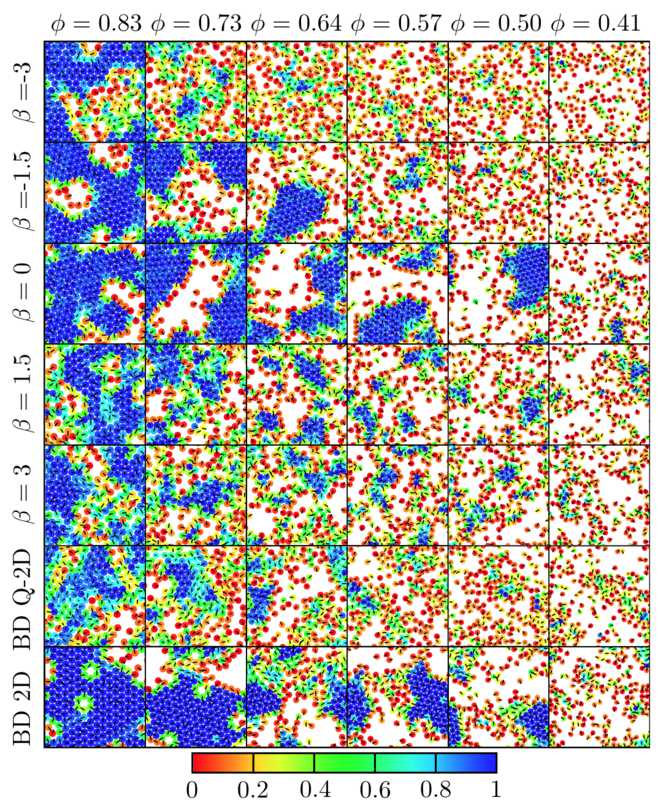}
\caption{\REV{Typical snapshots of the collective motion and aggregation 
of squirmers in a quasi-2D geometry, depending on the area fraction $\phi$ and 
the squirmer type ($\beta=B_2/B_1$, see Eq.~(\ref{eq:squirmvel}). 
Also shown are snapshots for active Brownian spheres moving in quasi-2D 
(BD Q-2D), and active Brownian disks moving in 2D (BD 2D). The colors indicate 
the local bond-orientational order. From \citet{zoet:14}.
}}
\label{fig:PD_ABP_hydro}
\end{figure}

\REV{Finally, we like to discuss the importance of hydrodynamic interactions 
for the phase behavior of spherical active colloids \cite{zoet:14,mata:14}. 
The study of two-dimensional systems of active discs suggest that 
aggregation is strongly suppressed by hydrodynamic interactions. 
On the other hand, a simulation study of a quasi-two-dimensional system 
of squirmer suspensions confined between two parallel walls, separated 
by a distance comparable to the swimmer size, indicates that hydrodynamic 
near-field interactions determine the phase behavior of active particles
\cite{zoet:14}. Near-field hydrodynamics implies an increase of the rotational 
diffusion, a slow-down of translational motion during collisions, and  
thereby leads to an enhanced self-trapping and the formation of crystalline 
clusters, see Fig.~\ref{fig:PD_ABP_hydro}. This indicates that dimensionality 
strongly affects the appearing structures in the presence of hydrodynamic 
interactions, as reflected by the significantly more ordered structures of 
active particles in a 2D system compared to a quasi-2D system. However, 
Fig.~\ref{fig:PD_ABP_hydro} suggests that hydrodynamic 
interactions {\em enhance} phase separation and structure formation compared to 
bare Brownian interactions, in particular for pullers.
} 

\subsection{Spermatozoa and Flagella}
\label{sec:cluster_sperm}

Experiments in recent years \cite{moor95,haya96,haya98,moor02,imml07,ried05}
have revealed an interesting swarm behavior of sperm at high concentration,
for example the distinctive aggregations or ``trains”
of hundreds of wood-mouse sperm \cite{moor02,imml07} or the vortex arrays of
swimming sea urchin sperm on a substrate \cite{ried05}. Thus, it is
interesting to study the clustering, aggregation, and vortex formation
of many sperm cells or flagella.

The results of Secs.~\ref{sec:attract} and \ref{sec:sperm_synchro} show
that when two sperm with the same beat frequency happen to get close
together and swim in parallel, they synchronize and attract through
hydrodynamic interactions. The collective behavior of sperm at finite
concentrations has mainly been studied numerically so far \cite{gg:gomp08l}.
Mesoscale simulations were performed for a two-dimensional system with a
number density of about three sperm per squared sperm length.
Considering that in real biological systems the beat frequency is
not necessary the same for all sperm, beat frequencies $\omega$
were selected from a  Gaussian distribution with variance
$\delta_f=\langle (\Delta \omega)^2 \rangle^{1/2}/\langle \omega \rangle$,
where $\langle (\Delta \omega)^2 \rangle$ is
the mean square deviation of the frequency distribution.

\begin{figure}
\centering
\includegraphics[width=0.3\textwidth]{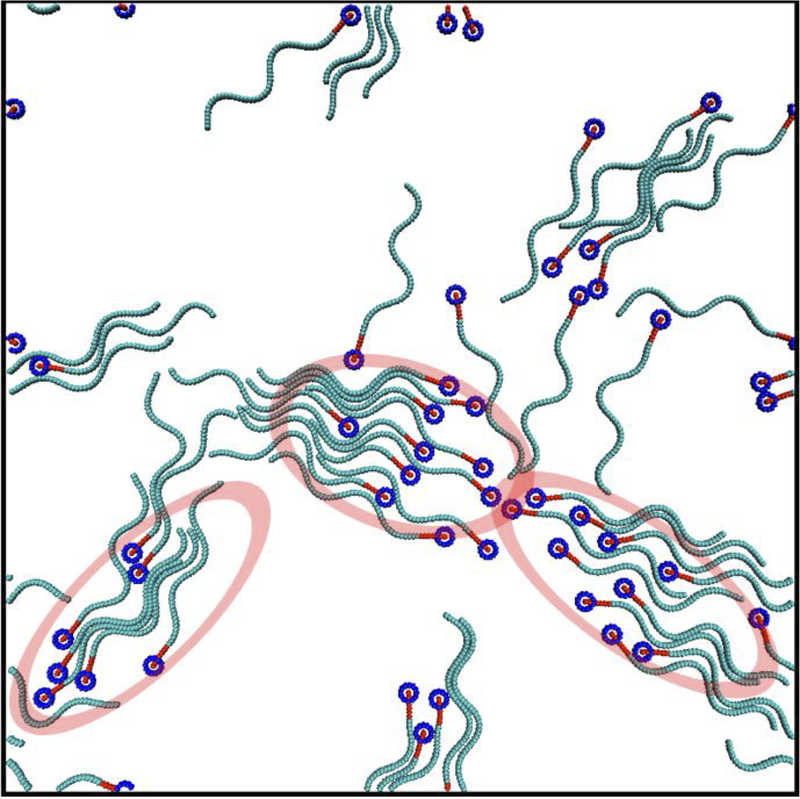}
\caption{Snapshot from simulations of 50 symmetric
sperm with width $\delta_f=0.9\%$ of a Gaussian distribution of
beating frequencies. The red ellipses indicate large
sperm clusters. The black frames show the simulation box.
Periodic boundary conditions are employed. From \citet{gg:gomp08l}.
}
\label{fig:multisperm}
\end{figure}

Fig.~\ref{fig:multisperm}
shows some snapshots of systems of symmetric sperm with different width
$\delta_f$ of the Gaussian frequency distribution.
For $\delta_f=0$, once a cluster has formed, it does not
disintegrate without a strong external force. A possible way of
break-up is by bumping head-on into another cluster.
For $\delta_f>0$, however, sperm cells can leave a cluster after
sufficiently long time, since the phase difference to other cells in
the cluster increases in time due to the different beat frequencies.
At the
same time, the cluster size can grow by collecting nearby free sperm or
by merging with other clusters. Thus, there is a balance between
cluster formation and break-up.
Obviously, the average cluster size
is smaller for large $\delta_f$ than for small $\delta_f$ (see
Fig.~\ref{fig:multisperm}).

\begin{figure}
\centering
\includegraphics[width=0.45\textwidth]{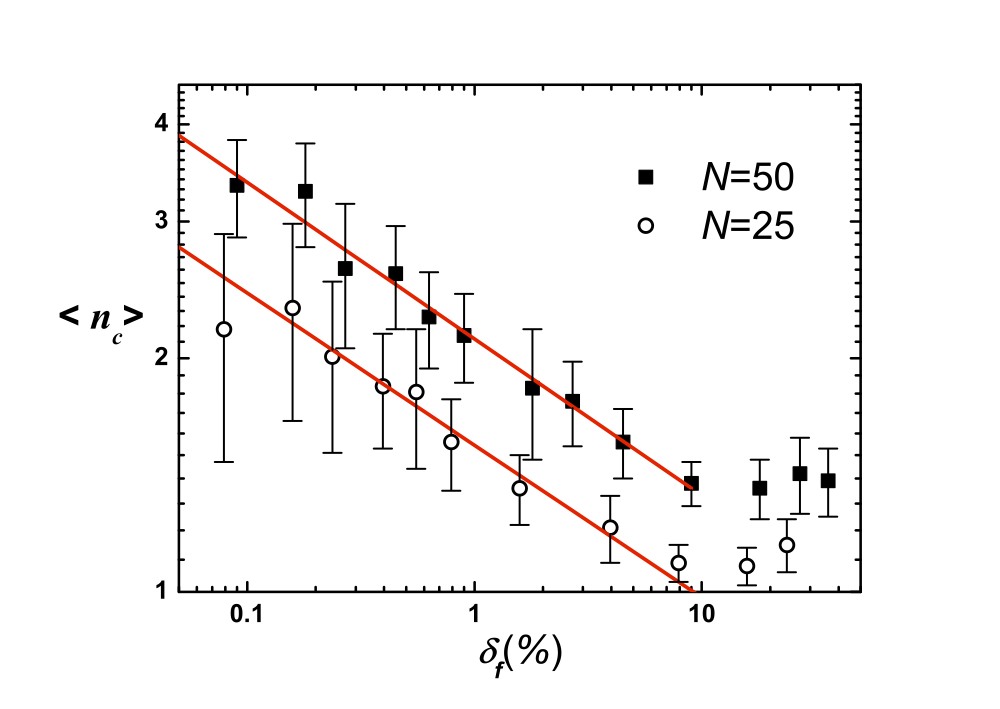}
\caption{Dependence of the average stationary cluster
size, $\langle n_c \rangle$, on the width of the frequency distribution
$\delta_f$. Data are shown for a 50-sperm system ($\blacksquare$)
and a 25-sperm system ($\circ$).  The lines indicate the power-law
decays $\langle rran_c \rangle=2.12\,\delta_f^{-0.201}$ (upper) and
$\langle n_c \rangle =1.55\,\delta_f^{-0.196}$ (lower). From \citet{gg:gomp08l}.}
\label{fig:sperm_clustersize}
\end{figure}

For
$\delta_f=0$, the average cluster size continues to increase with
time. Systems with $\delta_f>0$ reach
a stationary cluster size after about 50 beats.  The stationary
cluster size is plotted in Fig.~\ref{fig:sperm_clustersize} as a function of
the width $\delta_f$ of the frequency distribution. We find a decay
with a power law
\begin{equation}
\langle n_c \rangle \sim \delta_{f}^{-\gamma} ,
\end{equation}
where $\gamma=0.20\pm0.01$.  The negative
exponent indicates that the cluster size diverges when $\delta_f \to 0$.

Similar power-law behaviors of the cluster size as a function of the noise
level, or of the cluster-size distribution itself, have been found for
many systems of interacting microswimmers, from self-propelled
rods (see Sec.~\ref{sec:SPR}) to swimming flagella \cite{Yang2010}.
Such power laws reflect an underlying universal behavior of
self-propelled systems. Indeed, much simpler models with
a majority rule to align the velocity of a particle with its
neighbors predict such behavior \cite{visc95}.

\subsection{Sperm Vortices}
\label{sec:circle_swimmers}

\begin{figure}
\centering
\includegraphics[width=0.40\textwidth]{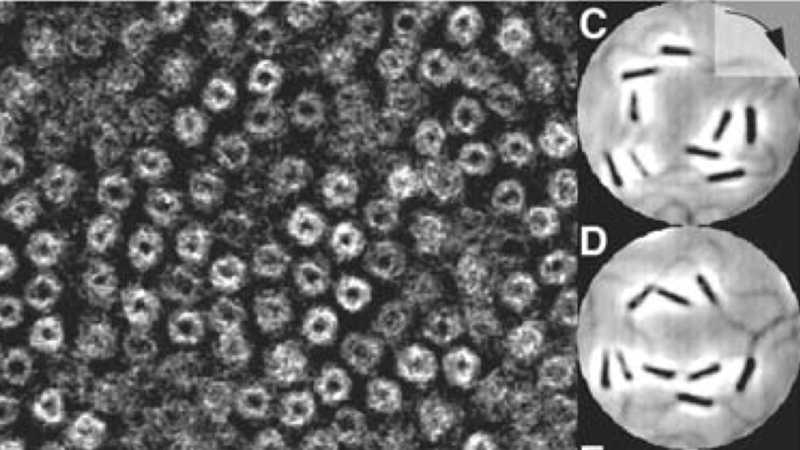}
\caption{Pattern of sperm vortices at a wall, which is observed
experimentally at high sperm densities. The inset shows the structure
of individual vortices, which consists of several sperm. The spatial
arrangement of the vortices has local hexagonal order.
From \citet{ried05}.
\label{fig:vortices_exp}
}
\end{figure}

Individual sperm cells of many species swim on helical trajectories in the bulk
and on circular trajectories at surfaces, as described in
Sec.~\ref{sec:sperm_hydro_wall}.  When the surface number density $\rho_0$ of
sperm increases beyond a value of about $2000/mm^2$, a new phenomenon is
observed \cite{ried05}, which is the formation of sperm vortices, in which
several sperm cells together form ring-like arrangements with a diameter $d_0$
of about $25 \mu m$. These vortices are themselves forming a fluid structure
with a local hexagonal order, see Fig.~\ref{fig:vortices_exp}.
The onset of this structure formation corresponds to a dimensionless surface
density of $\rho_0 d_0^2 \simeq 1$, which shows that this density is the
overlap density of circular trajectories. With increasing surface density,
the number of cells in each vortex increases.

\begin{figure}
\centering
\includegraphics[width=0.23\textwidth]{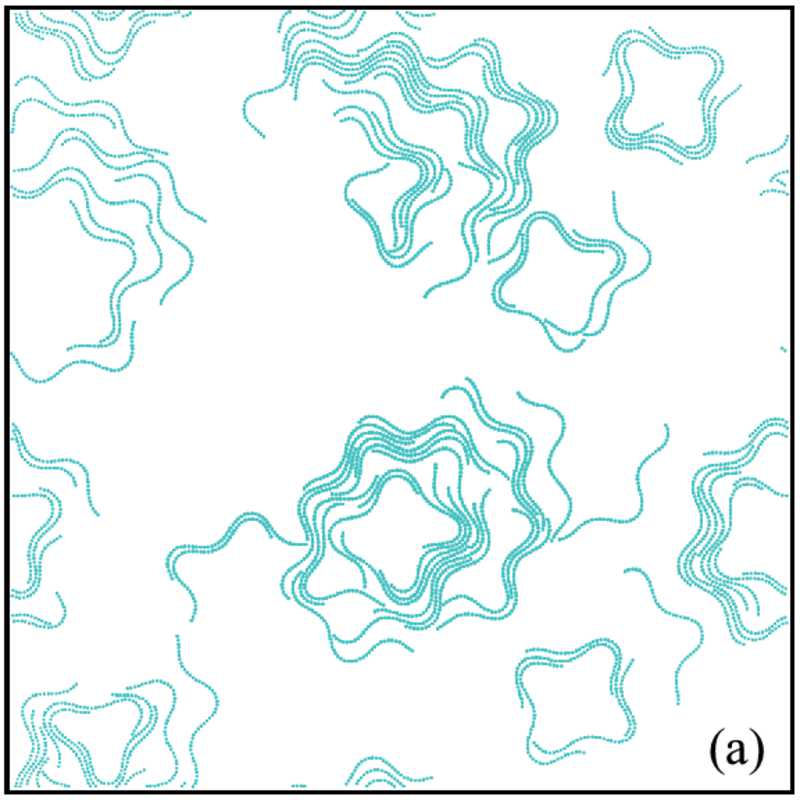}
\includegraphics[width=0.23\textwidth]{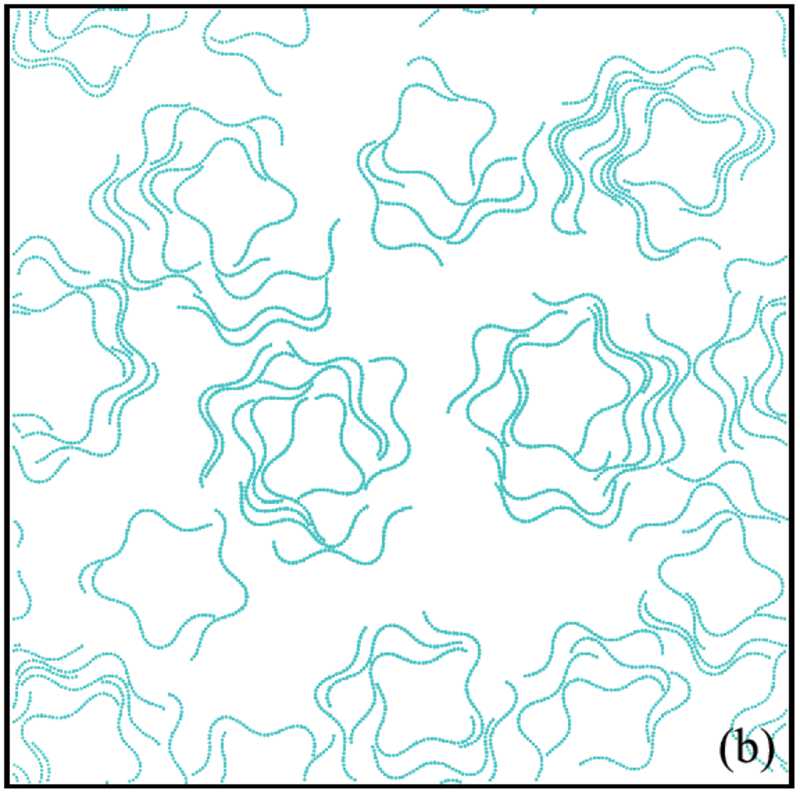}
\includegraphics[width=0.23\textwidth]{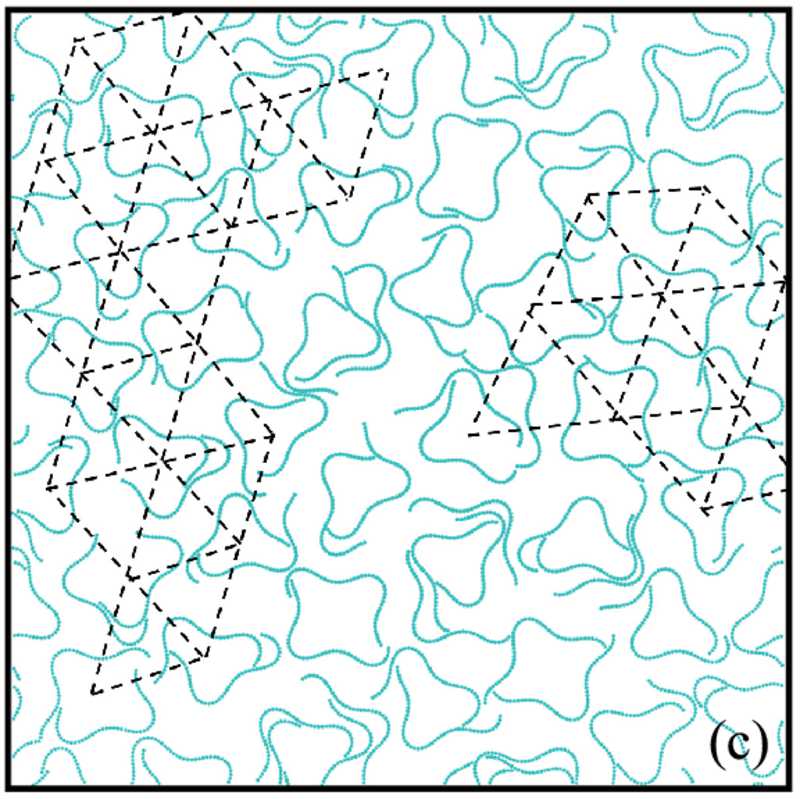}
\includegraphics[width=0.23\textwidth]{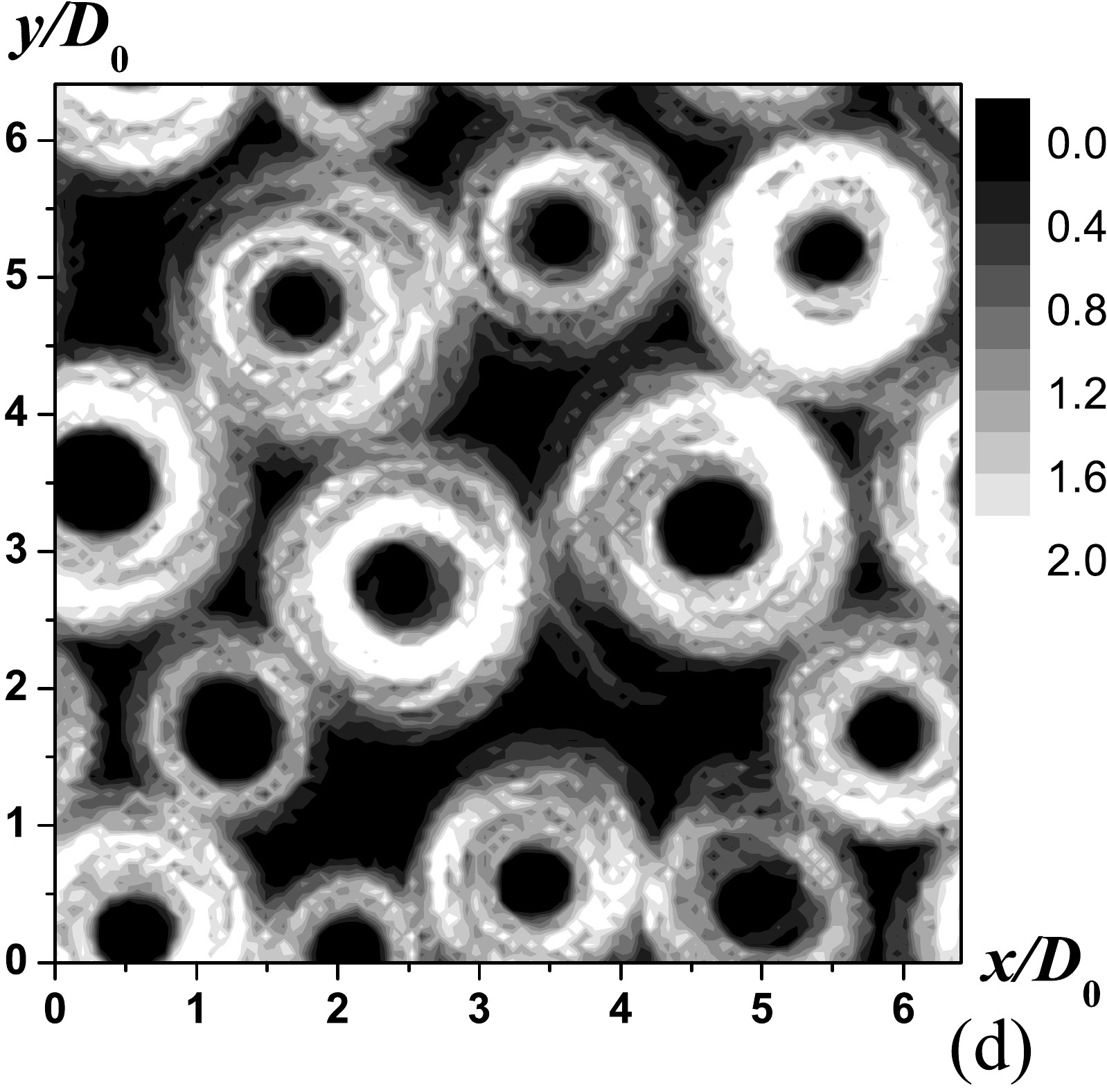}
\caption{
Snapshots of self-organized vortices of flagella (a) in a mesoscopic
fluid, and (b) with anisotropic friction, in both cases
with $d_0=0.614 L_{fl}$ and $\rho_0 d_0^2=2.36$,
and (c) with anisotropic friction, and $d_0=0.328 L_{fl}$ and
$\rho_0 d_0^2=0.67$. The dashed lines indicate the local hexagonal order.
Here, $L_{fl}$ is the length of the flagellum, $d_0$ the diameter of the
unperturbed circular motion, and $\rho_0$ the number density of flagella.
(d) Normalized flagellum density $\overline{\rho}_f({\bm r})$
averaged over a time interval of 30 flagellar beats of the system in (b).
From \citet{Yang14}.
\label{fig:sperm_vortices_sim}
}
\end{figure}

The study of this intriguing phenomenon by an extension of the three-dimensional
simulations for single sperm described in Sec.~\ref{sec:sperm_hydro_wall} is quite
difficult due to the large number of involved microswimmers. Therefore,
\citet{Yang14} have resorted to a two-dimensional description, in which each flagellum
is confined to the surface plane, and different flagella interact via excluded-volume
interactions. Furthermore, hydrodynamics is either taken into account by
resistive-force theory or by a two-dimensional mesoscale hydrodynamics approach.
Snapshots and averaged trajectories are shown in Fig.~\ref{fig:sperm_vortices_sim}.

\begin{figure}
\includegraphics[width=0.26\textwidth]{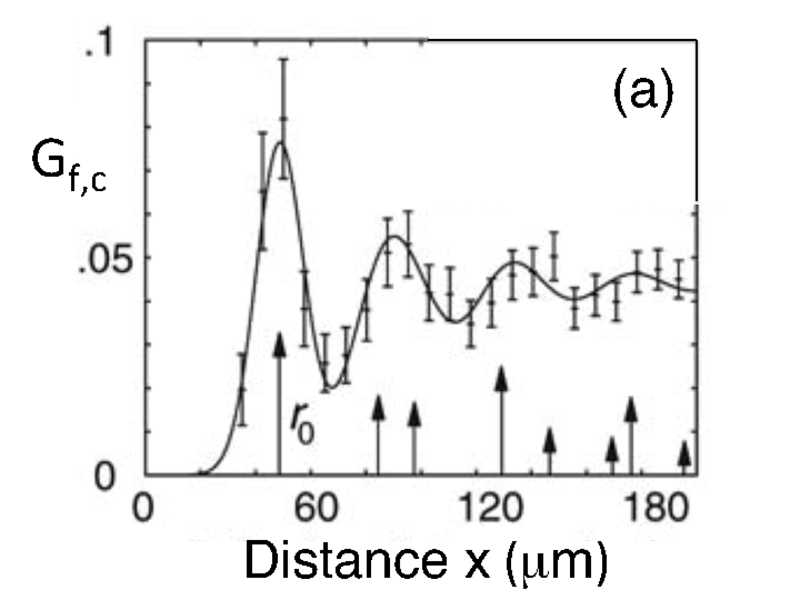}
\includegraphics[width=0.21\textwidth]{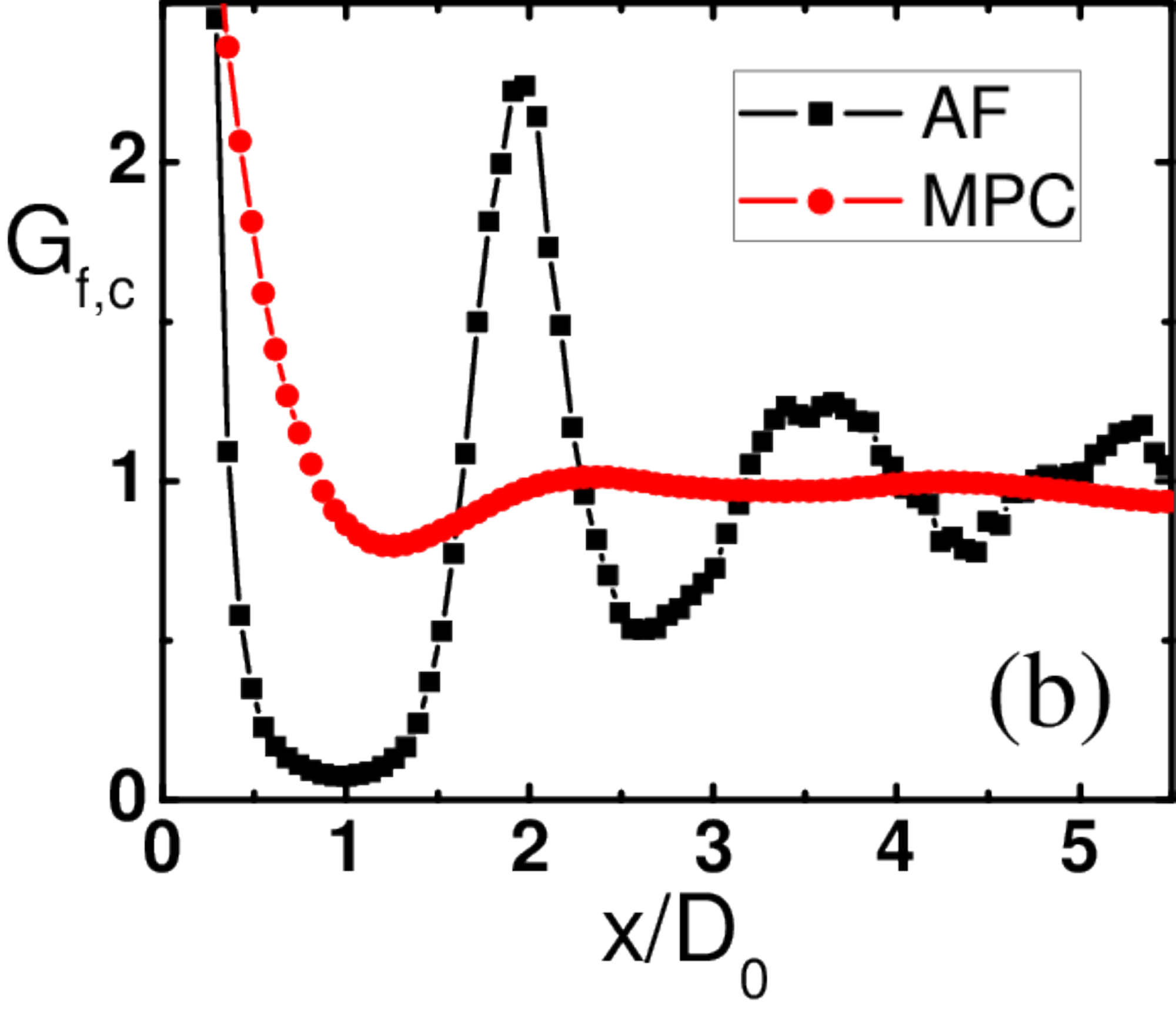}
\caption{
Pair correlation function of centers of swimming trails of circling sperm
near surfaces. (a) Experimental result for sperm number density of $6000/mm^2$,
corresponding to about $\rho_0 d_0^2=3.75$. Redrawn from \citet{ried05}.
(b) Simulation results for flagella for sperm density $\rho_0 d_0^2=2.36$,
both with (multi-particle collision dynamics, MPC) and without (anisotropic
friction, AF) two-dimensional hydrodynamic interactions.
Redrawn from \citet{Yang14}.
\label{fig:sperm_vortices_corr}
}
\end{figure}

Both, in experiments and simulations, the correlation function $G_{f,c}(r)$
of the instantaneous centers of the circular trajectories and the
variance $\Delta$ of the spatial distribution of these centers
have been determined \cite{ried05,Yang14}. A comparison of the results
leads to the following conclusions. First, the comparison of the simulation
results with and without hydrodynamic interactions reveals much weaker
correlations with hydrodynamic interactions, see
Fig.~\ref{fig:sperm_vortices_corr}b. The origin of this behavior is that
hydrodynamic interactions lead to the synchronization and attraction of
sperm swimming together, as discussed in Sec.~\ref{sec:sperm_synchro},
but also disrupt
the circular paths of sperm belonging to neighboring vortices, which move
in opposite directions. Second, since the agreement of the simulation results
with anisotropic friction only with experimental data is quite reasonable,
see Fig.~\ref{fig:sperm_vortices_corr}a,
we conclude that hydrodynamic interactions play a minor role in the
vortex formation. However, it should be noticed that both hydrodynamic
and excluded-volume interactions are stronger in two than in three
spatial dimensions. Therefore, it is conceivable that when the fluid
above the surface is taken into account, hydrodynamic interactions
contribute significantly to the stabilization of vortex arrays.
Third, the order parameter above the overlap concentration is found
in simulations and experiments to increase linearly with increasing
surface number density. This implies that the characteristic diameter of
a vortex hardly changes, but that its mass increases by increasing the
number of involved sperm or flagella.

A Vicsek-type model has been designed to capture the dynamics of structure
formation of ensembles of circle swimmers in two dimensions \cite{Yang14}.
Many properties are found to be very similar as for the explicit flagella model
described above. Thus, we expect a similar behavior for other circle
swimmers in two dimensions, such as the $L$-shapes colloidal microswimmers of
\citet{wens14}, or of other chiral microswimmers aggregating at surfaces.

\subsection{Swarming of Bacteria}
\label{sec:swarming_bacteria}

Flagellated microorganisms  exhibit collective behavior at a moist
surface or in a thin liquid film in form of swarming
\cite{cope:09,darn:10,kear:10,part:13}.
Swarming bacteria show a distinctively different motile behavior
than swimming cells \cite{hein:72,kear:10,darn:10}. Thereby
flagella are the most important
requirement, but cell-cell interactions also play a major role.
Chemotaxis is considered to be of minor importance for swarming
compared to physical interactions \cite{kear:10,part:13}.
During the transition from swimming to swarming cells, the number of flagella
increase and the cells become often more elongated by suppression
of cell division, i.e., cells become multi- or even hyperflagellated
\cite{stah:83,jone:04,kear:10,darn:10}. {\em E. coli} and
{\em Salmonella} bacteria double their  length and increase the
number of flagella, but the flagellar  density remains
approximately constant \cite{cope:10,kear:10,part:13,part:13.1}.
The changes for {\em P. mirabilis} are even more dramatic; their
length increases $10$ to $50$ times and an increase of  their
flagellum number from fewer than $10$ to $5,000$ is reported
\cite{cope:09,mcca:10,tuso:13}. As stated by \citet{kear:10},
neither is the reason known why swarming requires multiple
flagella on the cell surface nor is a significant cell elongation
required for many bacteria. Aside from a possible amplification of
swarming by shape-induced alignment of adjacent cells, elongation
associated with the increase in the number of flagella may help to
overcome surface friction \cite{part:13}.

The particular swarming behavior, specifically the role played by
flagella, seems to depend on the bacteria density and/or the
number of flagella as suggested by experiments on {\em E. coli}
\cite{turn:10,cope:10,darn:10} and {\em B. mirabilis} cells
\cite{stah:83,jone:04}. Studies of systems of lower density {\em
E. coli} bacteria reveal stable cohesive flagella bundles during
frequent collisions between cells \cite{cope:10}. Thereby, the bundles do
not need to be aligned with the body \cite{turn:10}. Moreover,
transient flagella bundles are formed with adjacent cells. For
higher density {\em B. mirabilis} systems, strong bundles between
adjacent cells are found.

Since the large-scale patterns in swarming bacteria colonies
\cite{cisn:07,wens12} are determined by cell-cell interactions,
it is of fundamental importance to unravel the interactions
between flagellated bacteria on the level of individual cells.


\section{Other Forms of Active Matter}

\subsection{Mixtures of Filaments and Motor Proteins --- Active Gels}

An active system, which shares many propeties with suspensions of
microswimmers, are mixtures of polar biological filaments with motor
proteins. The most prominent examples are actin-myosin and
microtubule-kinesin or microtubule-dynein systems. In both cases, the
filaments are polar, which
means that the motor proteins are walking only in one direction along
the filament. In the experiments \cite{nede97,surr01}, clusters of at least
two motorproteins are used, which can connect to two filaments
simultaneously. The motor proteins move along both filaments until they
reach the ends, where they either get stuck or drop off. This leads to
many interesting structures like asters, nematic phases, swirls, and
vortices.

On the theoretical side, mixtures of filaments and motor proteins
have been modeled both on the molecular and on the continuum level.
On the molecular level, semi-flexible polymers connected by motors with
attachment, detachment, and forward-stepping rates have been
considered \cite{surr01,Kruse2001,Head2011a,Head2011b,gord12,Head2014}.
On the continuum level, the starting point
is the nearly parallel arrangement of long passive filaments
in the nematic phase, for which a well-established liquid-crystal
description exists \cite{genn95}; additional contributions are then added
to this liquid-crytal model to capture the motor activity.
These hydrodynamic theories of active gels are equally suited to describe
suspensions of microswimmers.  Phenomena like swirls and vortices,
arrays of vortices and spontaneous collective orientation can all be
interpreted in the framework of these continuum
approaches \cite{Kruse2004,voit06,Elgeti2011,giom11,Fuerthauer2013}.
For a comprehensice discussion, we refer to the review articles by
\citet{tone05b}, \citet{Ramaswamy2010}, and \citet{marc:13}.

\subsection{Cellular Tissues}

Another interesting form of active matter are migrating cellular tissues.
A well-studied experimental model of such tissues are Madin-Darby
canine kidney (MDCK) cells, as explained in Sec.~\ref{sec:intro}.
Indeed, similar phenomena as in bacterial turbulence
(see Sec.~\ref{sec:SPR}) can be observed in
collectively migrating cellular tissues, however qualitatively
different phenomena also emerge. In particular, the adhesion between
cells and their active orientation strategy can lead to
novel behavior.

When plated on adhesive substrates, the cells move and grow.
Similar to the vortices observed in self-propelled rods or bacteria,
these cells display swirl structures even when already forming a
closed monolayer. As the density increases, this ``active liquid''
undergoes a glass-like transition, and eventually full dynamical arrest
\cite{Angelini2010,Angelini2011,Basan2013}.

A remarkable feature of growing cellular colonies is that they do not
spread by a pressure pushing cells outword, but rather as closed
monolayers under tension \cite{Trepat2009}. This tensile state can only be
understood if the motility forces against the substrate are considered.
Indeed a simple mechanism of alignment of the motility force with the
actual velocity can lead to tensile growing states \cite{Basan2013}.

Another feature emerging from the adhesion between the
``micro-crawlers'' is the formation of growth fingers. A classic experiment
is a wound-healing assay. Here, cells are plated
confluently around a rubber block. When the block is removed, cells
start to migrate into the newly available space. This leads to a
fingering instability with the formation of distinct leader cells
\cite{Petitjean2010,Poujade2007,Basan2013}.

Thus, while not microswimmers, motile epithelial tissues are
closely related in some aspects to swimming microorganisms. However, the
adhesion among themselves and with the substrate also creates novel
features not present in microswimmers.


\section{Summary and Conclusions}

In this review article, we have illustrated various physical aspects of locomotion of
microswimmers. After an overview of the basic propulsion concepts exploited by
biological cells and applied in synthetic systems, we have addressed low Reynolds number
hydrodynamics and its implication for the coordinated and concerted motion of
individual flagellated cells, and the interaction between cells and colloidal
microswimmers, respectively.  Alterations in the dynamical behavior due to restricting
surfaces have been illuminated along with physical mechanism for surface-capturing.
In addition,  the complex collective behavior of assemblies of microswimmers has been
addressed, which emerges from physical interactions such as hydrodynamics and
volume-exclusion interactions due to  shape anisotropies or even simply and foremost
by their propulsion.

Active systems comprise an exciting and broad range of phenomena. Not all of the aspects
could be addressed in this article. Some further characteristics are:

\begin{itemize}
\item Biological fluids are typically multicomponent systems containing polymers and
other colloid-like objects, which renders them viscoelastic rather than Newtonian.
The non-Newtonian environment strongly affects the behavior of the microswimmer  and
often leads to an enhanced swimming velocity
\REV{\cite{chau:79,fulf:98,laug09,fu09,elfr10,liu:11,zhu12,spag13,mino:11,rile:14}.}

By the presence of nonlinear constituitve equations, locomotion in complex fluids
overcomes limitations expressed by the scallop theorem \cite{laug09}. This is achieved
on the one hand by the  rate (velocity) dependence by the nonlinear evolution equations,
and on the other hand, by the nonlinear rheological properties of the fluid.

\item Since microswimmers in the absence of external forces are force- and torque-free,
their far-field flow profile is well described by a force dipole. There are however,
some notable exceptions, for example when a microswimmer is not neutrally buoyant or
when it is magnetically actuated. The far field of such a swimmers is usually dominated
by the Stokeslet.
Examples include Volvox algae, which are heavier than water, and thus react to gravity.
Their far field was indeed measured to be dominated by the Stockeslet \cite{dres10a}.
Paramecium is diamagnetic, and thus can be manipulated with magnetic fields,
causing external forces and torques \cite{Guevorkian2006a,Guevorkian2006}.

\item
Sedimentation in a gravitational field is also strongly affected by active propulsion
\cite{Tailleur2009,encu11}. While the density profile is still exponential, the
effective temperature is much larger than the real temperature \cite{Tailleur2009},
and the particles are polarized in the upwards direction \cite{encu11} --- even in
the absence of any hydrodynamic interactions.
\REV{Theoretical and experimental studies of $L$-shaped active colloids, i.e., of
particles with a chiral shape, demonstrate that the presence of a gravitational 
field leads to different classes of trajectories, from straight downward and 
upward motion (gravitaxis) to trochoid-like trajectories \cite{hage:14}. This 
suggests that gravitaxis of 
biological microswimmers opposite to the gravitational field can be entirely due
to the interplay of self-propulsion and shape asymmetry, and requires no 
active steering.}

\item Exposure to an external flow and confinement, such as in microfluid channels,
leads to particular, activity-induced effects.
A paradigmatic example is {\em E. coli} bacteria, which exhibit positive rheotaxis
in channel flow, i.e., a rapid and continuous upstream motility
\REV{\cite{hill:07,kaya:12,nash:10,cont:12,fu12,cost14}}.  Recent measurements 
for microalgae indicate
that they even swim along the vorticity direction in shear flow \cite{chen:13}.
\REV{Moreover, experiments indicate that flow gradients can lead to 
accumulation and layer formation of gyrotactic phytoplankton \cite{durh:09}. 
Strong shear flow can produce large spatial heterogeneities, characterized by 
cell depletion from low-shear regions due to ``trapping'' in high-shear regions 
\cite{rusc:14}. This impacts bacterial behavior by hampering chemotaxis and 
promoting surface attachment.}

\item The effective viscosity of suspensions of microswimmers in shear or extensional
flow depends on the swimming activity, and is therefore different from that of an
suspensions of passive particles
\cite{hatw04,soko09,rafa10,sain10,giom10,gyry11,muss13}.
Typically, it is found that the effective viscosity increases for pullers and
decreases for pushers.

\end{itemize}


Our understanding of the behavior of the ``classical" biological
microswimmers (such as sperm, bacteria and algae) dramatically advanced,
and many novel microswimmers have been designed
in recent years and decades. This opens a route for many new applications of
microswimmers , e.g., as microgears \cite{DiLeonardo2010}.
What is universal about the swimming behavior (and can thus be
described by the hydrodynamic far-field approximation), and what
is specific for a certain class of swimmers (like the synchronization
of different wave forms of flagella)? How can synthetic microswimmers
be designed to react to external stimuli and find their targets?
How can the ``biological complications" be incorporated into
theoretical models? The investigations of these and related questions are
challenging and exciting research topics in the future.

\vspace*{5mm}
\noindent {\bf  Acknowledgement:} \\[1ex]
With thank Masoud Abkenar, Luis Alvarez, Thorsten Auth, Ingo G\"otze,
Jinglei Hu, Jan Jikeli, U. Benjamin Kaupp, Kristian Marx, Shang Yik Reigh,
Marisol Ripoll, Mario Theers, Adam Wysocki, and Yingzi Yang
for enjoyable collaborations and stimulating discussions.
This work was supported in part by the VW Foundation (VolkswagenStiftung)
within the program {\em Computer Simulation of Molecular and Cellular
Bio-Systems as well as Complex Soft Matter}, 
\REV{by the German Science Foundation (DFG) within the priority
program {\em Microswimmers -- from single particle motion to collective
behavior} (SPP 1726)},  
and by the National Science Foundation under Grant No. NSF PHY11-2515.
GG thanks the Kavli Institute of Theoretical Physics (UCSB) for its
hospitality during the program ``Active Matter: Cytoskeleton, Cells,
Tissues and Flocks", where this work was nearly completed.



\newpage
\bibliography{microswimmer}

\end{document}